\documentclass[12pt,titlepage,letterpaper]{utarticle}

\usepackage{amsmath}
\usepackage{amsfonts}
\usepackage{amssymb}
\usepackage{amsthm}
\usepackage{mathtools}
\usepackage{graphicx}
\usepackage{color}
\usepackage{xparse}
\usepackage[utf8x]{inputenc}
\usepackage[titletoc,title]{appendix}
\usepackage{hyperref}
\usepackage{longtable}

\definecolor{aqua}{rgb}{0, 1.0, 1.0}
\definecolor{fuschia}{rgb}{1.0, 0, 1.0}
\definecolor{gray}{rgb}{0.502, 0.502, 0.502}
\definecolor{lime}{rgb}{0, 1.0, 0}
\definecolor{maroon}{rgb}{0.502, 0, 0}
\definecolor{navy}{rgb}{0, 0, 0.502}
\definecolor{olive}{rgb}{0.502, 0.502, 0}
\definecolor{purple}{rgb}{0.502, 0, 0.502}
\definecolor{silver}{rgb}{0.753, 0.753, 0.753}
\definecolor{teal}{rgb}{0, 0.502, 0.502}

\makeatletter
\newdimen\itex@wd%
\newdimen\itex@dp%
\newdimen\itex@thd%
\def\itexspace#1#2#3{\itex@wd=#3em%
\itex@wd=0.1\itex@wd%
\itex@dp=#2ex%
\itex@dp=0.1\itex@dp%
\itex@thd=#1ex%
\itex@thd=0.1\itex@thd%
\advance\itex@thd\the\itex@dp%
\makebox[\the\itex@wd]{\rule[-\the\itex@dp]{0cm}{\the\itex@thd}}}
\makeatother

\makeatletter
\newif\if@sup
\newtoks\@sups
\def\append@sup#1{\edef\act{\noexpand\@sups={\the\@sups #1}}\act}%
\def\reset@sup{\@supfalse\@sups={}}%
\def\mk@scripts#1#2{\if #2/ \if@sup ^{\the\@sups}\fi \else%
  \ifx #1_ \if@sup ^{\the\@sups}\reset@sup \fi {}_{#2}%
  \else \append@sup#2 \@suptrue \fi%
  \expandafter\mk@scripts\fi}
\def\tensor#1#2{\reset@sup#1\mk@scripts#2_/}
\def\multiscripts#1#2#3{\reset@sup{}\mk@scripts#1_/#2%
  \reset@sup\mk@scripts#3_/}
\makeatother

\makeatletter
\newbox\slashbox \setbox\slashbox=\hbox{$/$}
\def\itex@pslash#1{\setbox\@tempboxa=\hbox{$#1$}
  \@tempdima=0.5\wd\slashbox \advance\@tempdima 0.5\wd\@tempboxa
  \copy\slashbox \kern-\@tempdima \box\@tempboxa}
\def\slash{\protect\itex@pslash}
\makeatother

\def\clap#1{\hbox to 0pt{\hss#1\hss}}

\def\mathclap{\mathpalette\mathclapinternal}

\def\mathclapinternal#1#2{\clap{$\mathsurround=0pt#1{#2}$}}

\let\oldroot\root
\def\root#1#2{\oldroot #1 \of{#2}}
\renewcommand{\sqrt}[2][]{\oldroot #1 \of{#2}}

\DeclareSymbolFont{symbolsC}{U}{txsyc}{m}{n}
\SetSymbolFont{symbolsC}{bold}{U}{txsyc}{bx}{n}
\DeclareFontSubstitution{U}{txsyc}{m}{n}

\DeclareSymbolFont{stmry}{U}{stmry}{m}{n}
\SetSymbolFont{stmry}{bold}{U}{stmry}{b}{n}

\DeclareFontFamily{OMX}{MnSymbolE}{}
\DeclareSymbolFont{mnomx}{OMX}{MnSymbolE}{m}{n}
\SetSymbolFont{mnomx}{bold}{OMX}{MnSymbolE}{b}{n}
\DeclareFontShape{OMX}{MnSymbolE}{m}{n}{
    <-6>  MnSymbolE5
   <6-7>  MnSymbolE6
   <7-8>  MnSymbolE7
   <8-9>  MnSymbolE8
   <9-10> MnSymbolE9
  <10-12> MnSymbolE10
  <12->   MnSymbolE12}{}

\makeatletter
\def\re@DeclareMathSymbol#1#2#3#4{%
    \let#1=\undefined
    \DeclareMathSymbol{#1}{#2}{#3}{#4}}
\re@DeclareMathSymbol{\neArrow}{\mathrel}{symbolsC}{116}
\re@DeclareMathSymbol{\neArr}{\mathrel}{symbolsC}{116}
\re@DeclareMathSymbol{\seArrow}{\mathrel}{symbolsC}{117}
\re@DeclareMathSymbol{\seArr}{\mathrel}{symbolsC}{117}
\re@DeclareMathSymbol{\nwArrow}{\mathrel}{symbolsC}{118}
\re@DeclareMathSymbol{\nwArr}{\mathrel}{symbolsC}{118}
\re@DeclareMathSymbol{\swArrow}{\mathrel}{symbolsC}{119}
\re@DeclareMathSymbol{\swArr}{\mathrel}{symbolsC}{119}
\re@DeclareMathSymbol{\nequiv}{\mathrel}{symbolsC}{46}
\re@DeclareMathSymbol{\Perp}{\mathrel}{symbolsC}{121}
\re@DeclareMathSymbol{\Vbar}{\mathrel}{symbolsC}{121}
\re@DeclareMathSymbol{\sslash}{\mathrel}{stmry}{12}
\re@DeclareMathSymbol{\bigsqcap}{\mathop}{stmry}{"64}
\re@DeclareMathSymbol{\biginterleave}{\mathop}{stmry}{"6}
\re@DeclareMathSymbol{\invamp}{\mathrel}{symbolsC}{77}
\re@DeclareMathSymbol{\parr}{\mathrel}{symbolsC}{77}
\makeatother

\makeatletter
\def\Decl@Mn@Delim#1#2#3#4{%
  \if\relax\noexpand#1%
    \let#1\undefined
  \fi
  \DeclareMathDelimiter{#1}{#2}{#3}{#4}{#3}{#4}}
\def\Decl@Mn@Open#1#2#3{\Decl@Mn@Delim{#1}{\mathopen}{#2}{#3}}
\def\Decl@Mn@Close#1#2#3{\Decl@Mn@Delim{#1}{\mathclose}{#2}{#3}}
\Decl@Mn@Open{\llangle}{mnomx}{'164}
\Decl@Mn@Close{\rrangle}{mnomx}{'171}
\Decl@Mn@Open{\lmoustache}{mnomx}{'245}
\Decl@Mn@Close{\rmoustache}{mnomx}{'244}
\makeatother

\makeatletter
\DeclareRobustCommand\widecheck[1]{{\mathpalette\@widecheck{#1}}}
\def\@widecheck#1#2{%
    \setbox\z@\hbox{\m@th$#1#2$}%
    \setbox\tw@\hbox{\m@th$#1%
       \widehat{%
          \vrule\@width\z@\@height\ht\z@
          \vrule\@height\z@\@width\wd\z@}$}%
    \dp\tw@-\ht\z@
    \@tempdima\ht\z@ \advance\@tempdima2\ht\tw@ \divide\@tempdima\thr@@
    \setbox\tw@\hbox{%
       \raise\@tempdima\hbox{\scalebox{1}[-1]{\lower\@tempdima\box
\tw@}}}%
    {\ooalign{\box\tw@ \cr \box\z@}}}
\makeatother

\makeatletter
\NewDocumentCommand\mathraisebox{moom}{%
\IfNoValueTF{#2}{\def\@temp##1##2{\raisebox{#1}{$\m@th##1##2$}}}{%
\IfNoValueTF{#3}{\def\@temp##1##2{\raisebox{#1}[#2]{$\m@th##1##2$}}%
}{\def\@temp##1##2{\raisebox{#1}[#2][#3]{$\m@th##1##2$}}}}%
\mathpalette\@temp{#4}}
\makeatletter

\makeatletter
\def\udots{\mathinner{\mkern2mu\raise\p@\hbox{.}
\mkern2mu\raise4\p@\hbox{.}\mkern1mu
\raise7\p@\vbox{\kern7\p@\hbox{.}}\mkern1mu}}
\makeatother

\newcommand{\gt}{>}
\newcommand{\lt}{<}

\theoremstyle{plain}

\theoremstyle{definition}

\theoremstyle{remark}

\numberwithin{equation}{section}

\begin{document}

%-------------------------------------------------------------------

\preprint{
UTTG--24--13\\
TCC--020--13\\
ICTP--SAIFR/2013--011\\
}

\title{Tinkertoys for the Twisted D-Series}

\author{Oscar Chacaltana
    \address{
    ICTP South American Institute for\\ Fundamental Research,\\
    Instituto de F\'isica Te\'orica,\\Universidade Estadual Paulista,\\
    01140-070 S\~{a}o Paulo, SP, Brazil\\
    {~}\\
    \email{chacaltana@ift.unesp.br}\\
    },
    Jacques Distler ${}^\mathrm{b}$ and Anderson Trimm
     \address{
      Theory Group and\\
      Texas Cosmology Center\\
      Department of Physics,\\
      University of Texas at Austin,\\
      Austin, TX 78712, USA \\
      {~}\\
      \email{distler@golem.ph.utexas.edu}\\
      \email{atrimm@physics.utexas.edu}
      }
}
%\date{\today}
\date{September 9, 2013}

\Abstract{
We study 4D $\mathcal{N}=2$ superconformal field theories that arise from the compactification of 6D $\mathcal{N}=(2,0)$ theories of type $D_N$ on a Riemann surface, in the presence of punctures twisted by a $\mathbb{Z}_2$ outer automorphism. Unlike the untwisted case, the family of SCFTs is in general parametrized, not by $\mathcal{M}_{g,n}$, but by a branched cover thereof. The classification of these SCFTs is carried out explicitly in the case of the $D_4$ theory, in terms of three-punctured spheres and cylinders, and we provide tables of properties of twisted punctures for the $D_5$ and $D_6$ theories. We find realizations of $Spin(8)$ and $Spin(7)$ gauge theories with matter in all combinations of vector and spinor representations with vanishing $\beta$-function, as well as $Sp(3)$ gauge theories with matter in the 3-index traceless antisymmetric representation.
}

\maketitle

\thispagestyle{empty}
\tableofcontents
\vfill
\newpage
\setcounter{page}{1}

\section{Introduction}\label{introduction}

The study of four-dimensional $\mathcal{N}=2$ superconformal field theories (SCFTs) has benefited considerably in recent years from the construction of a class of such theories (sometimes called class S) as compactifications of the mysterious 6D (2,0) SCFTs on Riemann surfaces with a partial twist \cite{Gaiotto:2009we,Gaiotto:2009hg,Chacaltana:2010ks,Chacaltana:2011ze,Nanopoulos:2009uw,Nanopoulos:2010ga,Gaiotto:2011xs,Chacaltana:2012zy,Tachikawa:2009rb,Tachikawa:2010vg}. The realization of many Lagrangian theories whose Seiberg-Witten curves were previously unknown, the discovery of a multitude of interacting SCFTs that generalize the Minahan-Nemeschansky $E_N$ theories \cite{Minahan:1996fg,Minahan:1996cj}, and the understanding of S-duality \cite{Argyres:2007cn} are just a few of the remarkable features of this class of theories.

The key ingredient, greatly expanding the class of 4D theories one can obtain, is the possibility of adding codimension-two defects of the (2,0) theories localized at points on the Riemann surface $C$. Depending on our choice of these punctures on $C$, we get different 4D $\mathcal{N}=2$ SCFTs. A yet-wider class of theories can be obtained by including outer-automorphism twists \cite{Tachikawa:2009rb} on $C$, such that, when traversing an incontractible cycle on $C$ (either going around a handle of $C$, or circling a puncture on $C$) the ADE Lie algebra comes back to itself up to an outer-automorphism. In particular, this introduces a new class of codimension-two defects, which we refer to as ``twisted punctures'', and whose local properties were studied in \cite{Chacaltana:2012zy}.

In \cite{Chacaltana:2010ks}, we started our program of classifying the 4D N=2 SCFTs that arise from the 6D (2,0) theories by focusing on the $A_{N-1}$ series. In that paper, we constructed the possible ``fixtures'' (three-punctured spheres) and the cylinders that connect them, which are the basic building blocks for any pair-of-pants decomposition of a Riemann surface.  In \cite{Chacaltana:2011ze} we carried out a similar program for the $D_N$ theories, and in \cite{Chacaltana:2012ch} we studied the SCFTs that arise from incorporating outer-automorphism  twists in the $A_{2N-1}$ theories. In this paper, we want to continue our classification program by adding outer-automorphism twists to the theories of type $D_N$. Preliminary studies of the twisted $D_N$ series were made by Tachikawa in \cite{Tachikawa:2009rb,Tachikawa:2010vg}.

The $D_N$ Dynkin diagram is invariant under a $\mathbb{Z}_2$ outer automorphism group. Correspondingly, the possible twists are classified by giving an element $\gamma\in H^1(C-\{p_i\}, \mathbb{Z}_2)$. The forgetful map, which ``forgets'' the puncture, $p$, gives an inclusion $$
H^1(C-\{p_1,\dots\widehat{p},\dots\}, \mathbb{Z}_2)\hookrightarrow H^1(C-\{p_1,\dots p,\dots\}, \mathbb{Z}_2).
$$
If $\gamma$ descends to a nontrivial element of the quotient, $\tfrac{H^1(C-\{p_1,\dots p,\dots\}, \mathbb{Z}_2)}{H^1(C-\{p_1,\dots\widehat{p},\dots\}, \mathbb{Z}_2)}$, then we say that the puncture at $p$ is twisted (otherwise, untwisted). (For the $D_4$ theory, the $\mathbb{Z}_2$ enhances to a non-abelian $S_3$ group. The study of the 4D $\mathcal{N}=2$ SCFTs that arise from such enhancement is work in progress.)

For a given puncture, we explain how to compute all the local properties that contribute to determining the 4D $\mathcal{N}=2$ SCFT. Among these, are the contribution to the graded Coulomb branch dimensions, the global symmetry group, flavour-current central charges, the conformal-anomaly central charges $(a,c)$, and the ``pole structure'' and ``constraints'', which determine the contribution to the Seiberg-Witten curve. From this information, it is possible to determine gauge groups, hypermultiplet matter representations, and other properties.

As an application of our results, we are able to find realizations of $Spin(8)$ gauge theory with matter in the $6(8_v)$, or with matter in the $5(8_v) + 1(8_s)$. These two cases, of vanishing $\beta$-function for $Spin(8)$, were the ones that were not captured by the untwisted sector of the $D_N$ series. Similarly, for $Spin(7)$ gauge theory, we find the theory with matter in the $5(7)$, and in the $1(8)+4(7)$; the other combinations with vanishing $\beta$-function were already found in the untwisted sector of the $D_N$ series. We also study various realizations of $Sp(N)$ gauge theory, including $Sp(3)$ with matter in the $\tfrac{11}{2}(6)+\tfrac{1}{2}(14')$ and in the $3(6)+1(14')$, where the $14'$ is the 3-index traceless antisymmetric tensor representation.

\section{{The $\mathbb{Z}_2$-twisted $D_N$ Theory}}\label{the_twisted__theory}

The Coulomb branch geometry of the 4D $\mathcal{N}=2$ compactification \cite{Gaiotto:2009we,Gaiotto:2009hg} of the 6D $\mathcal{N}=(2,0)$ theories of type $D_N$ is governed by the Hitchin equations on $C$ with gauge algebra $\mathfrak{so}(2N)$. In particular, the Seiberg-Witten curve $\Sigma$ is a branched cover of $C$ described by the spectral curve \cite{Tachikawa:2009rb},

\begin{equation}
\Sigma: \det(\Phi - \lambda I) = \lambda^{2N}+\sum_{j=1}^{N-1} \phi_{2j}\lambda^{2N-2j} +\tilde{\phi}^2=0,
\label{swEquationSpectral}\end{equation}
where $\Phi$ is the $\mathfrak{so}(2N)$-valued Higgs field, while the $k$-differentials $\phi_{k}$ ($k=2,4,6,\dots,2N-2$) and the Pfaffian $N$-differential $\tilde{\phi}$ are associated with the Casimirs of the $D_N$ Lie algebra. In the rest of the paper, $N$ will always stand for the rank of $D_N$.

Introducing punctures on $C$ corresponds to imposing local boundary conditions on the Hitchin fields. We consider untwisted and twisted punctures under the action of the $\mathbb{Z}_2$ outer-automorphism group of the $\mathfrak{so}(2N)$ Lie algebra. Untwisted punctures are labeled by $\mathfrak{sl}(2)$ embeddings in $\mathfrak{so}(2N)$, or, equivalently, by nilpotent orbits in $\mathfrak{so}(2N)$, or by D-partitions\footnote{A D-partition of $2N$ is a partition of $2N$ where each even part appears with even multiplicity. However, ``very even" D-partitions --- those where all of the parts are even --- correspond to not one, but \emph{two}, nilpotent orbits. To distinguish between the two orbits, we assign a red or blue colour to the very-even Young diagrams.}  of $2N$. Instead of a compact curve, $C$, consider a semi-infinite cigar, with the puncture at the tip. Reducing along the circle action, we get 5D SYM on a half-space, with a Nahm-type boundary condition of the sort studied by Gaiotto and Witten in \cite{Gaiotto:2008ak}. For that reason, we call the D-partition that labels the untwisted puncture the \emph{Nahm pole}.

To describe the local Hitchin boundary condition for an untwisted puncture with Nahm-pole D-partition $p$, one must recall the Spaltenstein map\footnote{This Spaltenstein map consists in taking the ``D-collapse'' of the transpose of the D-partition. The D-collapse operation is explained in the untwisted D-series paper \cite{Chacaltana:2011ze}, as well as in the book \cite{CollingwoodMcgovern}.} , which takes $p$ into a new D-partition $d(p)$, called the \emph{Hitchin pole} of the puncture\footnote{When $p$ is \emph{non-special} (i.e., when it does not lie in the image of the Spaltenstein map), the information encoded in $d(p)$ must be supplemented by a nontrivial ``Sommers-Achar'' finite group, $C$, whose definition can be found in \cite{Chacaltana:2012zy}. This additional discrete information encodes the \emph{disconnected} part of the group of gauge transformations which we mod out by in constructing the solutions to the Hitchin system. In particular, it determines the presence (or absence) of the ``a-type" constraints, on the gauge-invariant $k$-differentials. This, in turn affects the local contributions to the graded Coulomb branch dimensions. In the Tables, we denote the Hitchin pole for non-special punctures as a pair $(d(p),C)$.} . Then, the local boundary condition corresponding to $p$ is

\begin{displaymath}
\Phi(z) = \frac{X}{z} + \mathfrak{so}(2N)
\end{displaymath}
where $X$ is an element of the nilpotent orbit\footnote{Using a nilpotent element $X$ in this equation amounts to writing the local boundary condition in the \emph{absence} of mass deformations. The mass-deformed boundary condition involves semisimple (diagonalizable) elements of $\mathfrak{so}(2N)$, whose eigenvalues take values in the Cartan subalgebra of the flavour Lie algebra for the puncture. For the untwisted $A$ series, a recipe for mass-deformed local boundary conditions was given in \cite{Nanopoulos:2009uw}. A general prescription is given in Sec. 2.4 of \cite{Chacaltana:2012zy}.}  associated to $d(p)$, and $\mathfrak{so}(2N)$ above denotes a generic regular function in $z$ valued in $\mathfrak{so}(2N)$.

On the other hand, we have a sector of twisted punctures, with monodromy given by the action of the nontrivial element $o$ of the $\mathbb{Z}_2$ outer automorphism group of $D_N$. The action of $o$ splits $\mathfrak{so}(2N)$ as

\begin{displaymath}
\mathfrak{so}(2N)=\mathfrak{so}(2N-1)\oplus o_{-1},
\end{displaymath}
where $\mathfrak{so}(2N-1)$ and $o_{-1}$ are the eigenspaces with eigenvalues +1 and -1, respectively. The action of $o$ on the $k$-differentials is also quite simple:

\begin{equation}
\begin{aligned}
o: \phi_{2k} & \mapsto \phi_{2k}\qquad (k=1,2,\dots, N-1)\\
\tilde{\phi} & \mapsto - \tilde{\phi}
\end{aligned}
\label{monodromyaction}\end{equation}
Following \cite{Chacaltana:2012zy}, the twisted punctures of the $D_N$ series are labeled by embeddings of $\mathfrak{sl}(2)$ in $\mathfrak{sp}(N-1)$ (the Langlands dual of $\mathfrak{so}(2N-1)$), or, equivalently, by nilpotent orbits in $\mathfrak{sp}(N-1)$, or by C-partitions\footnote{A C-partition of $2N$ is a partition of $2N$ where each odd part appears with even multiplicity. A B-partition of $2N-1$ is a partition of $2N-1$ where each even part appears with even multiplicity.}  of $2N-2$.

To describe the local boundary condition for a twisted puncture, we need to recall the relevant Spaltenstein map\footnote{This Spaltenstein map consists in adding a part ``1'' to a C-partition $p$, taking the transpose, and then doing a B-collapse. The result is always a B-partition. The ``B-collapse'' is discussed in \cite{Chacaltana:2012zy} and in \cite{CollingwoodMcgovern}.} . This is a map $d$ that takes a C-partition $p$ of $2N-2$ into a B-partition $d(p)$ of $2N-1$. A B-partition of $2N-1$ labels an $\mathfrak{sl}(2)$ embedding in $\mathfrak{so}(2N-1)$, or equivalently a nilpotent orbit in $\mathfrak{so}(2N-1)$. So, in our nomenclature, the Nahm pole $p$ of a twisted puncture is a C-partition of $2N-2$, and its Hitchin pole\footnote{Again, when the Nahm pole $p$ is non-special, the complete Hitchin pole information is not just $d(p)$, but a pair $(d(p),C)$, with $C$ the Sommers-Achar group \cite{Chacaltana:2012zy}.}  is a B-partition $d(p)$ of $2N-1$. The local boundary condition for the Higgs field is then:

\begin{displaymath}
\Phi(z) = \frac{X}{z} + \frac{o_{-1}}{z^{1/2}} + \mathfrak{so}(2N-1)
\end{displaymath}
Here $X$ is an element of the $\mathfrak{so}(2N-1)$ nilpotent orbit $d(p)$, while $o_{-1}$ and $\mathfrak{so}(2N-1)$ in the equation above denote generic regular functions in $z$ valued in these linear spaces, respectively.

\subsection{{Local Properties of Punctures}}\label{local_properties_of_punctures}

\subsubsection{{Global Symmetry Group and Central Charges}}\label{global_symmetry_group_and_central_charges}

The local properties of a puncture that we list in our tables are the pole structure (with constraints), the flavour group (with flavour-current central charges for each simple factor) and the contributions $(\delta n_h, \delta n_v)$ to, respectively, the effective number of hypermultiplets and vector multiplets (or, equivalently, to the conformal-anomaly central charges $(a,c)$). We will discuss how to compute pole structures and constraints in \S\ref{pole_structures},\ref{constraints}. Here we will briefly focus on the other properties.

Given the Nahm partition, for every part $l$, let its multiplicity  be $n_l$. Then, the flavour group of untwisted and twisted punctures are, respectively,

\begin{displaymath}
\begin{aligned}
G_{\text{flavour}}&=\prod_{l\;\; \text{even}} Sp\left(\tfrac{n_l}{2}\right)\times \prod_{l\;\; \text{odd}}SO(n_l)\qquad (\text{untwisted})\\
G_{\text{flavour}}&=\prod_{l\;\; \text{even}} SO(n_l)\times \prod_{l\;\; \text{odd}}Sp\left(\tfrac{n_l}{2}\right)\qquad (\text{twisted})
\end{aligned}
\end{displaymath}
The flavour-current central charges for each simple factor above can be computed using the formulas in Section 3 of \cite{Chacaltana:2012zy}. In that reference, one can also see how to compute $\delta n_h$ and $\delta n_v$. Instead of reviewing the general formulas, we find it more useful to discuss an example.

Consider the $D_6$ twisted puncture with Nahm pole C-partition $[3^2,1^4]$. The flavour group is $G_{\text{flavour}}=Sp(2)\times SU(2)$. To compute the central charges, we need to know how the adjoint representation of $Sp(5)$ decomposes under the subgroup $SU(2)\times G_{\text{flavour}}$ (the first factor being the embedding of $SU(2)$, corresponding to this partition). The C-partition itself tells us that the fundamental of $Sp(5)$ decomposes as $10=(1;4,1)+(3;1,2)$. The embedding indices of each factor of $SU(2)\times G_{\text{flavour}}=SU(2)\times Sp(2)\times SU(2)$ in $Sp(5)$ are 8,1 and 3, respectively. With this information, it is not hard to see that the adjoint representation of $Sp(5)$ decomposes as

\begin{equation}
55=(1;10,1)+(1;1,3)+(3;1,1)+(3;4,2)+(5;1,3).
\label{decompexample}\end{equation}
Now, to find $\delta n_h$ and $\delta n_v$, we use eq. (3.19) of \cite{Chacaltana:2012zy}. In the notation of that paper, we have $\mathfrak{j}=\mathfrak{so}(6)$, $\mathfrak{g}=\mathfrak{sp}(5)$, and, in their respective usual root bases, the Weyl vectors $\rho_{Spin(6)}=(5,4,3,2,1,0,0,0,0,0,0,0)$, $\rho_{Sp(5)}=(5,4,3,2,1,0,0,0,0,0)$. We also find $h/2=(1,1,0,0,0,-1,-1,0,0,0)$ using, say, the formulas of Section 5.3 of \cite{CollingwoodMcgovern}. Since the adjoint representation of $Sp(5)$ decomposes under the Nahm-pole $SU(2)$ as $55=13(1)+9(3)+3(5)$, we have $\dim\;\mathfrak{g}_0=13+9+3=25$ and $\dim\;\mathfrak{g}_{1/2}=0$. Thus, eq. (3.19) of \cite{Chacaltana:2012zy} yields $\delta n_h=368$ and $\delta n_v=\tfrac{717}{2}$.

Finally, from \eqref{decompexample} above as well as eq. (3.20) of \cite{Chacaltana:2012zy}, we compute the flavour-current central charges for each simple factor of $G_{\text{flavour}}$,

\begin{displaymath}
\begin{aligned}
k_{Sp(2)} &= 1\times l_{Sp(2)}(10) + 2\times l_{Sp(2)}(4) =8\\
k_{SU(2)} &= 1\times l_{SU(2)}(3)+1\times l_{SU(2)}(3)+4\times l_{SU(2)}(2) =12
\end{aligned}
\end{displaymath}
where $l_{\mathfrak{h}}(R)$ denotes the index of the representation $R$ of $\mathfrak{h}$.

\subsubsection{{Pole Structures}}\label{pole_structures}

The pole structure of a puncture is the set of leading pole orders $\{p_2,p_4,p_6,\dots,p_{2N-2};\tilde{p}\}$ in the expansion of the $k$-differentials $\phi_k(z)$ ($k=2,4,6,\dots,2N-2$) and the Pfaffian $\tilde{\phi}(z)$ around the position of the puncture on $C$. Knowing the pole structures of the various punctures allows us to write down the Seiberg-Witten curve \eqref{swEquationSpectral} of a theory. The pole orders are all integers, except for $\tilde{p}$ in a twisted puncture, which must be a half-integer because of the monodromy \eqref{monodromyaction}.

We already saw in \cite{Chacaltana:2011ze} how to read off the pole structure of an untwisted puncture from its Hitchin-pole D-partition $p$. Basically, regard $p$ as a partition in the untwisted A-series, use the procedure to write down the pole structure \cite{Chacaltana:2010ks}, and discard the pole orders that would correspond to $\phi_k$ with odd $k$. Finally, divide the pole order $p_{2N}$ of $\phi_{2N}$ by two, to obtain the pole order $\tilde{p}$ of the Pfaffian $\tilde{\phi}$. $p_{2N}$ will always be even, so that $\tilde{p}$ will come out to be an integer, as expected for an untwisted puncture.

To compute the pole structure of a twisted puncture, we use its Hitchin B-partition $p$. Simply, add 1 to the first (i.e., the largest) part in $p$, and use the same procedure to compute the pole structure as for an untwisted D-series puncture. Notice that upon adding 1 to the largest part, the B-partition becomes a partition of $2N$, and one can show that the pole order $p_{2N}$ of $\phi_{2N}$ is always odd, so that the pole order $\tilde{p}$ of the Pfaffian is a half-integer, as it should be.

For instance, consider the $D_6$ twisted puncture with Nahm-pole C-partition $[4^2,1^2]$. The Hitchin B-partition is $[5,2^2,1^2]$. Following our prescription, we add 1 to the largest part, so we get $[6,2^2,1^2]$, and read off the pole structure as in the untwisted A-series. We thus get $\{1,2,3,4,5,5,6,6,7,7,7\}$ (corresponding to scaling dimensions $2,3,4,\dots, 11,12$). We discard the pole orders at odd dimensions, and divide the pole order of $\phi_{12}=\tilde{\phi}^2$ by two, and we are left with the correct pole structure, $\{1,3,5,6,7;\tfrac{7}{2}\}$.

\subsubsection{{Constraints}}\label{constraints}

In the untwisted D-series, punctures featured ``constraints'', which are either: 1) relations among leading coefficients in the $k$-differentials (``c-constraints''); or 2) expressions defining \emph{new} parameters $a^{(k)}$ of scaling dimension $k$ as, roughly, the square roots of a leading coefficient $c^{(2k)}$ of dimension $2k$ (``a-constraints''). Both kinds of constraints affect the counting of graded Coulomb branch dimensions of the theory, as well as the Seiberg-Witten curve. As expected, we find a-constraints and c-constraints also in the twisted sector. The pole structure and the constraints provide a ``fingerprint'' \cite{Gukov:2008sn} that allows us to identify the puncture uniquely.

Let us briefly review our nomenclature. For a puncture at $z=0$, we consider the coefficients $c^{(2k)}_l$ and $\tilde{c}_l$ of the leading singularities in the expansion in $z$ of the $2k$-differentials ($2k=2,4,\dots, 2N-2$) and the Pfaffian $\tilde{\phi}$, respectively,

\begin{displaymath}
\begin{aligned}
\phi_{2k}(z) &= \tfrac{c^{(2k)}_l}{z^l}+\dots\\
\tilde{\phi}(z) &= \tfrac{\tilde{c}_l}{z^l}+\dots
\end{aligned}
\end{displaymath}
where $\dots$ denotes less singular terms. (The pole orders $l$ above are, of course, the same as those in the pole structure, so we have $l=p_{2k}$ or $l=\tilde{p}$, respectively; in this subsection we just write $l$ to keep expressions simple.)

An \emph{a-constraint} of scaling dimension $2k$ is an expression linear in $c^{(2k)}_l$ that defines (up to sign) a new parameter $a^{(k)}_{l/2}$ of dimension $k$,

\begin{displaymath}
c^{(2k)}_l = \left(a^{(k)}_{l/2}\right)^2 + \dots,
\end{displaymath}
where $\dots$ stands for a polynomial in leading coefficients (of dimension less than $2k$) as well as new coefficients $a^{(j')}_{l'}$ (which would themselves be defined by other a-constraints). This polynomial is homogeneous in dimension and pole order, i.e., in every term in the polynomial, the sum of the scaling dimensions of every factor must be $2k$, and the sum of pole orders must be $l$. The existence of an $a$-constraint implies that, in counting graded Coulomb branch dimensions, a parameter of scaling dimension $2k$ is to be replaced by one of dimension $k$.

A \emph{c-constraint} of dimension $2k$ is an expression linear in $c^{(2k)}_l$, which relates it to other leading coefficients, and perhaps also to new parameters $a^{j}_l$ defined by a-constraints,

\begin{displaymath}
c^{(2k)}_l = \dots
\end{displaymath}
where, again, the ellipsis denotes a homogeneous polynomial in leading coefficients and new parameters. For even $N$, if the puncture is very-even, a ``very-even'' c-constraint, which is linear in the leading coefficients of both $\phi_N$ and the Pfaffian, may appear,

\begin{displaymath}
c^{(N)}_l \pm 2\tilde{c}_l= \dots
\end{displaymath}
Unlike an a-constraint, a c-constraint does not define any new parameters; it simply tells us that $c^{(2k)}_l$ (or, say, $c^{(N)}$ for a very-even c-constraint) is not independent, and so it should not be considered when counting Coulomb branch dimensions.

Finally, at every scaling dimension $2k$, we find at most one constraint, which can be either an a-constraint or a c-constraint.

Below, we present algorithms to compute the scaling dimensions $2k$ at which a-constraints and c-constraints appear for a given puncture. This information is enough to compute the local contribution to the graded Coulomb branch dimensions.

\paragraph{\underline{Untwisted punctures}}\label{untwisted_punctures}

Let $p$ be the Nahm pole D-partition of an untwisted puncture. Also, let $q=\{q_1,q_2,\dots\}$ be the transpose partition, and $s=\{s_1,s_2,\dots\}$ the sequence of partial sums of $q$ ($s_i=q_1+q_2+\dots+q_i$). Below, $s_1$ denotes the first element of $s$, and $p_{\text{last}}$, the last element of the D-partition $p$. (By the conditions that define a D-partition, $s_1$ is always an even number.)

Then, an a-constraint of dimension $2k$ exists if the following conditions are met:

\begin{enumerate}%
\item $2k$ belongs to $s$, say, $s_j=2k$.
\item $j$ is even.
\item If $s_j$ is a multiple of $s_1$, say, $s_j=rs_1$, one has $r\geq 2\left\lfloor\tfrac{p_{\text{last}}}{2}\right\rfloor+1$.
\item $s_j$ is not the last element of $s$.
\end{enumerate}

On the other hand, a c-constraint of scaling dimension $2k$ exists if the following conditions are met:

\begin{enumerate}%
\item $2k$ belongs to $s$, say, $s_j=2k$.
\item If $j$ is even, one has that: a) $s_j$ is a multiple of $s_1$, say, $s_j=rs_1$; b) $\left\lfloor\tfrac{p_{\text{last}}}{2}\right\rfloor+1\leq r\leq 2\left\lfloor\tfrac{p_{\text{last}}}{2}\right\rfloor$; c) $s_j$ is not the last element of $s$.
\item If $j$ is odd, one has that: a) $s_j$ is neither the first nor the last element of $s$; b) both $s_{j-1}$ and $s_{j+1}$ are even; c) $s_j=\tfrac{s_{j-1}+s_{j+1}}{2}$ ; d) if $s_j$ is divisible by $s_1$, say, $s_j=rs_1$, one has $r\geq \left\lfloor\tfrac{p_{\text{last}}}{2}\right\rfloor+1$.
\end{enumerate}

Finally, if $p$ is very even, an additional, ``very-even'', c-constraint exists at $2k=N$ if $N$ belongs to $s$ and $N=\tfrac{s_1p_{\text{last}}}{2}$. As already mentioned, this very-even c-constraint is linear in \emph{both} leading coefficients $c^{(N)}_l$ and $\tilde{c}_l$. (The pole orders of $\phi_N$ and $\tilde{\phi}$ are the same if the conditions just mentioned hold, so such a linear constraint is possible.) A generic very-even puncture may or may not have this very-even c-constraint. In particular, a very-even puncture could have a c-constraint of dimension $N$ which is not very even (in the sense that it is not linear in both $c^{(N)}_l$ and $\tilde{c}_l$).

\paragraph{\underline{Twisted punctures}}\label{twisted_punctures}

Suppose we have a twisted puncture labeled by the Nahm-pole C-partition $p$. Let $q$ be the transpose partition, and $s$ the sequence of partial sums of $q$. It is convenient to define another sequence $s'$, obtained by adding 2 to every element in $s$. (As a check, the last element of $s'$ must be $2N$.) Let $s'=\{s'_1,s'_2,\dots\}$.

Then, an a-constraint of scaling dimension $2k$ exists if the following conditions are met:

\begin{enumerate}%
\item $2k$ belongs to $s'$, say, $s'_j=2k$.
\item $j$ is odd.
\item $s'_j$ is not the last element of $s'$.
\end{enumerate}

On the other hand, a c-constraint of scaling dimension $2k$ exists if the following conditions are met:

\begin{enumerate}%
\item $2k$ belongs to $s'$, say, $s'_j=2k$.
\item $j$ is even.
\item $s'_j$ is not the last element of $s'$
\item Both $s'_{j-1}$ and $s'_{j+1}$ are even, and $s'_j=\tfrac{s'_{j-1}+s'_{j+1}}{2}$.
\end{enumerate}

\paragraph{\underline{Constraint structure}}\label{constraint_structure}

The constraints of twisted punctures are very simple. c-constraints are always ``cross-terms'' between a-constraints, or between an a-constraint and the Pfaffian (where $\phi_{2N}=\tilde{\phi}^2$ is seen as another ``a-constraint''). As a schematic example, $c^{(k+m)}$ below is a cross-term for the ``squares'' at dimensions $2k$ and $2m$:

\begin{equation}
%\begin{aligned}
c^{(2k)}=\left(a^{(k)}\right)^2,\qquad
c^{(k+m)}=2a^{(k)}a^{(m)},\qquad
c^{(2m)}=\left(a^{(k)}\right)^2
%\end{aligned}
\label{crossTerms}\end{equation}
(In an actual example, $k+m$ would always turn out to be even). a-constraints also generically contain cross-terms, in addition to the quadratic term in the new parameter. Many examples can be found in the Tables.

The constraints of untwisted punctures are slightly more complicated, but they resemble very much the constraints of twisted punctures in the $A_{2N-1}$ series \cite{Chacaltana:2012ch}, so we refrain from repeating the details. To be brief, there is a sequence of c-constraints (illustrated below in an example), all related to each other, and which are associated to the first terms in the set of partial sums $s$. c-constraints outside this sequence are simply cross-terms between a-constraints and/or the Pfaffian, as in \eqref{crossTerms}. For a very-even puncture, the very-even c-constraint, if it exists, becomes part of the sequence just mentioned. As usual, a-constraints can include cross-terms in addition to the quadratic term that defines the new parameter.

Let us discuss the constraints of a $D_6$ very-even puncture, $[6^2]$. In this case, $q=[2^6]$ and $s=[2,4,6,8,10,12]$. Also, $p_{\text{last}}=6$ and $s_1=2$. So, there are c-constraints at $2k=rs_1$ with $4\leq r\leq 6$, that is, at $2k=8,10$. There is also a very-even c-constraint (at $2k=6$). All c-constraints in this case constitute the sequence mentioned in the previous paragraph. There are no a-constraints. We can also compute the pole structure to be $\{1,2,3,4,5;6\}$. Let us see the structure of these c-constraints by writing:
\begin{displaymath}
\begin{aligned}
c^{(0)}_0 & = 1,&\qquad
c^{(8)}_4 &=\tfrac{1}{4}\left(t^{(4)}_2\right)^2 + \tfrac{1}{2}t^{(6)}_3t^{(2)}_1,\\
c^{(2)}_1 &\equiv t^{(2)}_1,&\qquad
c^{(10)}_5 &=t^{(6)}_3 t^{(4)}_2,\\
c^{(4)}_2 &\equiv \tfrac{1}{4} \left(t^{(2)}_1\right)^2 + t^{(4)}_2,&\qquad
c^{(12)}_6 &\equiv \left(\tilde{c}^{(6)}_3\right)^2 = 
\tfrac{1}{4}\left(t^{(6)}_3\right)^2.\\
c^{(6)}_3 &\equiv\tfrac{1}{2}t^{(2)}_1 t^{(4)}_2 + t^{(6)}_3,&&
\end{aligned}
\end{displaymath}
The first line above is trivial, but it facilitates the construction of the other expressions. Disregarding the very-even c-constraint at $2k=6$ for a moment, the expressions at $2k=2,4,6$ provide definitions for the quantities $t^{(2)}_1$, $t^{(4)}_2$ and $t^{(6)}_3$. Besides, each term in the equations above can be interpreted as either a cross-term or a square of 1, $t^{(2)}_1$, $t^{(4)}_2$ and $t^{(6)}_3$. For example, the term $t^{(2)}_1$ is not a square, so it has to be a cross-term (for 1 and $\tfrac{1}{4}\left(t^{(2)}_1\right)^2$), which is why we include the term $\tfrac{1}{4}\left(t^{(2)}_1\right)^2$ in $c^{(4)}_2$. Since $c^{(4)}_2$ cannot be equal to $\tfrac{1}{4}\left(t^{(2)}_1\right)^2$ (since that would be a c-constraint at $k=4$), we introduce the new quantity $t^{(4)}_2$. Notice that we have also written $c^{(12)}_6$ as a square of $t^{(6)}_3$. Since $\phi_{12}$ is the square of the Pfaffian, we must have $t^{(6)}_3=\pm 2\tilde{c}_3$, and we recover the very-even constraint at $2k=6$. Solving for $t^{(2)}_1$ and $t^{(4)}_2$, we find our actual c-constraints:

\begin{displaymath}
\begin{aligned}
c^{(6)}_3 \mp 2\tilde{c}_3 &= \tfrac{1}{2}c^{(2)}_1\left(c^{(4)}_2 - \tfrac{1}{4}\left(c^{(2)}_1\right)^2\right),\\
c^{(8)}_4 &= \tfrac{1}{4}\left(c^{(4)}_2 - \tfrac{1}{4}\left(c^{(2)}_1\right)^2\right)^2 \pm  \tilde{c}_4 c^{(2)}_1,\\
c^{(10)}_5 &= \pm \tilde{c}_3 \left(c^{(4)}_2 - \tfrac{1}{4}\left(c^{(2)}_1\right)^2\right).
\end{aligned}
\end{displaymath}
Flipping the sign of $\tilde{c}_3$ switches between the constraints for the red and the blue versions of this puncture.

\subsection{{Collisions}}\label{collisions}

When two punctures collide, a new puncture appears. This process can be described at the level of the Higgs field, using the local boundary conditions discussed in \S\ref{the_twisted__theory}, or at the level of the $k$-differentials, using the pole structures and the constraints of \S\ref{pole_structures} and \S\ref{constraints}. Of course, both mechanisms are quite related, because the $k$-differentials are, essentially, the trace invariants of the Higgs field. These procedures are analogous to those for the twisted $A_{2N-1}$ series described in \cite{Chacaltana:2012ch}.

Let us start by discussing collisions using the Higgs field. Consider two untwisted punctures at $z=0$ and $z=x$ on a plane. The respective local boundary conditions are:
\begin{displaymath}
\begin{aligned}
\Phi(z) &= \frac{X_1}{z} + \mathfrak{so}(2N),\\
\Phi(z) &= \frac{X_2}{z-x} + \mathfrak{so}(2N),
\end{aligned}
\end{displaymath}
where $X_1$ and $X_2$ are representatives of the respective Hitchin-pole orbits for the punctures. Then, in the collision limit, $x\to 0$, a new untwisted puncture appears at $z=0$,

\begin{displaymath}
\Phi(z) = \frac{X_1+X_2}{z} + \mathfrak{so}(2N).
\end{displaymath}
Here, $X_1+X_2$ is an element of the mass-deformed Hitchin-pole orbit for the new puncture, and the mass deformations correspond to the VEVs of the decoupled gauge group. Taking the mass deformations to vanish, $X_1+X_2$ becomes the Hitchin-pole nilpotent orbit for the new puncture. The fact that the new residue is $X_1+X_2$ also follows from the residue theorem applied to the three-punctured sphere that appears in the degeneration limit; another derivation ensues from an explicit ansatz for the Higgs field on the plane with two punctures \cite{Chacaltana:2012ch}, where the limit $x\to 0$ can be taken.

Now consider an untwisted and a twisted puncture, at $z=0$ and $z=x$, respectively. The respective local boundary conditions are:
\begin{displaymath}
\begin{aligned}
\Phi(z) &= \frac{X}{z} + \mathfrak{so}(2N),\\
\Phi(z) &= \frac{Y}{z-x} + \frac{o_{-1}}{(z-x)^{1/2}} + \mathfrak{so}(2N-1).
\end{aligned}
\end{displaymath}
Then, the local boundary condition for the new twisted puncture is:

\begin{displaymath}
\Phi(z) = \frac{X|_{\mathfrak{so}(2N-1)} +Y}{z} + \frac{o_{-1}}{z^{1/2}} + \mathfrak{so}(2N-1),
\end{displaymath}
where $X|_{\mathfrak{so}(2N-1)}$ is the restriction of $X\in \mathfrak{so}(2N)$ to the subalgebra $\mathfrak{so}(2N-1)$.

Finally, consider two twisted punctures at $z=0$ and $z=x$,

\begin{displaymath}
\begin{aligned}
\Phi(z) &= \frac{Y_1}{z} + \frac{o_{-1}}{z^{1/2}} + \mathfrak{so}(2N-1),\\
\Phi(z) &= \frac{Y_2}{z-x} + \frac{o_{-1}}{(z-x)^{1/2}} + \mathfrak{so}(2N-1).
\end{aligned}
\end{displaymath}
Then, the local boundary condition for the new untwisted puncture is:

\begin{displaymath}
\Phi(z) = \frac{Y_1+Y_2 + o_{-1}}{z} + \mathfrak{so}(2N),
\end{displaymath}
where $o_{-1}$ denotes a generic element in such space.

The procedure to collide punctures using $k$-differentials is explained in \cite{Chacaltana:2012ch} for the case of the twisted  $A_{2N-1}$ series. The discussion is entirely analogous, so we leave the details to that paper. Here we will just give an example of how to use it.

Consider the collision of three punctures,
\begin{displaymath}
[2(N-r)-1,2r+1] \times [2(N-r)-1,2r+1] \times [2(N-1)],
\end{displaymath}
which yields the $[2(N-2r-1),1^{4r}]$ puncture with an $Sp(r)\times Sp(r)$ gauge group. We will use this result in \S\ref{atypical_degenerations_and_ramification}. Let us show how to derive it for the particular case $r=3$.

The puncture $[2N-7,7]$ has pole structure $\{1,2,3,4,5,6,6,6,\dots,6;3\}$, no a-constraints, and three c-constraints at $2k=8,10,12$:

\begin{equation}
\begin{aligned}
c^{(8)}_4 =& \tfrac{1}{4}\left(c^{(4)}_2-\tfrac{1}{4}\left(c^{(2)}_1\right)^2\right)^2 + \tfrac{1}{2}c^{(2)}_1\left(c^{(6)}_3-\tfrac{1}{2}c^{(2)}_1\left(c^{(4)}_2-\tfrac{1}{4}\left(c^{(2)}_1\right)^2\right)\right),\\
c^{(10)}_5 =& \tfrac{1}{2}\left(c^{(4)}_2-\tfrac{1}{4}\left(c^{(2)}_1\right)^2\right)\left(c^{(6)}_3-\tfrac{1}{2}c^{(2)}_1\left(c^{(4)}_2-\tfrac{1}{4}\left(c^{(2)}_1\right)^2\right)\right),\\
c^{(12)}_6 =&\tfrac{1}{4}\left(c^{(6)}_3-\tfrac{1}{2}c^{(2)}_1\left(c^{(4)}_2-\tfrac{1}{4}\left(c^{(2)}_1\right)^2\right)\right)^2.
\end{aligned}
\label{cConstraintsExample}\end{equation}
On the other hand, the puncture $[2(N-1)]$, which is the ``minimal'' twisted puncture, has pole structure $\{1,1,1,\dots,1;\tfrac{1}{2}\}$, and no constraints.

First, consider two $[2N-7,7]$ punctures on the plane, at positions $z=0$ and $z=x$, and write down the $k$-differentials:

\begin{displaymath}
\begin{aligned}
\phi_{2k}(z) &= \frac{u_{2k}+v_{2k}z+\dots}{z^k(z-x)^k}\,\,\, & &(2k=2,4,6,8,10,12)\\
\phi_{2k}(z) &= \frac{u_{2k}+\dots}{z^6(z-x)^6} & &(2k=14,16,\dots,2N-2)\\
\tilde{\phi}(z) &= \frac{\tilde{u}+\dots}{z^3(z-x)^3} & &
\end{aligned}
\end{displaymath}
Then, in the $x\to 0$ limit, which corresponds to the collision, we find the pole orders $\{2,4,6,8,10,12,12,12,\dots,12;6\}$. So, at first sight, we would have gauge-group Casimirs at $2k=2,4,6,8,10,12$. However, the c-constraints \eqref{cConstraintsExample} from the two $[2N-7,7]$ punctures imply that the leading and subleading coefficients $u_{2k}$ and $v_{2k}$ for $2k=8,10,12$ are dependent on the coefficients $u_2,u_4,u_6$, and furthermore vanish when we take $u_2,u_4,u_6\to 0$. Thus, the only independent gauge-group Casimirs are $u_2,u_4,u_6$, and the massless puncture has pole structure $\{1,3,5,6,8,10,12,12,\dots,12;6\}$, with no constraints. These properties single out the puncture $[2N-13,2^6,1]$, which has $Sp(3)$ flavour symmetry. Thus, the gauge group must be $Sp(3)$.

Colliding the new puncture $[2N-13,2^6,1]$ with the minimal twisted puncture is much easier, because none is constrained. So all we need to do is add up pole orders, and identify gauge-group Casimirs. The sum of the pole structures is $\{2,4,6,7,9,11,13,13,\dots,13;\tfrac{13}{2}\}$. Hence, we have again a gauge group with Casimirs $2,4,6$, and a new puncture with pole structure $\{1,3,5,7,9,11,13,13,\dots,13;\tfrac{13}{2}\}$, with no constraints. These properties correspond to the puncture $[2N-14,1^{12}]$, which has flavour symmetry $Sp(6)$. Thus, we are gauging an $Sp(3)$ gauge group out of the $Sp(6)$. Actually, since the two new punctures we find in the subsequent collisions are not maximal, it must be that an $Sp(3)\times Sp(3)$ subgroup (each factor from each of the two cylinders) of $Sp(6)$ is being gauged. (We saw multiple examples of this phenomenon in \cite{Chacaltana:2011ze,Chacaltana:2012ch}.)

Let us derive the same result by doing the collisions in a different order: first, we collide a $[2N-7,7]$ puncture (at $z=0$) with the minimal twisted puncture (at $z=x$). We use the $k$-differentials\footnote{In this subsection, we use generic names for Coulomb branch parameters such as $u_{2k}, v_{2k}, r_k$, etc. They are understood to be different variables in different collisions.}

\begin{displaymath}
\begin{aligned}
\phi_{2k}(z) &= \frac{u_{2k}+\dots}{z^k(z-x)}\,\,\, & &(2k=2,4,6,8,10,12),\\
\phi_{2k}(z) &= \frac{u_{2k}+\dots}{z^6(z-x)}\qquad & &(2k=14,16,\dots,2N-2),\\
\tilde{\phi}(z) &= \frac{\tilde{u}+\dots}{z^3(z-x)^{1/2}} & &
\end{aligned}
\end{displaymath}
This time, solving the c-constraints is less simple. The constraints are not solvable unless one introduces parameters $r_2,r_4,r_6$ of dimension 2,4,6 such that:

\begin{displaymath}
%\begin{aligned}
u_2=r_2 x^{1/2},\qquad
u_4=-\tfrac{(r_2)^2}{4}+r_4 x^{1/2},\qquad
u_6=-r_2r_4 + r_6 x^{1/2}
%\end{aligned}
\end{displaymath}
(See Sec. 4.1.3 of \cite{Chacaltana:2012ch} for a similar example in more detail.) Then, the constraints imply:

\begin{displaymath}
%\begin{aligned}
u_8=-\tfrac{1}{4}((r_4)^2+2r_2r_6),\qquad
u_{10}=-\tfrac{1}{2}r_6r_4,\qquad
u_{12}=-\tfrac{1}{4}(r_6)^2
%\end{aligned}
\end{displaymath}
and in the limit $x\to 0$, we get a pole structure $\{1,3,4,5,6,7,7,7,\dots,\tfrac{7}{2}\}$, with constraints

\begin{displaymath}
\begin{aligned}
c^{(4)}_3&=-\tfrac{(r_2)^2}{4},\qquad \qquad\qquad& c^{(10)}_6&=-\tfrac{r_4r_6}{2}\\
c^{(6)}_4&=-r_2r_4,\qquad & c^{(12)}_7&=-\tfrac{(r_6)^2}{4}\\
c^{(8)}_5&=-(r_4)^2-\tfrac{r_2r_6}{2} & &\\
\end{aligned}
\end{displaymath}
that is, we have a-constraints at $2k=4,8,12$ and c-constraints at $2k=6,10$. These properties uniquely identify the twisted puncture $[2N-8,6]$. Notice that there are no gauge-group Casimirs, so our interpretation is that the cylinder is ``empty''. This is an example of an ``atypical degeneration'', as we will recall in \S\ref{atypical_degenerations_and_ramification}.

Let us now collide the new puncture $[2N-8,6]$ (at $z=0$) with the remaining untwisted puncture $[2N-7,7]$ (at $z=x$). We have the $k$-differentials

\begin{displaymath}
\begin{aligned}
\phi_2(z)&=\frac{u_2+\dots}{z(z-x)} &  &\\
\phi_{2k}(z)&=\frac{xu_{2k}+v_{2k} z+\dots}{z^{k+1}(z-x)^{k}}\qquad& &(2k=4,6,8,10,12)\\
\phi_{2k}(z)&=\frac{u_{2k}+\dots}{z^7 (z-x)^6}\qquad & &(2k=14,16,18,\dots,2N-2)\\
\tilde{\phi}(z)&=\frac{\tilde{u}+\dots}{z^{7/2}(z-x)^3}& &
\end{aligned}
\end{displaymath}
Taking the collision limit $x\to 0$, we get the pole orders $\{2,4,6,8,10,12,13,13,\dots,13;\tfrac{13}{2}\}$. So, in principle, the gauge-group VEVs are $u_2,v_4,v_6,v_8,v_{10}, v_{12}$. However, $v_8,v_{10},v_{12}$  are polynomials in $u_2,v_4,v_6$ and in three new parameters $r_2,r_4,r_6$, of respective dimensions 2,4,6, which arise from combining the a-/c-constraints of $[2N-8,6]$ with the c-constraints of $[2N-7,7]$. So the actual gauge-group VEVs are $u_2,v_4,v_6,r_2,r_4,r_6$. These VEV dimensions are consistent with an $Sp(3)\times Sp(3)$ gauge group, as before, except that now both $Sp(3)$ factors are supported on a single cylinder. Setting to zero the gauge-group VEVs, we get the massless pole orders $\{1,3,5,7,9,11,13,13,\dots,13;\tfrac{13}{2}\}$, with no constraints, which, as before, correspond to the $[2N-14,1^{12}]$ puncture.

\subsection{{Gauge Couplings}}\label{gauge_couplings}

Consider an $\mathcal{N}=2$ supersymmetric gauge theory, with simple gauge group, $G$, and matter content chosen so that the $\beta$-function vanishes. This gives rise to a family of SCFTs, parametrized by

\begin{displaymath}
\tau = \frac{\theta}{\pi} + \frac{8\pi i}{g^2}
\end{displaymath}
A rich class of (though not all) such theories can be realized as compactifications of the $(2,0)$ theory on a sphere with four \emph{untwisted} punctures. If the four puncture are distinct, then the S-duality group, $\Gamma(2)\subset PSL(2,\mathbb{Z})$, is generated by

\begin{displaymath}
T^2:\,\tau\mapsto \tau+2,\quad S T^2 S:\,\tau\mapsto \frac{\tau}{1-2\tau}
\end{displaymath}
The fundamental domain for $\Gamma(2)$ is isomorphic to $\mathcal{M}_{0,4}\simeq \mathbb{C}\mathrm{P}^1$. In particular, the coordinate on the complex plane, $f$, is given by\footnote{Our $\theta$-function conventions are

\begin{displaymath}
\begin{split}
\theta_2(0,\tau)&= \sum_{n\in\mathbb{Z}} q^{(n+1/2)^2 /2}\\
\theta_3(0,\tau)&= \sum_{n\in\mathbb{Z}} q^{n^2 /2}\\
\theta_4(0,\tau)&= \sum_{n\in\mathbb{Z}} {(-1)}^n q^{n^2 /2}\\
\end{split}
\end{displaymath}
where $q=e^{2\pi i \tau}$.} 

\begin{displaymath}
\begin{split}
   f(\tau) &= -\frac{\theta_2^4(0,\tau)}{\theta_4^4(0,\tau)}\\
           &= -\left(16 q^{1/2} + 128 q +704 q^{3/2}+\dots\right)
\end{split}
\end{displaymath}
Since $\Gamma(2)$ is index-6 in $PSL(2,\mathbb{Z})$, the generators of the latter group act on $\mathcal{M}_{0,4}$ as

\begin{displaymath}
T:\, f\mapsto \frac{f}{f-1},\quad S:\, f\mapsto \frac{1}{f}
\end{displaymath}
These generate an $S_3$ action on $\mathcal{M}_{0,4}$, as depicted in the figure

\begin{displaymath}
 \includegraphics[width=239pt]{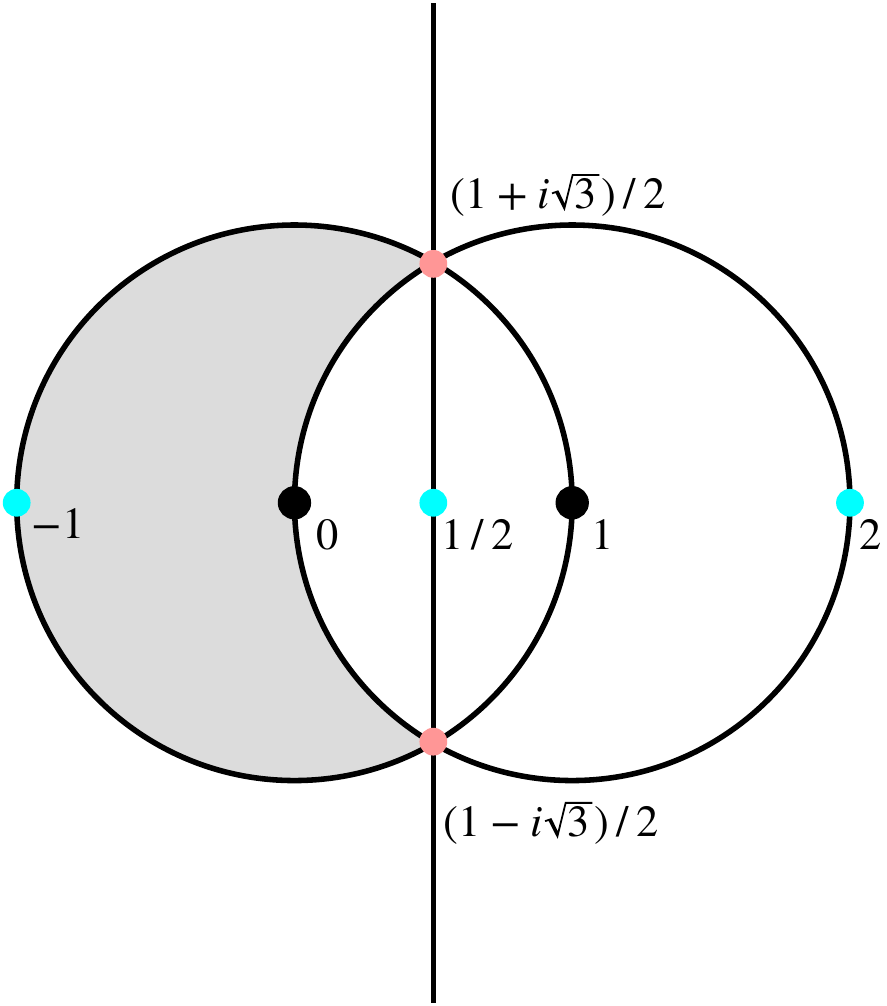}
\end{displaymath}
The points, $\{0,1,\infty\}$, of the compactification divisor, are fixed points with stabilizer group $\mathbb{Z}_2$. The points $\{-1,1/2,2\}$ are also fixed points with stabilizer group $\mathbb{Z}_2$. Finally, the points $(1\pm i\sqrt{3})/2$ are fixed points with stabilizer group $\mathbb{Z}_3$. The $j$-invariant (invariant under the action of $PSL(2,\mathbb{Z})$) is

\begin{displaymath}
\begin{split}
    j(\tau) &= 256 \frac{{(1-f+f^2)}^3}{f^2{(1-f)}^2}\\
   &= \frac{1}{q}+744+196884q+\dots
\end{split}
\end{displaymath}
Of course, while the $j$-invariant is invariant under the full $PSL(2,\mathbb{Z})$, the physics generically is not

If two of the punctures are identical, then $\tau\mapsto -1/\tau$ leaves the physics unchanged. The S-duality group is $\Gamma_0(2)\subset PSL(2,\mathbb{Z})$, generated by $T^2:\, \tau\mapsto \tau+2$ and $S:\, \tau\mapsto -1/\tau$, whose fundamental domain is the $\mathbb{Z}_2$ quotient of $\mathcal{M}_{0,4}$ by $f\mapsto 1/f$. The physics at $f=0$ and at $f=\infty$ are both that of a weakly-coupled $G$ gauge theory. The other boundary point, $f=1$, and the interior point, $f=-1$ are fixed-points of the $\mathbb{Z}_2$ action.

If \emph{three} of the punctures (or all four) are identical, then the S-duality group is the full $PSL(2,\mathbb{Z})$, the physics at all three boundary points is that of a weakly-coupled $G$-gauge theory and the fundamental domain is just the shaded region in the figure.

How this picture gets modified, in the presence of twisted punctures, will be one of our main themes in this paper.

\subsection{{Very-even Punctures}}\label{veryeven_punctures}

In the $A_{2N-1}$ series, the outer automorphism twists acted trivially on the \emph{set} of nilpotent orbits. So the identities of the untwisted punctures were unaffected by the introduction of twisted punctures. By contrast, in the $D_N$ series (for $N$ even), the outer automorphism twists act by exchanging the ``red'' and ``blue'' very-even punctures. Dragging an untwisted very-even puncture around a twisted puncture turns it from red to blue, or vice-versa.

To illustrate the phenomenon, let us look at an example in the twisted $D_4$ theory.

\begin{displaymath}
 \includegraphics[width=123pt]{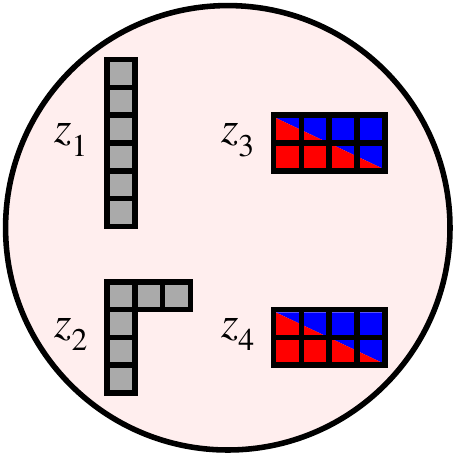}
\end{displaymath}
Here, it is useful to recall \cite{Chacaltana:2011ze} that the very-even puncture\footnote{As in\cite{Chacaltana:2010ks,Chacaltana:2011ze,Chacaltana:2012ch}, a Nahm-pole partition $p$ is represented by a Young diagram such that the column heights are equal to the parts of $p$. (So $\includegraphics[width=32pt]{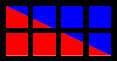}$ is the puncture with Nahm pole D-partition $[2^4]$.) In this paper we do not use Young diagrams to represent Hitchin-pole partitions.} $\includegraphics[width=32pt]{D4untwisted2222}$ has only one constraint, which is a very-even c-constraint,

\begin{displaymath}
c^{(4)}_3 \pm 2\tilde{c}_3 = 0,
\end{displaymath}
where the top (bottom) sign corresponds to a red (blue) puncture.

The Higgs field (with Coulomb branch parameters $u_2,u_4,\tilde{u},u_6$) yields the differentials

\begin{displaymath}
\begin{split}
\phi_2(z)&=\frac{u_2 z_{1 2} z_{3 4}{(d z)}^2}{(z-z_1)(z-z_2)(z-z_3)(z-z_4)}\\
\phi_4(z)&=\frac{
z_{2 4}z_{3 4}{(d z)}^4}{(z-z_1){(z-z_2)}^3{(z-z_3)}^3{(z-z_4)}^3}\\
&\times \left[
u_4 (z-z_3)(z-z_4) z_{1 2}z_{2 3}
+2\tilde{u}(z_2-z)
\left(
(z-z_3){(z_{1 3}z_{2 3}z_{1 4}z_{2 4})}^{1/2}
+(z-z_4)z_{1 3}z_{2 3}
\right)
\right]\\
%\end{split}
%\end{displaymath}
%\begin{displaymath}
%\begin{split}
\phi_6(z)&=\frac{u_6 z_{1 2} z_{2 3} z_{2 4} z_{3 4}^3 {(d z)}^6}{(z-z_1){(z-z_2)}^3{(z-z_3)}^4{(z-z_4)}^4}\\
\tilde{\phi}(z)&=\frac{\tilde{u} z_{2 4} z_{3 4}^2 {\left(z_{1 3} z_{2 3} \right)}^{1/2}{(d z)}^4}{{(z-z_1)}^{1/2}{(z-z_2)}^{3/2}{(z-z_3)}^3{(z-z_4)}^3}
\end{split}
\end{displaymath}
The powers of $z_{i j}\equiv z_i-z_j$ have been introduced to make the above expressions M\"obius-invariant\footnote{To minimize the number of ensuing branch cuts, we have chosen not to preserve the obvious $z_3\leftrightarrow z_4$ symmetry. We can restore it by redefining the Coulomb branch parameter

\begin{displaymath}
\hat{\tilde{u}} = \tilde{u} {\left(\frac{z_{1 3} z_{2 4}}{ z_{1 2}z_{3 4}}\right)}^{1/2}
\end{displaymath}
The resulting theory lives naturally on the 4-fold branched cover of $\mathcal{M}_{0,4}$.} , and hence well-defined on the moduli space. However, the (unavoidable) square-roots mean that moduli space is, itself, a double-cover (in fact, a 4-fold cover, but the SW geometry factors through a $\mathbb{Z}_2$ quotient) of the moduli space of the 4-punctured sphere.

Whether a very-even puncture is red or blue depends on the \emph{relative sign} of the residues of the cubic poles of $\phi_4(z)$ and $\tilde{\phi}(z)$ at the location of the puncture. But the square-roots are such that if we drag the very-even puncture (say, the one located at $z_3$) around one of the twisted punctures (say, the one located at $z_1$), the relative sign changes, indicating that the puncture has changed from red to blue, or vice versa.

Since the formulae are a little bit formidable-looking in their fully M\"obius-invariant form, it helps to fix the M\"obius invariance by setting

\begin{displaymath}
(z_1,z_2,z_3,z_4) \to (0,\infty,w^2,1)
\end{displaymath}
The expressions for $\phi_4(z),\tilde{\phi}(z)$ (which are all we need for the present discussion) simplify to

\begin{displaymath}
\begin{split}
\phi_4(z)&=\frac{(w^2-1)\left[u_4 (z-w^2)(z-1)
+2\tilde{u}
\left(w (z-w^2)+ w^2 (z-1)\right)
\right]{(d z)}^4}{z{(z-w^2)}^3 {(z-1)}^3}\\
\tilde{\phi}(z)&=\frac{\tilde{u} w{(w^2-1)}^{2}{(d z)}^4}{z^{1/2} {(z-w^2)}^3(z-1)^3}
\end{split}
\end{displaymath}
Dragging the point $z_3 =w^2$ around the origin changes the sign of $w$ in the above expressions. This changes the \emph{relative} sign of the residues of $\phi_4$ and $\tilde{\phi}$ at $z=w^2$, whilst preserving the relative sign of the residues at $z=1$.

Of course, the Seiberg-Witten geometry is invariant under the operation of simultaneously flipping all of the colours of all of the very-even punctures. This gives a $\mathbb{Z}_2$ which acts freely on the gauge theory moduli space. We will often find it useful to work on the quotient, fixing the colour of \emph{one} of the very-even punctures.

Having seen the phenomenon is global example, let us recover the same result, working locally on the plane, with the Higgs field itself (rather than the gauge-invariant $k$-differentials). Consider a very-even Higgs-field residue $B\in \mathfrak{so}(2N)$, which belongs to a, say, red nilpotent orbit. We can write $B=B|_{\mathfrak{so}(2N-1)}+B|_{o_{-1}}$, corresponding to the splitting $\mathfrak{so}(2N)=\mathfrak{so}(2N-1)\oplus o_{-1}$. Then, one can check that the map $B|_{o_{-1}}\mapsto -B|_{o_{-1}}$ puts the residue $B$ in the other (blue) nilpotent orbit. This map defines an isomorphism between the elements of the red and the blue nilpotent orbits.

Now suppose that the twisted puncture (with residue $A\in \mathfrak{so}(2N-1)$ is at $z=0$ and the very-even puncture (with residue $B\in\mathfrak{so}(2N)$) is at $z=x$. Then, the Higgs field for this system is:

\begin{displaymath}
\Phi(z) = \tfrac{(z-x)A+z B|_{\mathfrak{so}(2N-1)}}{z(z-x)} + \tfrac{x^{1/2}B|_{o_{-1}}+(z-x)D+\dots}{z^{1/2}(z-x)}+\dots
\end{displaymath}
where $D$ is a generic element in $o_{-1}$, and the \ldots{} denote regular terms. The factor of $x^{1/2}$ is necessary to make $\Phi$ well-defined as a one-form. Then, $x$ parametrizes the distance between the very-even puncture and the twisted puncture, and if $x$ circles the origin, $x^{1/2}\to -x^{1/2}$, it enforces $B|_{o_{-1}}\to -B|_{o_{-1}}$, so our red puncture becomes blue, or vice versa.

\subsection{{Atypical Degenerations}}
\subsubsection{{Atypical Punctures}}\label{atypical_punctures}

As an application of the formulas in \S\ref{local_properties_of_punctures}, let us find the series of punctures with contribution $n_2=2$. We will call these ``atypical punctures'', as they give rise to theories where the number of simple factors in the gauge group is not equal to the dimension of the moduli space of the punctured Riemann surface, $C$. We have seen this phenomenon already in the twisted $A_{2N-1}$ series \cite{Chacaltana:2012ch}.

From our rules for a-constraints, it is easy to see that there are \emph{no} untwisted atypical punctures, and that for a twisted puncture to be atypical, its Nahm pole C-partition must consist of exactly \emph{two} parts. Hence, the atypical punctures are

\begin{displaymath}
\stackrel{\mathclap{[2(N-r-1),2r]}}{ \includegraphics[width=9pt]{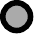}}\quad,\qquad\text{for}\,\,r= 1,2,\dots, \left\lfloor \frac{N-1}{2} \right\rfloor
\end{displaymath}
with the addition of

\begin{displaymath}
\stackrel{\mathclap{[N-1,N-1]}}{ \includegraphics[width=9pt]{twistedPuncture}}
\qquad\text{if}\,\,N\,\,\text{is even}.
\end{displaymath}
These arise, respectively, as the coincident limit of

\begin{itemize}%
\item[a)] $\quad\stackrel{\mathclap{[2(N-1)]}}{ \includegraphics[width=9pt]{twistedPuncture}}\quad$ and $\,\qquad\stackrel{\mathclap{[2(N-r)-1,2r+1]}}{ \includegraphics[width=9pt]{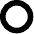}}\qquad$
\item[b)] $\quad\stackrel{\mathclap{[2(N-1)]}}{ \includegraphics[width=9pt]{twistedPuncture}}\quad$ and $\quad\stackrel{[N,N]}{ \includegraphics[width=9pt]{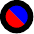}}\quad$ (for $N$ even)

\end{itemize}
Normally, the OPE of two (regular) punctures, $p$ and $p'$, yields a third (regular) puncture, $p''$, \emph{coupled} to a gauge theory, $(X,H)$, where

\begin{itemize}%
\item The gauge group, $H$, is a subgroup of the global symmetry group of $p''$.
\item In the coincident limit, the gauge coupling of $H$ goes to zero.

\end{itemize}
Here, when $p''$ is atypical, the would-be gauge theory is \emph{empty}: $(X,H)=(\emptyset,\emptyset)$. Instead, the theory with an insertion of $p''$ has one more simple factor in the gauge group than the ``expected'' $3g-3+n$.

For a surface, $C$, with $n$ punctures, $m$ of which are atypical, the number of simple factors in the gauge group is $3g-3+n+m$. ``Resolving'' each atypical puncture by the pair of punctures, above, yields a surface with $n+m$ punctures and the moduli space of the gauge theory is a branched cover of $\mathcal{M}_{g,n+m}$. In contrast to the usual case, where each component of the boundary of the moduli space corresponds to one simple factor in the gauge group becoming weakly-coupled, the boundaries of $\mathcal{M}_{g,n+m}$, where an atypical puncture arises in the OPE, do not typically correspond to any gauge coupling becoming weak (that is, under the branched covering, they are the image of loci in the interior of the gauge theory moduli space).

\subsubsection{{Gauge Theory Fixtures}}\label{gauge_theory_fixtures}

In particular, for $n=3$, $m=1$ (or $2$), we have a ``gauge theory fixture.'' Resolving the atypical puncture yields a gauge theory moduli space which is branched cover of $\mathcal{M}_{0,4}$. We may well ask, ``Where, in the gauge theory moduli space, have we landed, in the coincident limit which yields the atypical puncture?'' The answer is that we are at the interior point, ``$f(\tau)=-1$'', though the mechanics of how this happens varies between the cases.

Let us resolve

\begin{displaymath}
\begin{matrix} \includegraphics[width=114pt]{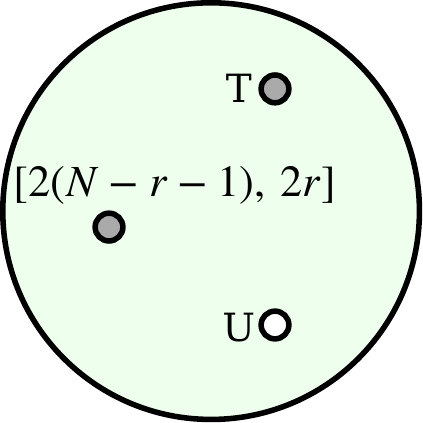}\end{matrix}\quad\text{or}\quad\begin{matrix} \includegraphics[width=114pt]{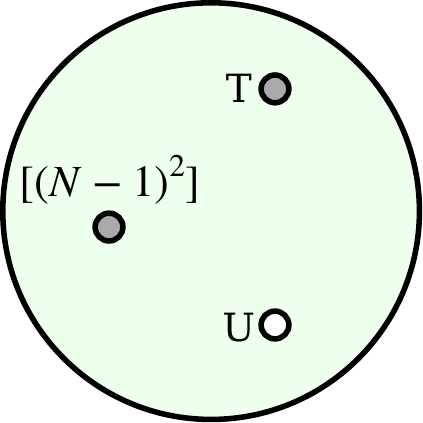}\end{matrix}
\end{displaymath}
to

\begin{displaymath}
\begin{matrix} \includegraphics[width=114pt]{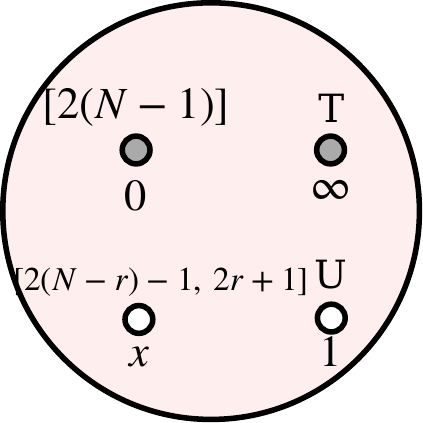}\end{matrix}\quad\text{and}\quad
\begin{matrix} \includegraphics[width=114pt]{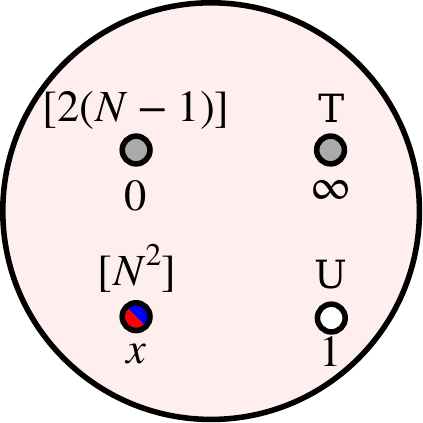}\end{matrix}
\end{displaymath}
respectively. We have parametrized $\mathcal{M}_{0,4}$ by $x$, but the gauge theory moduli space is a branched cover, parametrized by $w$, with

\begin{displaymath}
w^2=x
\end{displaymath}
The gauge coupling

\begin{equation}
f(\tau) = \frac{w-1}{w+1}
\label{ramifiedGaugeCoupling}\end{equation}
so that $f=0$ and $f=\infty$ both map to $x=1$, while $f=1$ maps to $x=\infty$. Our gauge-theory fixture is whatever lies over the point $x=0$. From \eqref{ramifiedGaugeCoupling}, this is the interior point, $f(\tau)=-1$, of the gauge theory moduli space.

As an example, let us consider the $D_4$ gauge theory fixture

\begin{displaymath}
 \includegraphics[width=114pt]{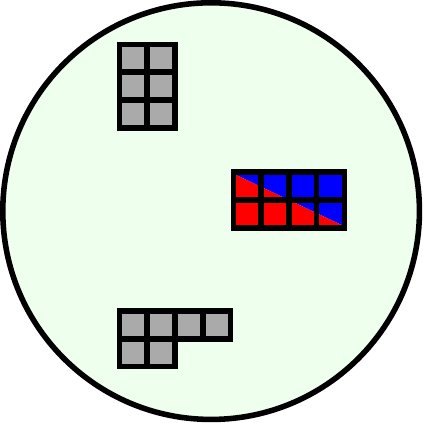}
\end{displaymath}
whose resolution is

\begin{displaymath}
 \includegraphics[width=114pt]{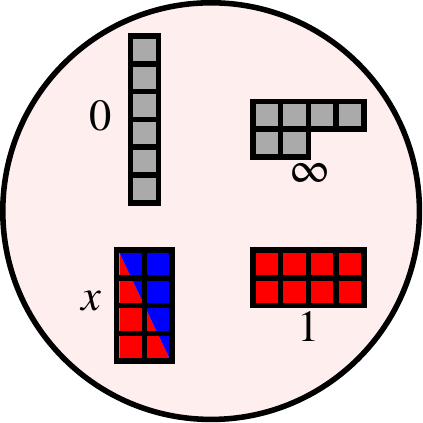}
\end{displaymath}
Actually, since we have two very-even punctures, the full moduli space is a 4-sheeted cover of $\mathcal{M}_{0,4}$. The SW geometry is invariant under simultaneously flipping the colours of both punctures, so we can consistently work on the quotient by that $\mathbb{Z}_2$, and take the colour of the $ \includegraphics[width=32pt]{D4untwisted2222}$ puncture to be red.

$SU(4)$ gauge theory, with matter in the $1(6)+4(4)$ was studied in \cite{Chacaltana:2010ks}. Near $f(\tau)=0$, the weakly-coupled description is the Lagrangian field theory. Near $f(\tau)=1$, the weakly-coupled description is an $SU(2)$ gauging of the ${SU(8)}_8\times {SU(2)}_6$ SCFT, $R_{0,4}$. Near $f(\tau)=\infty$, the weakly-coupled description is $SU(3)$, with two hypermultiplets in the fundamental, coupled to the ${(E_7)}_8$ SCFT.

In the present case, the $f\to1$ theory arises as $x\to\infty$

\begin{displaymath}
 \includegraphics[width=262pt]{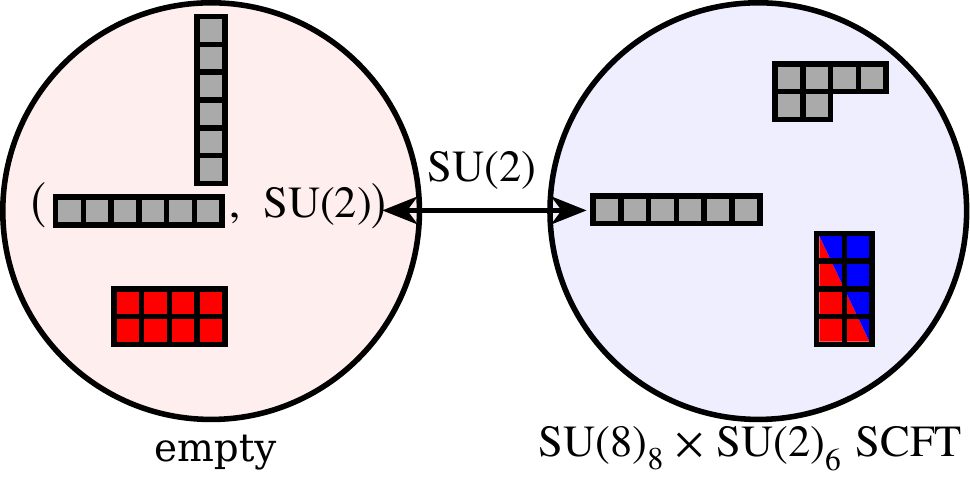}
\end{displaymath}
Over $x=1$, we have two distinct degenerations, which are exchanged by dragging the $ \includegraphics[width=16pt]{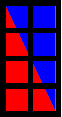}$ puncture around the origin and returning it to its original position: the Lagrangian field theory ($f=0$)

\begin{displaymath}
 \includegraphics[width=262pt]{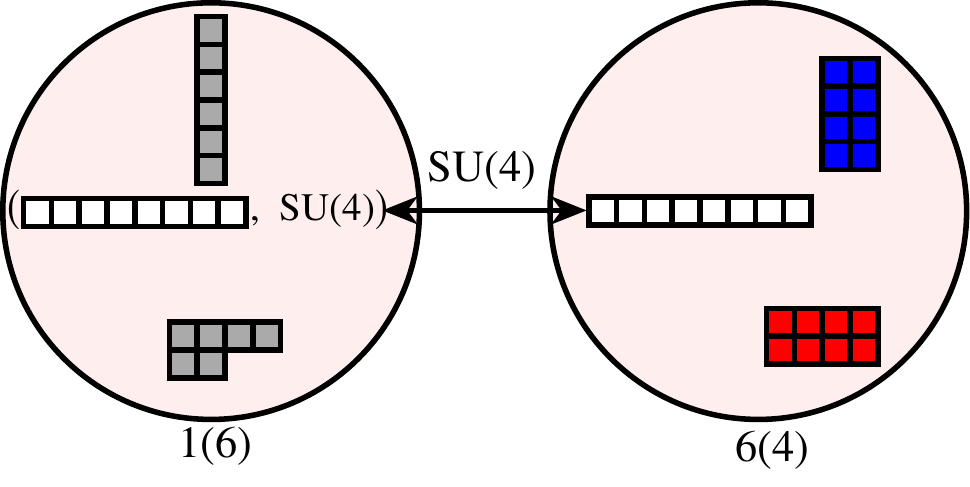}
\end{displaymath}
and the theory at $f=\infty$

\begin{displaymath}
 \includegraphics[width=262pt]{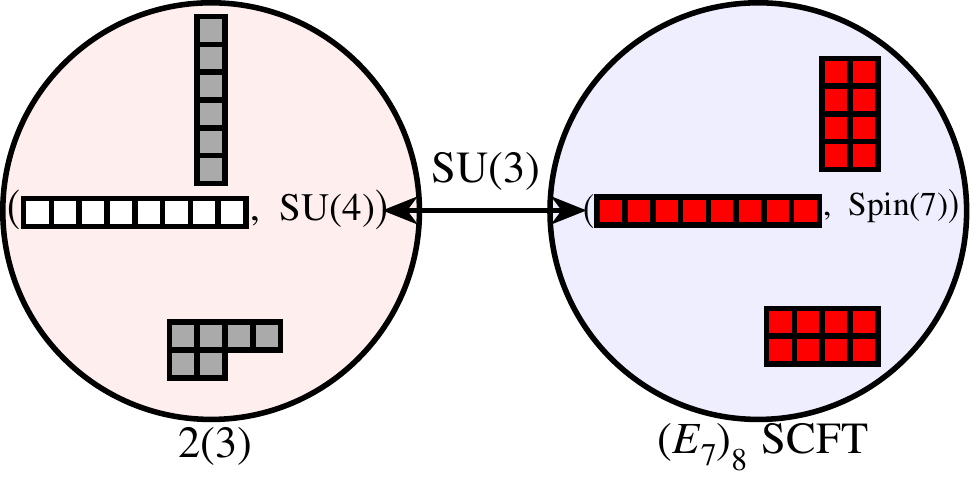}
\end{displaymath}
Having fixed the behaviour of $f$ over this two-sheeted cover of $\mathcal{M}_{0,4}$, by reproducing the correct asymptotics as $x\to 1$ and $x\to\infty$, we can now take $x\to 0$

\begin{displaymath}
 \includegraphics[width=262pt]{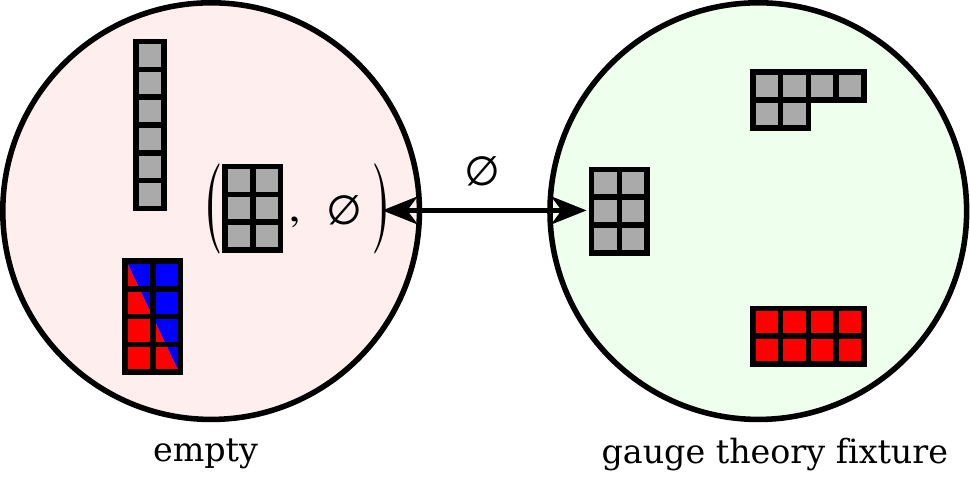}
\end{displaymath}
and recover that the gauge theory fixture is the aforementioned $SU(4)$ gauge theory at $f(\tau)=-1$.

\subsubsection{{Gauge Theory Fixtures with Two Atypical Punctures}}\label{gauge_theory_fixtures_with_two_atypical_punctures}

When we resolve the gauge theory fixtures with \emph{two} atypical punctures, we obtain a branched covering of $\mathcal{M}_{0,5}$.

The geometry of $\mathcal{M}_{0,5}$, and the relevant branched covering thereof, were discussed in detail in section 5.1.2 of \cite{Chacaltana:2012ch}. Here, we will simply borrow the relevant results.

The (compactified) $\mathcal{M}_{0,5}$ is a rational surface. The boundary divisor consists of ten $(-1)$-curves ($\mathbb{C}\mathrm{P}^1$s with normal bundle $\mathcal{O}(-1)$). We label these curves as $D_{i j}$, corresponding to the locus where the punctures $p_i$ and $p_j$ collide. The $D_{i j}$, in turn, intersect in $15$ points.

The moduli space of the $(2,0)$ compactification is a branched covering, $\tilde{\mathcal{M}}\to \mathcal{M}_{0,5}$, which is branched over the boundary divisor.

The $D_4$ gauge theory fixture

\begin{displaymath}
 \includegraphics[width=114pt]{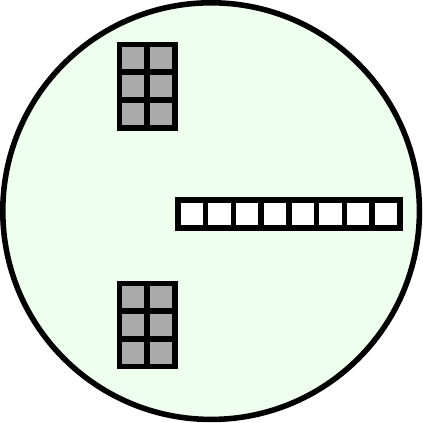}
\end{displaymath}
is an $Sp(2)\times SU(2)$ gauge theory, with matter in the $6(4,1)+4(1,2)$, with gauge couplings $(f_{Sp(2)},f_{SU(2)}) = (-1,-1)$. Resolving the atypical punctures, we have a 5-punctured sphere,

\begin{displaymath}
 \includegraphics[width=129pt]{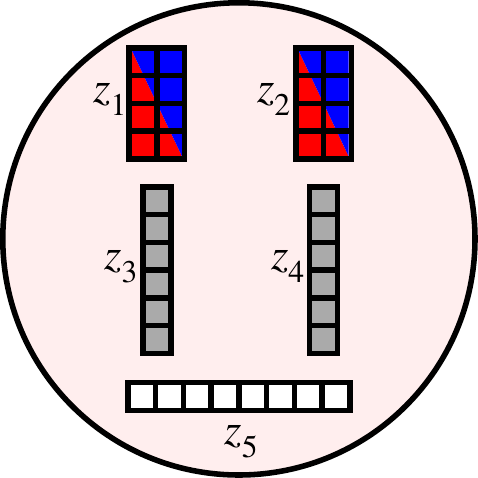}
\end{displaymath}
Since the resolution has two very-even punctures, $\tilde{\mathcal{M}}$ is an 8-sheeted branched cover of $\mathcal{M}_{0,5}$. However since the gauge couplings (and the rest of the physics) are invariant under simultaneously flipping the colours of both very-even punctures, we can pass to the quotient, $X=\tilde{\mathcal{M}}/\mathbb{Z}_2$, and it is the geometry of 4-sheeted branched cover, $X\to \mathcal{M}_{0,5}$, that was studied in detail in \cite{Chacaltana:2012ch}.

Meromorphic functions on $\mathcal{M}_{0,5}$ are rational functions of the cross-ratios

\begin{displaymath}
s_1=\frac{z_{1 3}z_{2 5}}{z_{1 5}z_{2 3}},\quad
  s_2=\frac{z_{1 4}z_{2 5}}{z_{1 5}z_{2 4}}
\end{displaymath}
$X$ is a branched 4-fold cover of $\mathcal{M}_{0,5}$, whose ring of meromorphic functions is generated by rational functions of $w_1,w_2$

\begin{displaymath}
w_1^2 = s_1,\quad w_2^2 = s_2
\end{displaymath}
The gauge couplings are meromorphic functions on $X$, given by

\begin{equation}
f_{Sp(2)}=\frac{w_1-1}{w_1+1}\frac{w_2+1}{w_2-1},\quad f_{SU(2)}=\frac{w_1-1}{w_1+1}\frac{w_2-1}{w_2+1}
\label{sp2su2couplings}\end{equation}
There is a natural action of the dihedral group, $D_4$, on $X$. The $\mathbb{Z}_2\times \mathbb{Z}_2$ subgroup is generated by the deck transformations,

\begin{displaymath}
\begin{split}
\alpha &:\, w_1\to -w_1,\, w_2\to  w_2\\
\beta  &:\, w_1\to  w_1,\, w_2\to -w_2\\
\end{split}
\end{displaymath}
which act on the gauge couplings as

\begin{displaymath}
\begin{split}
\alpha &:\, f_{Sp(2)}\to 1/f_{SU(2)},\, f_{SU(2)}\to  1/f_{Sp(2)}\\
\beta  &:\, f_{Sp(2)}\leftrightarrow f_{SU(2)}\\
\end{split}
\end{displaymath}
Both $\alpha$ and $\beta$ change the relative colour of the two very-even punctures. The additional generator of $D_4$,

\begin{displaymath}
\gamma :
, w_1\leftrightarrow w_2
\end{displaymath}
acts as S-duality for the $Sp(2)$,

\begin{displaymath}
\gamma :\, f_{Sp(2)}\to 1/f_{Sp(2)},\, f_{SU(2)}\to f_{SU(2)}
\end{displaymath}
At the boundary, various sheets come together, and the behaviour of the gauge couplings is

\begin{itemize}%
\item Over $D_{1 5}$ and $D_{2 5}$, both couplings go to $f=1$, but the ratio $\frac{f_{Sp(2)}-1}{f_{SU(2)}-1}$ is arbitrary.
\item Over $D_{3 5}$, both couplings are weak ($f=0$ or $f=\infty$), but the ratio $\frac{f_{Sp(2)}}{f_{SU(2)}}$ is arbitrary.
\item Over $D_{4 5}$, both couplings are weak ($f_{Sp(2)}=0$, $f_{SU(2)}=\infty$ or vice-versa), but the product $f_{Sp(2)}\cdot f_{SU(2)}$ is arbitrary.
\item Over $D_{1 2}$, one coupling is weak ($f=0$ or $\infty$), while the other is arbitrary.
\item Over $D_{3 4}$, one coupling is $f=1$, while the other is arbitrary.
\item Over $D_{1 3}$ and $D_{2 3}$, $f_{Sp(2)}=1/f_{SU(2)}$.
\item Over $D_{1 4}$ and $D_{2 4}$, $f_{Sp(2)}=f_{SU(2)}$.

\end{itemize}
Over the \emph{intersections} of these divisors, we see the various S-duality frames of the gauge theory.

Over $D_{1 2}\cap D_{3 4}$, we have

\begin{displaymath}
 \includegraphics[width=407pt]{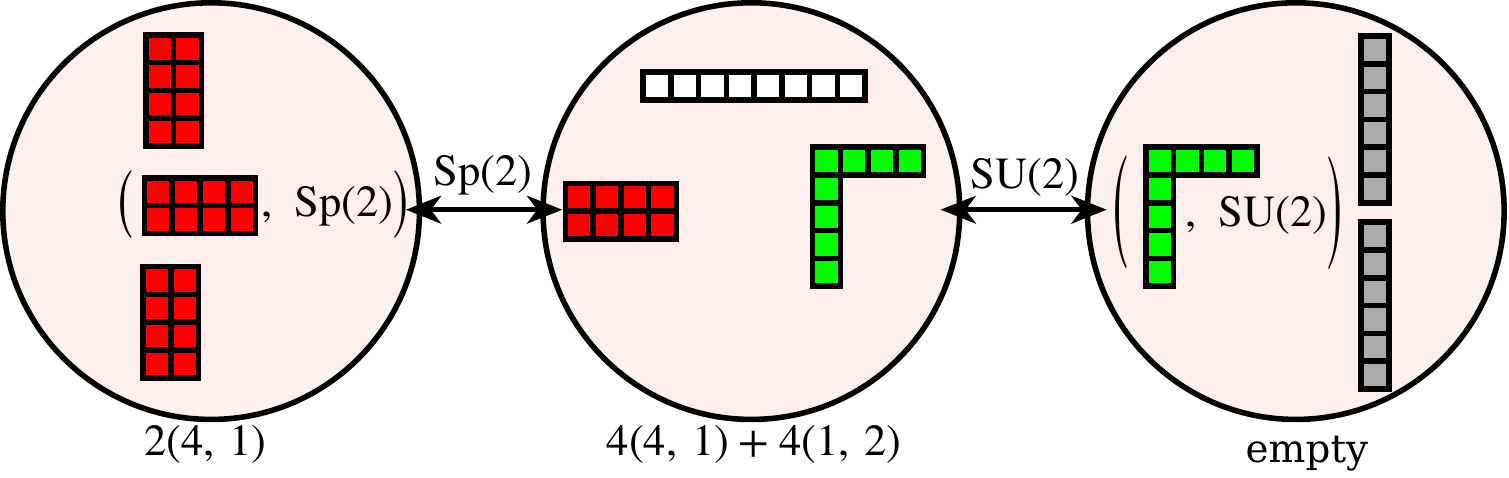}
\end{displaymath}
and

\begin{displaymath}
 \includegraphics[width=407pt]{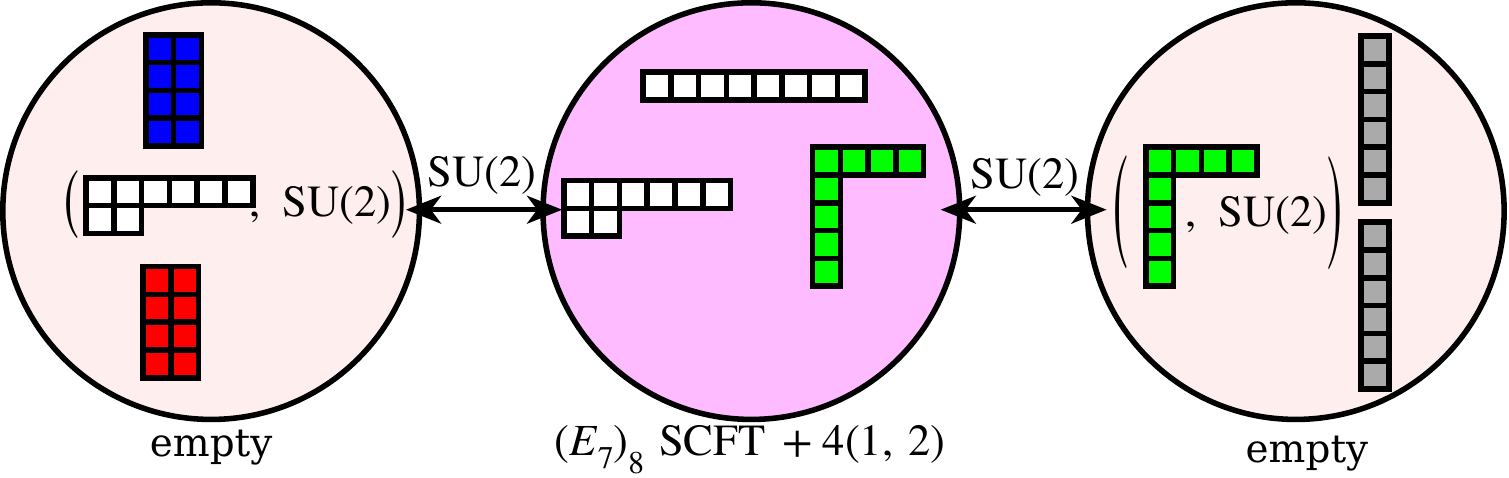}
\end{displaymath}
In the first case, $f_{Sp(2)}=0$ or $\infty$ and $f_{SU(2)}=1$; in the latter, $f_{Sp(2)}=1$ and $f_{SU(2)}=0$ or $\infty$.

Over $D_{1 2}\cap D_{3 5}$ and $D_{1 2}\cap D_{4 5}$, we have

\begin{displaymath}
 \includegraphics[width=497pt]{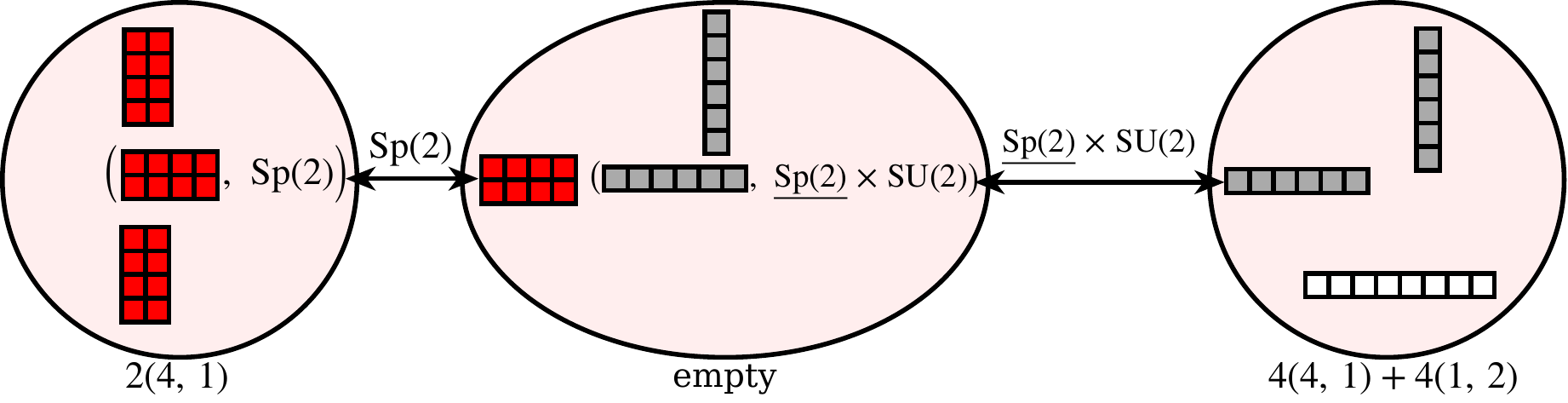}
\end{displaymath}
and

\begin{displaymath}
 \includegraphics[width=495pt]{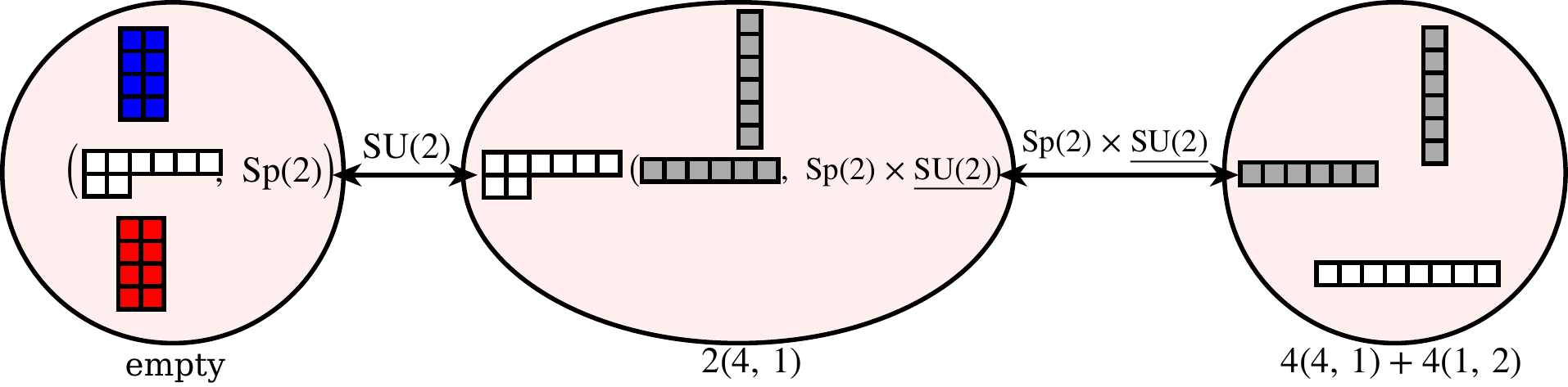}
\end{displaymath}
In both cases, the underlined gauge group on the right-hand cylinder is \emph{identified} with the gauge group on the left-hand cylinder. The notation, which we introduced in \cite{Chacaltana:2012ch}, indicated that when the cylinder on the right pinches off, \emph{both} factors in the gauge group become weakly-coupled ($f\to 0$ or $\infty$). When the cylinder on the left pinches off, only one of the gauge group factors becomes weakly-coupled.

Over $D_{3 4}\cap D_{1 5}$ and $D_{3 4}\cap D_{2 5}$, $f_{Sp(2)}=f_{SU(2)}=1$. So we have

\begin{displaymath}
 \includegraphics[width=497pt]{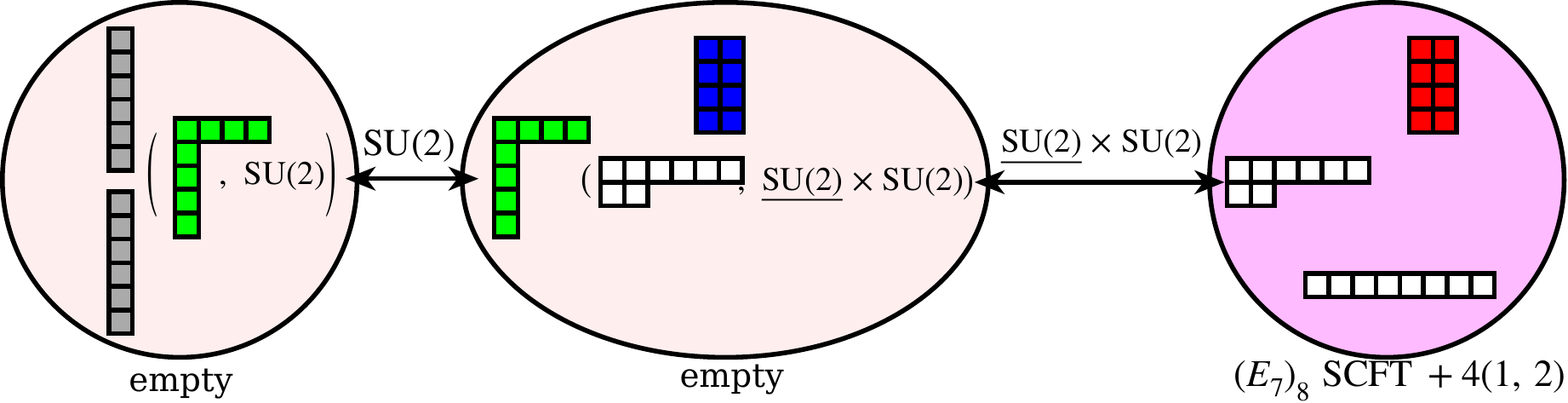}
\end{displaymath}
and

\begin{displaymath}
 \includegraphics[width=497pt]{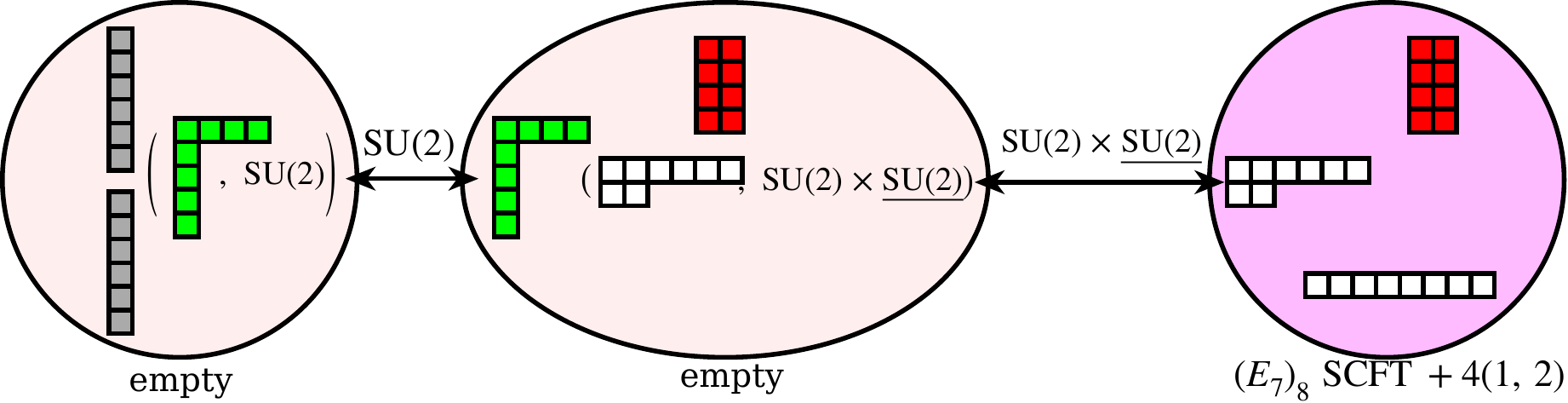}
\end{displaymath}
These differ only very subtly, as to ``which'' $SU(2)$ gauge coupling is controlled by the cylinder on the left. In the first case, it is the $SU(2)$ which couples to the ${(E_7)}_8$ (i.e., the one which becomes weakly-coupled at $f_{Sp(2)}=1$); in the second case, it is the $SU(2)$ which couples to the 4 fundamental hypermultiplets.

Over $D_{1 3}\cap D_{4 5}$, $D_{2 3}\cap D_{4 5}$, $D_{1 4}\cap D_{3 5}$ and $D_{2 4}\cap D_{3 5}$, we have

\begin{displaymath}
 \includegraphics[width=497pt]{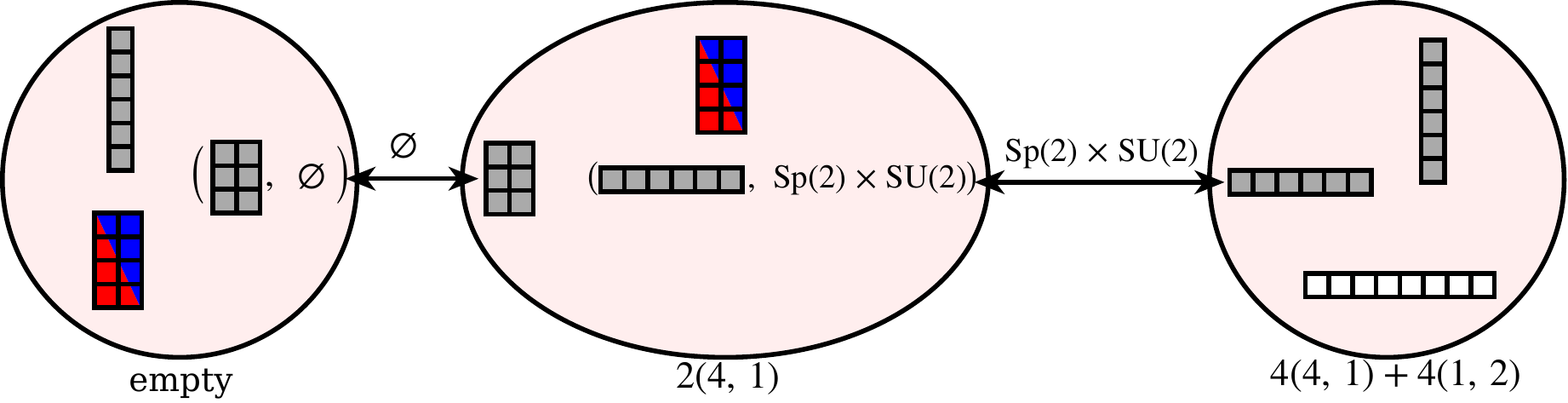}
\end{displaymath}
Over $D_{1 3}\cap D_{2 5}$, $D_{1 4}\cap D_{2 5}$, $D_{2 3}\cap D_{1 5}$ and $D_{2 4}\cap D_{1 5}$, we have $f_{Sp(2)}=1$, $f_{SU(2)}=1$:

\begin{displaymath}
 \includegraphics[width=497pt]{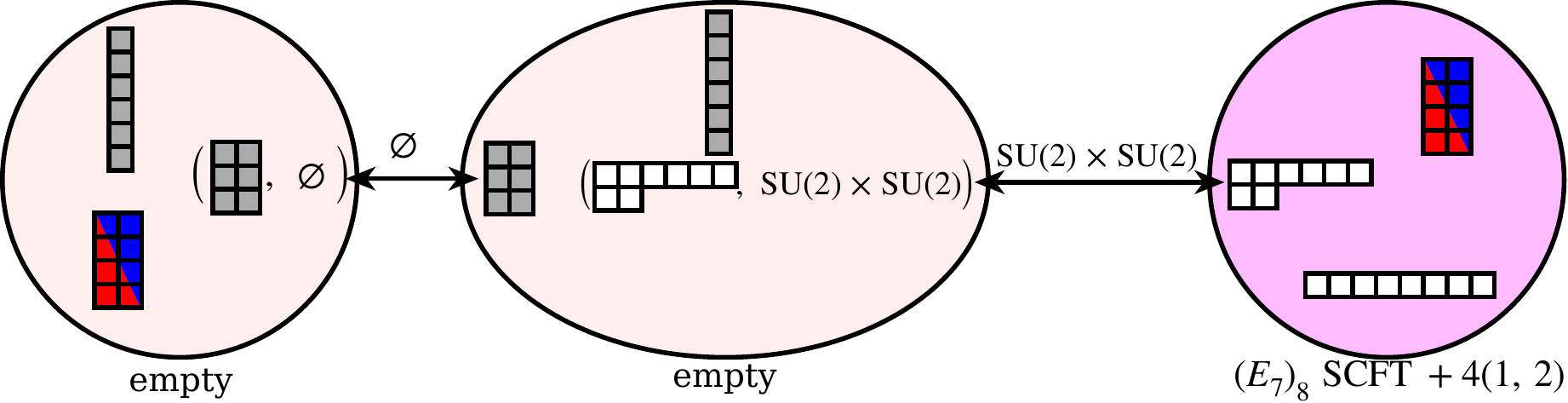}
\end{displaymath}
Finally, over $D_{1 3}\cap D_{2 4}$ and $D_{1 4}\cap D_{2 3}$, we recover our gauge theory fixture, and read off that its gauge theory couplings are $f_{Sp(2)}=f_{SU(2)}=-1$

\begin{displaymath}
 \includegraphics[width=402pt]{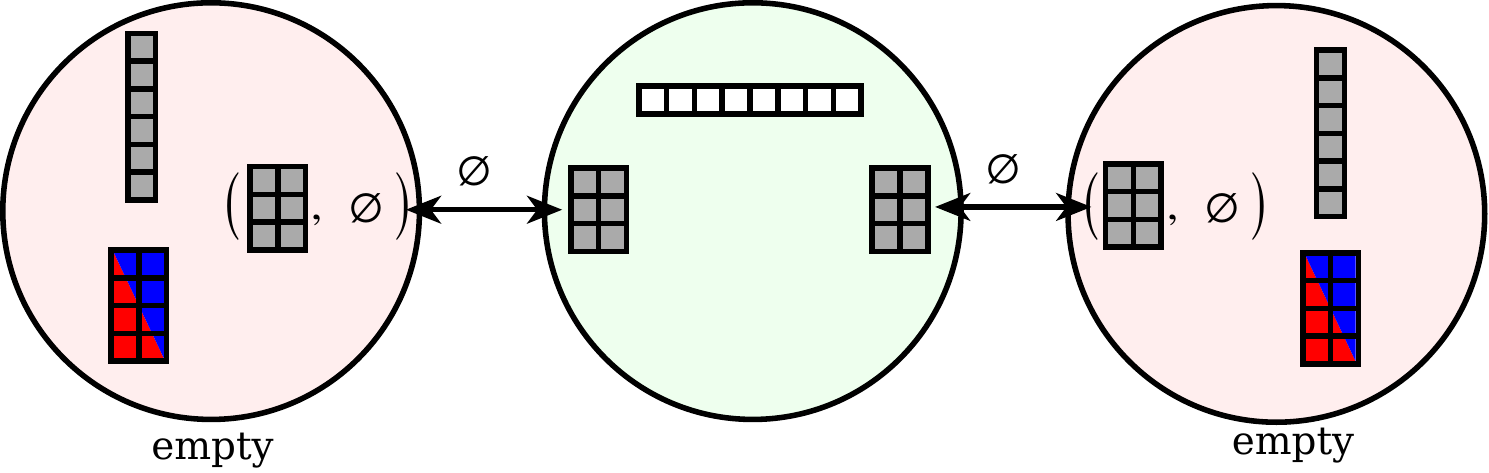}
\end{displaymath}

\subsubsection{{Atypical Degenerations and Ramification}}\label{atypical_degenerations_and_ramification}

Once we introduce outer-automorphism twists, the moduli space of the gauge theory no longer coincides with $\mathcal{M}_{g,n}$, the moduli space of punctured curves. As we saw, in \S\ref{atypical_punctures}, even the dimensions don't agree, until we ``resolve'' each atypical puncture, replacing $\mathcal{M}_{g,n}$ by $\mathcal{M}_{g,n+m}$ (for $m$ atypical punctures). Even then, the moduli space of the gauge theory is a branched covering of $\mathcal{M}_{g,n+m}$, branched over various components of the boundary.

Over a generic point on ``most'' of the components of the boundary, the covering is unramified, and the gauge couplings behave ``normally'': one (and only one) gauge coupling becomes weak at that irreducible component of the boundary. Here, we would like to catalogue the exceptions: those components of the boundary where

\begin{itemize}%
\item the covering is ramified
\item an ``unexpected'' (either 0 or 2, in the cases at hand) number of gauge couplings become weak
\item both

\end{itemize}
Let us denote, by $D_{p_1,p_2,\dots p_l}$, the component of the boundary of $\mathcal{M}_{g,n+m}$ where the punctures $p_1,p_2,\dots p_l$ collide, bubbling off an $(l+1)$-punctured sphere. All of our exceptional cases will involve either $D_{p_1,p_2}$ or $D_{p_1,p_2,p_3}$.

\paragraph{\underline{${D_{T,V}}$}}\label{DTV}

The first source of ramification, as we saw in \S\ref{veryeven_punctures}, is that the outer automorphism changes the colour of a very even puncture from red to blue and vice versa. In general, this changes the physics of the gauge theory. So, for a theory with $v$ very-even punctures, we get a $2^v$ sheeted cover of the moduli space of curves, ramified (with ramification index 2) over $D_{T,V}$ where ``$T$'' denotes any twisted-sector puncture and ``$V$'' represents any very-even. As already noted, simultaneously changing the colour of all of the very-even punctures leads to isomorphic physics so we can (and usually will) pass to the $\mathbb{Z}_2$ quotient.

Generically, the gauge couplings behave ``normally,'' with one gauge coupling becoming weak at $D_{T,V}$.

\paragraph{\underline{${D_{[2(N-1)], [N^2]}}$}}\label{NN}

When $N$ is even, there is one such collision where, in addition to ramification, no gauge coupling becomes weak. Instead, the two punctures fuse (in non-singular fashion) into an atypical puncture.

\begin{displaymath}
 \includegraphics[width=198pt]{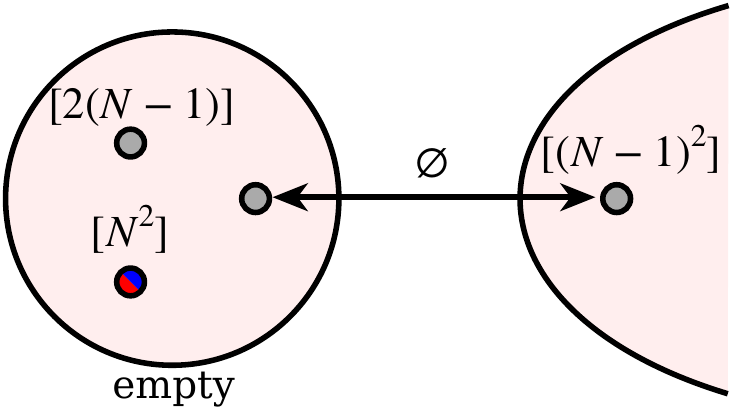}
\end{displaymath}

\paragraph{\underline{${D_{[2(N-1)],[2(N-r)-1,2r+1]}}$}}\label{_3}

For $r=1,2,\dots,\left\lfloor\frac{N-1}{2}\right\rfloor$, we again obtain an atypical puncture as the OPE. No gauge coupling become weak, but the moduli space is ramified (with ramification index 2).

\begin{displaymath}
 \includegraphics[width=251pt]{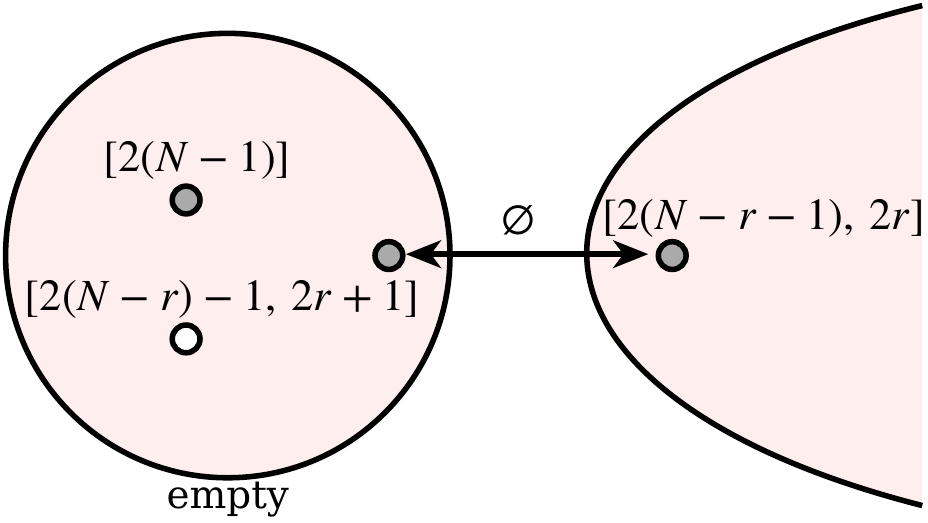}
\end{displaymath}

\paragraph{\underline{${D_{[2(N-1)],[2(N-1)],[2(N-r)-1,2r+1]}}$}}\label{_4}

The moduli space is \emph{unramified} over this component of the boundary. Nonetheless, \emph{two} gauge couplings become weak.

\begin{displaymath}
 \includegraphics[width=347pt]{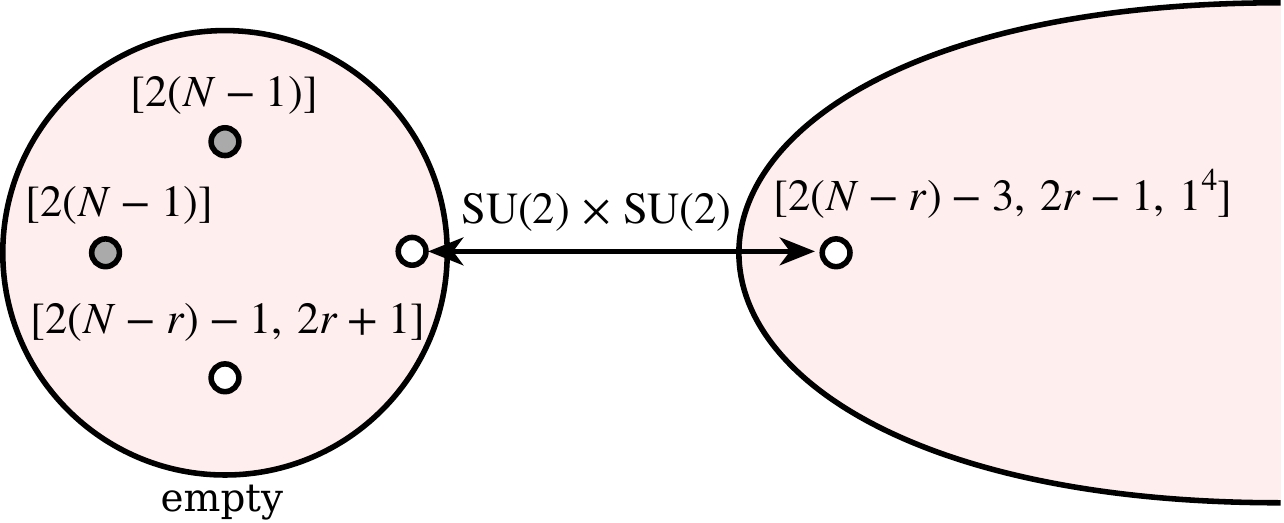}
\end{displaymath}

\paragraph{\underline{${D_{[2(N-1)],[2(N-1)],[N^2]}}$}}\label{_5}

Here, again, an $SU(2)\times SU(2)$ gauge group becomes weak, but now the moduli space is also ramified (with ramification index 2)

\begin{displaymath}
 \includegraphics[width=333pt]{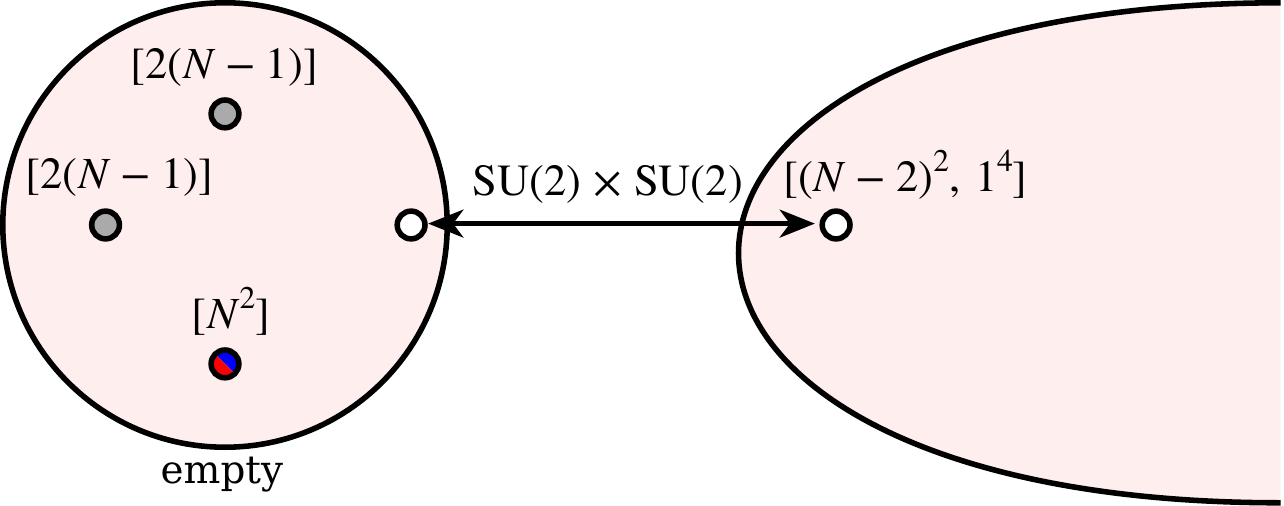}
\end{displaymath}

\paragraph{\underline{${D_{t,u,u'}}$}}\label{_6}

In all of the remaining cases, the moduli space is ramified (with ramification index 2) and two gauge couplings become weak.

Over $D_{[2(N-1)],[2(N-r)-1,2r+1],[2(N-r)-1,2r+1]}$ (with the same untwisted puncture), we have an $Sp(r)\times Sp(r)$ gauge group becoming weak

\begin{displaymath}
 \includegraphics[width=347pt]{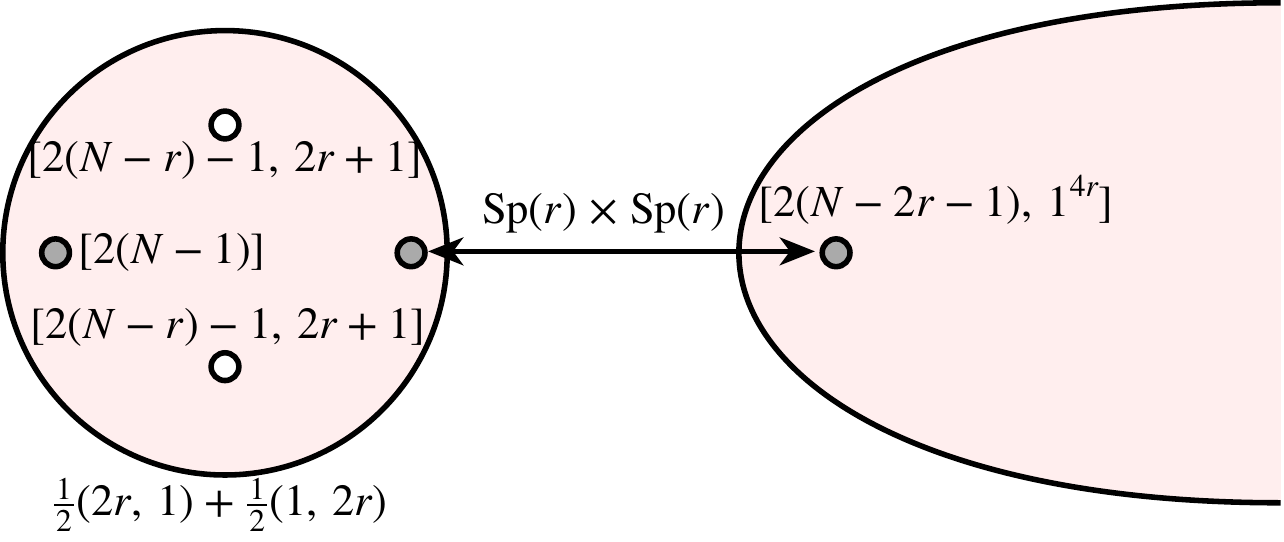}
\end{displaymath}
and, for $N$ even, the gauge group which becomes weak is $Sp\left(\frac{N}{2}\right)\times Sp\left(\frac{N-2}{2}\right)$

\begin{displaymath}
 \includegraphics[width=309pt]{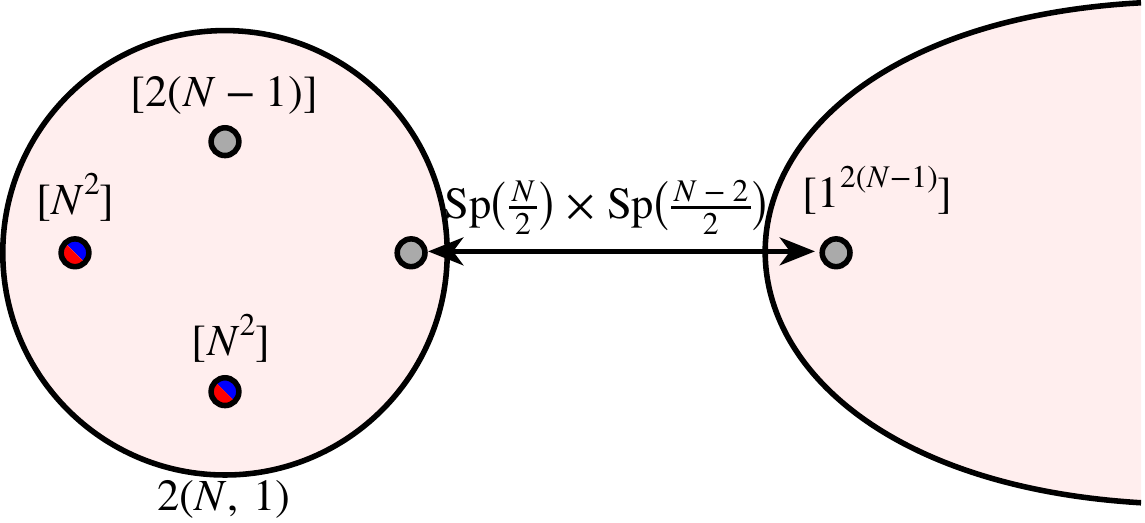}
\end{displaymath}
Over $D_{t,u,u'}$, with $r',r=1,2,\dots,\left\lfloor\frac{N-1}{2}\right\rfloor$ (and, without loss of generality, $r'\gt r$)

\begin{displaymath}
 \includegraphics[width=347pt]{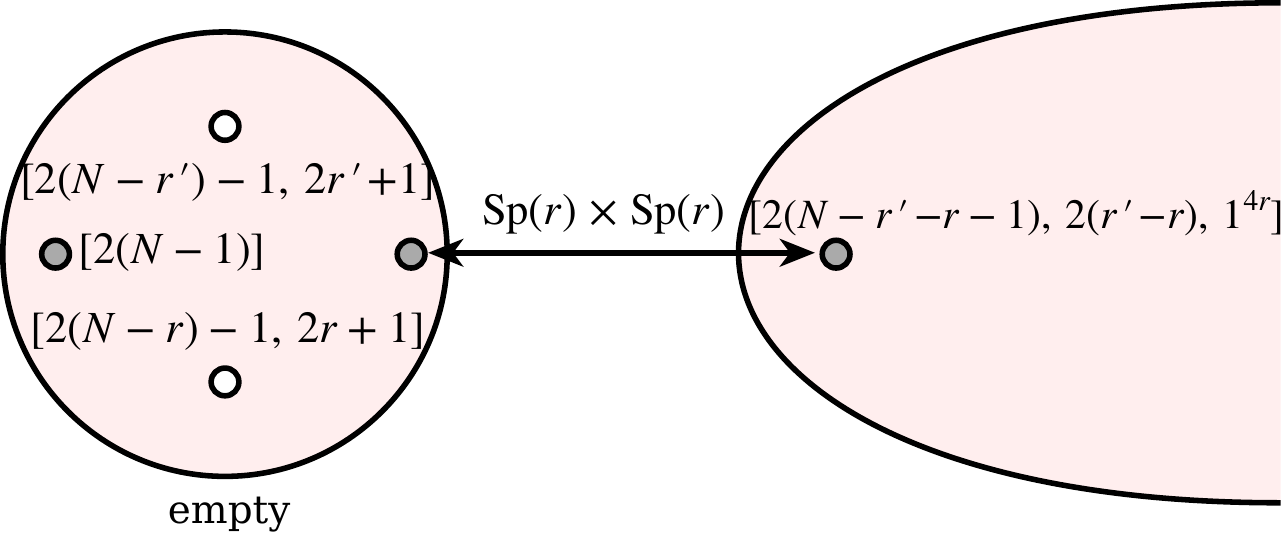}
\end{displaymath}
and, for $N$ even,

\begin{displaymath}
 \includegraphics[width=309pt]{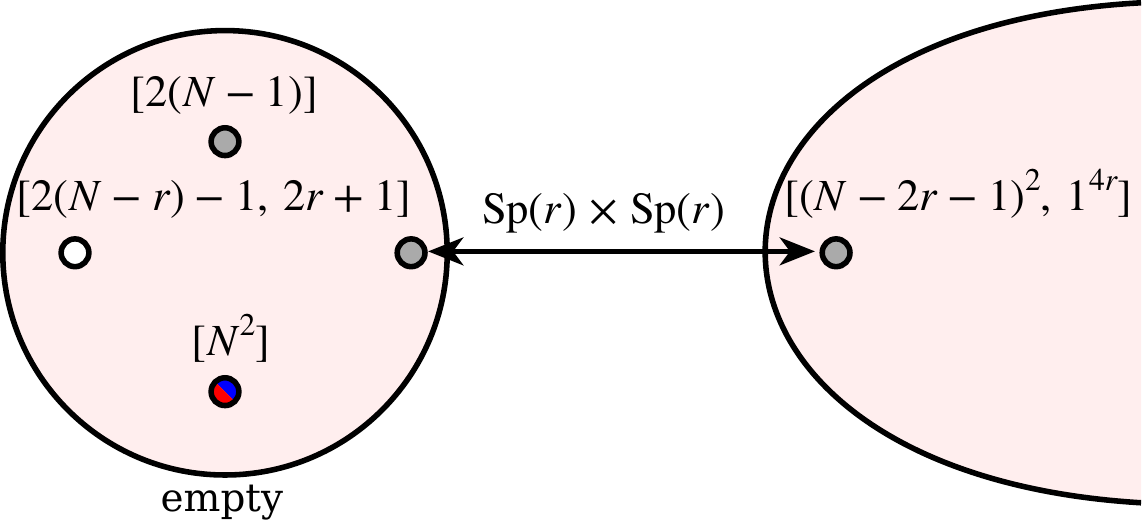}
\end{displaymath}

\subsection{{Global Symmetries and the Superconformal Index}}\label{global_symmetries_and_the_superconformal_index}

\subsubsection{Computing the Index in the Hall-Littlewood Limit}\label{enhancement}
Each puncture has a ``manifest'' global symmetry associated to it. The global symmetry group of the SCFT associated to a fixture contains the product of the ``manifest'' global symmetry groups, associated to each of the punctures, as a subgroup. But, in general, it is larger. Here, we will outline how to use the superconformal index \cite{Kinney:2005ej, Gadde:2009kb, Gadde:2011ik, Gadde:2011uv} to determine the global symmetry group of the fixture \emph{and} (in the case of a mixed fixture) the number of free hypermultiplets that it contains.

The prescription to compute the superconformal index of an interacting SCFT defined by a $D_N$-series fixture was given in \cite{Lemos:2012ph}. For a $D_N$ $\mathbb{Z}_2$-twisted sector fixture with punctures $(\tilde{\Lambda}_1,\tilde{\Lambda}_2,\Lambda_3)$, where $\tilde{\Lambda}$ denotes a twisted puncture and $\Lambda$ an untwisted puncture, the index is given by \footnote{In the following, we need only consider the ``Hall-Littlewood'' limit of the index, where we restrict to the one-parameter slice in the space of superconformal fugacities given by ($p=0,\, q=0,\, t^{1/2}\equiv\tau$) \cite{Gadde:2011uv}.} 

\begin{equation}
\begin{split}
\mathcal{I}(\mathbf{a}, \mathbf{b}, \mathbf{c})&=\mathcal{A}(\tau)\mathcal{K}(\mathbf{a}(\tilde{\Lambda}_1))\mathcal{K}(\mathbf{b}(\tilde{\Lambda}_2))\mathcal{K}(\mathbf{c}(\Lambda_3))\\
&\times \sum_{\lambda'}\frac{P^{\lambda'}_{Sp(N-1)}(\mathbf{a}(\tilde{\Lambda}_1)|\tau)P^{\lambda'}_{Sp(N-1)}(\mathbf{b}(\tilde{\Lambda}_2)|\tau)P^{\lambda=\lambda'}_{SO(2N)}(\mathbf{c}(\Lambda_3)|\tau)}{P^{\lambda=\lambda'}_{SO(2N)}(1,\tau,\tau^2,\dots,\tau^{N-1}|\tau)}.
\end{split}
\label{index}\end{equation}
The various elements of this formula are summarized below. Detailed explanations can be found in \cite{Lemos:2012ph}:

\begin{itemize}%
\item $\mathcal{A}(\tau)$ is the overall (fugacity-independent) normalization, given by\begin{displaymath}
\mathcal{A}(\tau)=\frac{(1-\tau^{2N})}{(1-\tau^2)^\frac{N}{2}}\prod_{j=1}^{N-1}(1-\tau^{4j}).
\end{displaymath}

\item $P^\lambda$ are the Hall-Littlewood polynomials of type $SO(2N)$ and $Sp(N)$, given by\begin{displaymath}
\begin{split}
P^\lambda_{SO(2N)}(x_1,\dots,x_N)&=W_\lambda(\tau)^{-1}\sum_{\sigma \in S_N}\sum_{\stackrel{s_1,\dots, s_N=\pm 1}{\prod s_i=+1}}x_{\sigma(1)}^{s_1\lambda_1}\cdots x_{\sigma(N)}^{s_N\lambda_N}\prod_{i\lt j}\frac{1-\tau^2x_i^{-s_i}x_j^{\pm s_j}}{1-x_i^{-s_i}x_j^{\pm s_j}}, \\
P^\lambda_{Sp(N)}(x_1,\dots,x_{N})&=W_\lambda(\tau)^{-1}\sum_{\sigma \in S_{N}}\sum_{s_1,\dots, s_{N}=\pm 1}x_{\sigma(1)}^{s_1\lambda_1}\cdots x_{\sigma(N)}^{s_{N}\lambda_{N}}\prod_{i\lt j}\frac{1-\tau^2x_i^{-s_i}x_j^{\pm s_j}}{1-x_i^{-s_i}x_j^{\pm s_j}}\\
&\times \prod_{i=1}^{N}\frac{1-\tau^2x_i^{-2s_i}}{1-x_i^{-2s_i}}, 
\end{split}
\end{displaymath}
where
\begin{displaymath}
W_\lambda(\tau)=\left(\sum_{\stackrel{w \in W}{w\lambda=\lambda}}\tau^{2\ell(w)}\right)^\frac{1}{2}
\end{displaymath}
with $\ell(w)$ denoting the length of the Weyl group element $w$.

\item The prescription for writing the $\mathcal{K}$-factors can be found in \cite{Lemos:2012ph}. Their precise form will not be important here.
\item The sum runs over all partitions $\lambda'=(\lambda_1',\dots,\lambda_{N-1}')$ corresponding to the highest weight of a finite-dimensional irreducible representation of $Sp(N-1)$ (in the standard orthonormal basis); ``$\lambda=\lambda'$'' means that we only sum over representations of $SO(2N)$ of the form $\lambda=(\lambda_1',\dots,\lambda_{N-1}',0)$.
\item The fugacities $\mathbf{a}_I$ dual to the Cartan subalgebra of the flavor symmetry group of the puncture $\Lambda_I$ ($\tilde{\Lambda}_I$) are assigned by setting the character of the fundamental representation of $SO(2N)$ ($Sp(N-1)$) equal to the sum of $SU(2)$ characters corresponding to the decomposition determined by the puncture, with $SU(2)$ fugacity equal to $\tau$. The multiplicity of each $SU(2)$ representation is then replaced by the character of the fundamental representation of the flavor symmetry determined by that multiplicity. From this equation, one can simply read off the fugacities. \footnote{If the puncture is not ``very even'', different choices of fugacities are related by a Weyl transformation, under which the Hall-Littlewood polynomials are invariant. For ``very even'' punctures there are two inequivalent choices, which are permuted by the $\mathbb{Z}_2$ outer-automorphism, corresponding to the red and blue coloring. For examples, see \cite{Lemos:2012ph}.} 
\end{itemize}

For example, the $D_4$ twisted puncture $\begin{matrix} \includegraphics[width=39pt]{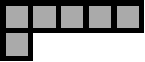}\end{matrix}$ corresponds to the $SU(2)$ embedding under which the $6$ of $Sp(3)$ decomposes as $2+4(1)$. So setting

\begin{displaymath}
\begin{split}
\chi^\mathbf{6}_{Sp(3)}(x_1,x_2,x_3)&=1\cdot\chi^\mathbf{2}_{SU(2)}(\tau)+\chi^\mathbf{4}_{Sp(2)}(a_1,a_2) \cdot 1 \\
\sum_{i=1}^3(x_i+x_i^{-1})&=\tau+\tau^{-1}+\sum_{i=1}^2(a_i+a_i^{-1})
\end{split}
\end{displaymath}
we can take fugacities $x_1=\tau, x_2=a_1, x_3=a_2$.

To determine the global symmetry, as well as any decoupled sector, of an interacting SCFT fixture from its superconformal index, we need only compute \eqref{index} to order $\tau^2$: as explained in \cite{Gaiotto:2012uq}, the contribution at order $\tau$ is due to free hypermultiplets while the contribution at order $\tau^2$ is due to moment map operators of flavor symmetries.

Computing the index to order $\tau^2$ while keeping only the term $\lambda'=0$ in the sum over representations gives the contribution
\begin{displaymath}
1+(\chi^\mathbf{adj}_{G_1}+\chi^\mathbf{adj}_{G_2}+\chi^\mathbf{adj}_{G_3})\tau^2,
\end{displaymath}
encoding the manifest global symmetry. The global symmetry of the SCFT is enhanced if there are additional terms contributing at order $\tau^2$ coming from the sum over $\lambda'\gt 0$.

As an example, consider the fixture

\begin{displaymath}
 \includegraphics[width=114pt]{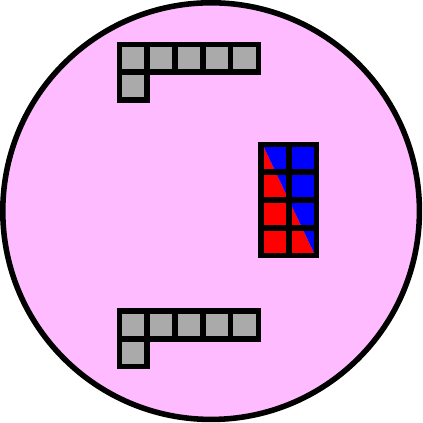}\quad.
\end{displaymath}
Letting $(a_1,a_2), (b_1,b_2)$ be $Sp(2)$ fugacities and $c$ an $SU(2)$ fugacity, from \eqref{index} we find

\begin{displaymath}
\begin{split}
\mathcal{I}&=1+\chi^\mathbf{2}_{SU(2)}(c)\tau+[2\chi^\mathbf{3}_{SU(2)}(c)+\chi^\mathbf{10}_{Sp(2)}(a_1,a_2)+\chi^\mathbf{10}_{Sp(2)}(b_1,b_2)+\chi^\mathbf{4}_{Sp(2)}(a_1,a_2)\chi^\mathbf{4}_{Sp(2)}(b_1,b_2)\\
&+\chi^\mathbf{2}_{SU(2)}(c)(\chi^\mathbf{4}_{Sp(2)}(a_1,a_2)+\chi^\mathbf{4}_{Sp(2)}(b_1,b_2))]\tau^2+\dots \\
&=1+\chi^\mathbf{2}_{SU(2)}(c)\tau+[2\chi^\mathbf{3}_{SU(2)}(c)+\chi^\mathbf{36}_{Sp(4)}(a_1,a_2,b_1,b_2)+\chi^\mathbf{2}_{SU(2)}(c) \chi^\mathbf{8}_{Sp(4)}(a_1,a_2,b_1,b_2)]\tau^2+\dots \\
\end{split}
\end{displaymath}
The order $\tau$ term signals the contribution of a free hypermultiplet in the $\frac{1}{2}(1,1,2)$ of $Sp(2) \times Sp(2) \times SU(2)$, the index of which is given by

\begin{displaymath}
\mathcal{I}_\text{free}=PE[\tau\chi^\mathbf{2}_{SU(2)}(c)]=1+\chi^\mathbf{2}_{SU(2)}(c)\tau+\chi^\mathbf{3}_{SU(2)}(c)\tau^2+\dots,
\end{displaymath}
where $PE$ denotes the plethystic exponential \cite{Lemos:2012ph}. Removing the contribution of the free hypermultiplet, the index of the interacting SCFT is given by

\begin{displaymath}
\begin{split}
\mathcal{I}_{SCFT}&=\mathcal{I} / \mathcal{I}_\text{free}\\
&=1+[\chi^\mathbf{3}_{SU(2)}(c)+ \chi^\mathbf{36}_{Sp(4)}(a_1,a_2,b_1,b_2)+\chi^\mathbf{2}_{SU(2)}(c)\chi^\mathbf{8}_{Sp(4)}(a_1,a_2,b_1,b_2)]\tau^2+\dots \\
&=1+\chi^\mathbf{55}_{Sp(5)}(a_1,a_2,b_1,b_2,c)\tau^2+\dots
\end{split}
\end{displaymath}
and hence this SCFT has an enhanced $Sp(5)$ global symmetry.

We can also use the second order expansion of \eqref{index} as a check on our identifications for the gauge theory fixtures. For example, the fixture

\begin{displaymath}
\begin{matrix} \includegraphics[width=114pt]{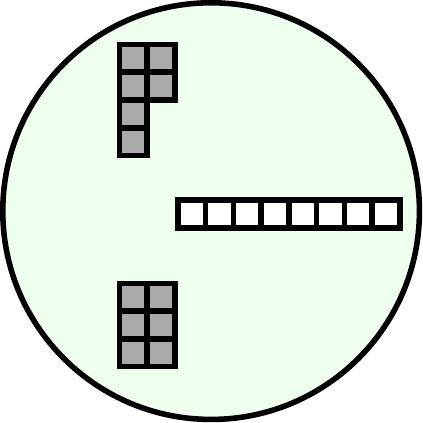}\end{matrix}
\end{displaymath}
is an $SU(2) \times SU(2)$ gauge theory with 4 hypermultiplets in the $(2,1)$, 4 hypermultiplets in the $(1,2)$, and 8 free hypermultiplets transforming in the $\frac{1}{2}(2,8_v)$ of the manifest $SU(2)_8 \times SO(8)_{12}$ global symmetry. Thus the manifest global symmetry of this fixture should be enhanced to $SO(8)^2 \times Sp(8)$. Choosing $(b;c_1,c_2,c_3,c_4)$ as fugacities for the manifest global symmetry, indeed we find the expansion of the index is given by

\begin{displaymath}
\mathcal{I}=1+\chi^\mathbf{2}_{SU(2)}(b)\chi^\mathbf{8_v}_{SO(8)}(c_1,c_2,c_3,c_4)\tau+(2 \chi^\mathbf{28}_{SO(8)}(c_1,c_2,c_3,c_4)+\chi^\mathbf{136}_{Sp(8)}(b,c_1,c_2,c_3,c_4))\tau^2+\dots
\end{displaymath}
where
\begin{displaymath}
\chi^\mathbf{136}_{Sp(8)}(b,c_1,c_2,c_3,c_4)=\chi^\mathbf{3}_{SU(2)}(b)+\chi^\mathbf{28}_{SO(8)}(c_1,c_2,c_3,c_4)+\chi^\mathbf{3}_{SU(2)}(b)\chi^\mathbf{35_v}_{SO(8)}(c_1,c_2,c_3,c_4).
\end{displaymath}
We have used this technique to check the global symmetries and the number of free hypermultiplets in our tables of fixtures for the $\mathbb{Z}_2$-twisted $D_4$ theory.

\subsubsection{{The ${Sp(4)}_6\times {SU(2)}_8$ SCFT}}\label{the__scft}

Here we use the superconformal index to argue that the $D_4$ interacting fixture

\begin{displaymath}
 \includegraphics[width=114pt]{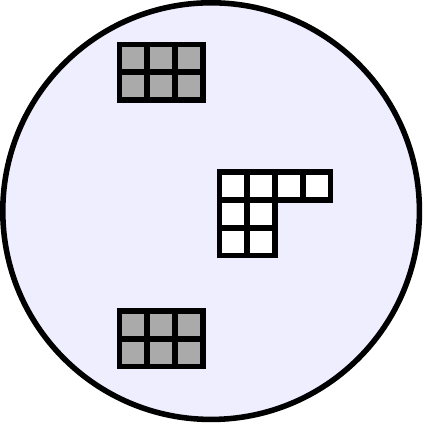}
\end{displaymath}
gives rise to the $Sp(4)_6 \times SU(2)_8$ SCFT. For this fixture, we cannot use any S-dualities to study its properties as none of the flavor symmetries carried by the punctures can be gauged.

The $Sp(4)_6 \times SU(2)_8$ SCFT first appeared in \cite{Chacaltana:2012ch} as the twisted-sector fixture

\begin{displaymath}
 \includegraphics[width=114pt]{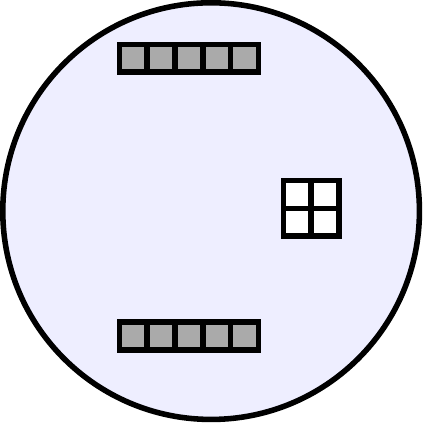}
\end{displaymath}
in the $A_3$ theory. It also appears, accompanied by six free hypermultiplets, as

\begin{equation}
 \begin{matrix}\includegraphics[width=114pt]{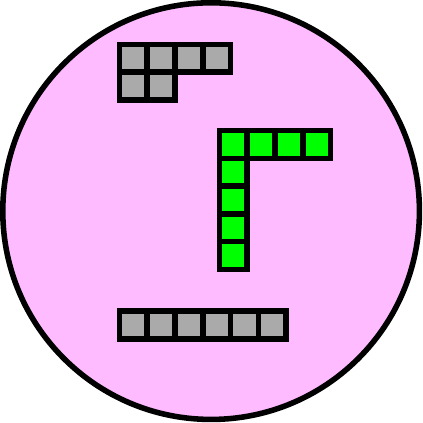}\end{matrix}
\label{D4mixed2211_111111_5111}\end{equation}
in our list of twisted-sector mixed fixtures in the $D_4$ theory. In those cases, we are able to use various S-dualities to study it.

Letting $a$ and $b$ be $SU(2)$ fugacities and $c_1^2, c_2^2$ $U(1)$ fugacities, the expansion of the index of this fixture is given by

\begin{equation}
\begin{split}
\mathcal{I}&=1+(\chi^\mathbf{3}_{SU(2)}(a)+\chi^\mathbf{3}_{SU(2)}(b)+(1+c_1^2+c_1^{-2})+\chi^\mathbf{3}_{SU(2)}(a)\chi^\mathbf{3}_{SU(2)}(b)(1+c_1^2+c_1^{-2})  \\
&+( 1+c_2^2+c_2^{-2}))\tau^2+\dots \\
&=1+(\chi^\mathbf{3}_{SU(2)}(a)+\chi^\mathbf{3}_{SU(2)}(b)+\chi^\mathbf{3}_{SU(2)}(c_1)+\chi^\mathbf{3}_{SU(2)}(a)\chi^\mathbf{3}_{SU(2)}(b) \chi^\mathbf{3}_{SU(2)}(c_1)+\chi^\mathbf{3}_{SU(2)}(c_2))\tau^2+\dots \\
&=1+(\chi^\mathbf{36}_{Sp(4)}(a,b,c_1)+ \chi^\mathbf{3}_{SU(2)}(c_2))\tau^2+\dots,  \\
\end{split}
\label{fixture73}\end{equation}
indicating that the manifest $SU(2)_{24}^2 \times U(1)^2$ global symmetry is enhanced to $Sp(4) \times SU(2)$. This, along with the other numerical invariants of this fixture agree with our previous results for the $Sp(4)_6 \times SU(2)_8$ SCFT.

Since $A_3 \cong D_3$, we can use \eqref{index} to compute the index of the twisted $A_3$ fixture by appropriately identifying fugacities and replacing $P^\lambda_{SO(6)} (P^{\lambda'}_{Sp(2)}) \to P^{\mu}_{SU(4)} (P^{\mu'}_{SO(5)})$ where $\mu$ ($\mu'$) is the highest weight of the $SU(4)$ ($SO(5)$) representation corresponding to $\lambda$ ($\lambda'$). Letting $a$ be an $SU(2)$ fugacity and $(b_1,b_2), (c_1,c_2)$ $SO(5)$ fugacities, the expansion of the index of the twisted $A_3$ fixture is

\begin{displaymath}
\begin{split}
\mathcal{I}&=1+(\chi^\mathbf{3}_{SU(2)}(a)+\chi^\mathbf{10}_{Sp(2)}(\sqrt{b_1b_2},\sqrt{\frac{b_1}{b_2}})+\chi^\mathbf{10}_{Sp(2)}(\sqrt{c_1c_2},\sqrt{\frac{c_1}{c_2}})\\
&+\chi^\mathbf{4}_{Sp(2)}(\sqrt{b_1b_2},\sqrt{\frac{b_1}{b_2}}))\chi^\mathbf{4}_{Sp(2)}(\sqrt{c_1c_2},\sqrt{\frac{c_1}{c_2}})\tau^2+\dots \\
&=1+(\chi^\mathbf{3}_{SU(2)}(a)+\chi^\mathbf{36}_{Sp(4)}(\sqrt{b_1b_2},\sqrt{\frac{b_1}{b_2}}, \sqrt{c_1c_2},\sqrt{\frac{c_1}{c_2}}))\tau^2+\dots,
\end{split}
\end{displaymath}
in agreement with \eqref{fixture73}. We have checked further that the unrefined indices (obtained by setting all flavor fugacities to ``$1$") of these two fixtures agree to tenth order in $\tau$. The unrefined index of each fixture is given by

\begin{displaymath}
\mathcal{I}=1+39\tau^2+878\tau^4+13396\tau^6+152412\tau^8+1370975\tau^{10}+\dots.
\end{displaymath}
We can also compare with the mixed fixture \eqref{D4mixed2211_111111_5111}. After removing the contribution to the index of a free hypermultiplet in the $6$ of $Sp(3)$, the index of this fixture is given by

\begin{displaymath}
\begin{split}
\mathcal{I}&=1+(\chi^\mathbf{3}_{SU(2)}(a_2)+\chi^\mathbf{21}_{Sp(3)}(b_1,b_2,b_3)+\chi^\mathbf{2}_{SU(2)}(a_2)\chi^\mathbf{6}_{Sp(3)}(b_1,b_2,b_3)+\chi^\mathbf{3}_{SU(2)}(c))\tau^2+\dots \\
&=1+(\chi^\mathbf{36}_{Sp(4)}(a_2,b_1,b_2,b_3)+\chi^\mathbf{3}_{SU(2)}(c))\tau^2+\dots.
\end{split}
\end{displaymath}
Again, the numerical invariants of this fixture imply the SCFT is the $Sp(4)_6\times SU(2)_8$ theory. We have computed the unrefined index of this fixture to fourth order in $\tau$; removing the contribution of the free hypermultiplet, we find agreement with the fixtures above.

\section{{The $\mathbb{Z}_2$-twisted $D_4$ Theory}}\label{the_twisted__theory_2}

\subsection{{Punctures and Cylinders}}\label{punctures_and_cylinders}

\subsubsection{{Regular Punctures}}\label{regular_punctures}

The untwisted sector of regular punctures was discussed in \cite{Chacaltana:2011ze}. The $\mathbb{Z}_2$-twisted regular punctures are shown in the Table below.

{
\renewcommand{\arraystretch}{1.5}
\begin{longtable}{|c|c|c|c|c|c|}
\hline
\begin{tabular}{c}Flavour\\C-partition\end{tabular}&\begin{tabular}{c}Hitchin\\B-partition\end{tabular}&\begin{tabular}{c}Pole\\ structure\end{tabular}&Constraints&\begin{tabular}{c}Flavour\\ group\end{tabular}&$(\delta n_{h},\delta n_{v})$\\
\hline
\endhead 
$\begin{matrix}\includegraphics[width=47pt]{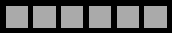}\end{matrix}$&$[7]$&$\{1,3,5;\frac{7}{2}\}$&$-$&${Sp(3)}_8$&$(112,\tfrac{207}{2})$\\
\hline
$\begin{matrix}\includegraphics[width=39pt]{D4twisted21111}\end{matrix}(\text{ns})$&$([5,1^{2}],\mathbb{Z}_2)$&$\{1,3,5;\frac{5}{2}\}$&$-$&${Sp(2)}_7$&$(102,\tfrac{193}{2})$\\
\hline
$\begin{matrix}\includegraphics[width=32pt]{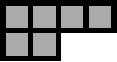}\end{matrix}$&$[5,1^{2}]$&$\{1,3,5;\frac{5}{2}\}$&$c^{(6)}_{5}=(a^{(3)})^{2}$&${SU(2)}_6\times U(1)$&$(94,\tfrac{181}{2})$\\
\hline
$\begin{matrix}\includegraphics[width=24pt]{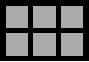}\end{matrix}$&$[3^{2},1]$&$\{1,3,4;\frac{5}{2}\}$&$-$&$SU(2)_{24}$&$(88,\tfrac{171}{2})$\\
\hline
$\begin{matrix}\includegraphics[width=17pt]{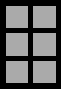}\end{matrix}$&$[3,2^{2}]$&$\{1,3,4;\frac{5}{2}\}$&$\begin{gathered}c^{(4)}_{3}=(a^{(2)})^{2}\\c^{(6)}_4=2a^{(2)}\tilde{c}_{5/2}\end{gathered}$&$SU(2)_8$&$(72,\tfrac{141}{2})$\\
\hline
$\begin{matrix} \includegraphics[width=24pt]{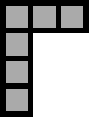}\end{matrix}(\text{ns})$&$([3,1^{4}],\mathbb{Z}_2)$&$\{1,3,3;\frac{3}{2}\}$&$-$&${SU(2)}_5$&$(69,\tfrac{135}{2})$\\
\hline
$\begin{matrix}\includegraphics[width=17pt]{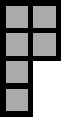}\end{matrix}$&$[3,1^{4}]$&$\{1,3,3;\frac{3}{2}\}$&$c^{(4)}_{3}=(a^{(2)})^{2}$&none&$(64,\tfrac{127}{2})$\\
\hline
$\begin{matrix}\includegraphics[width=9pt]{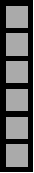}\end{matrix}$&$[1^{7}]$&$\{1,1,1;\frac{1}{2}\}$&$-$&none&$(24,\tfrac{49}{2})$\\
\hline
\end{longtable}
}

\subsubsection{{Irregular Punctures}}\label{irregular_punctures}

A fairly lengthy list of irregular untwisted punctures, arising from the OPE of untwisted punctures, was discussed in \cite{Chacaltana:2011ze}. Additional ones arise from considering the OPE of two $\mathbb{Z}_2$-twisted punctures. Moreover, twisted-sector irregular twisted punctures arise from the OPE of an untwisted puncture and a $\mathbb{Z}_2$-twisted puncture. These two sets of new irregular punctures are listed in the Tables below.

\paragraph{{Untwisted}}\label{untwisted}{~}
\medskip

{
\renewcommand{\arraystretch}{1.5}
\begin{longtable}{|c|c|c|}
\hline
Irregular puncture&$(n_h,n_v)$&Flavour Symmetry\\
\endhead
\hline 
$\bigl(\begin{matrix} \includegraphics[width=62pt]{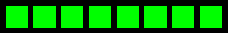}\end{matrix}\,,Sp(2)\bigr)$&$(112,118)$&${Sp(2)}_0$\\
\hline 
$\Bigl(\begin{matrix} \includegraphics[width=47pt]{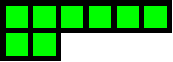}\end{matrix},SU(2)\times SU(2)\Bigr)$&$(128,133)$&${SU(2)}_0\times {SU(2)}_0$\\
\hline 
$\biggl(\begin{matrix} \includegraphics[width=47pt]{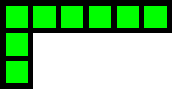}\end{matrix},SU(2)\times SU(2)\biggr)$&$(136,140)$&${SU(2)}_0\times {SU(2)}_0$\\
\hline 
$\Biggl(\begin{matrix} \includegraphics[width=32pt]{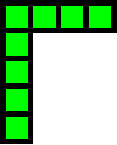}\end{matrix},SU(2)\Biggr)$&$(176,179)$&${SU(2)}_0$\\
\hline 
\end{longtable}
}

As was the case in \cite{Chacaltana:2011ze}, there are three inequivalent embeddings of $Sp(2)\hookrightarrow Spin(8)$, exchanged by triality, under which one of the 8-dimensional representations decomposes as $5+3(1)$ while the other two decompose as $2(4)$. To indicate which we mean, we assign a green/red/blue colour to $ \includegraphics[width=62pt]{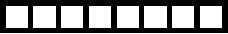}$. The same remark applies to the three index-1 embeddings of $SU(2)\times SU(2)$ in the ${SU(2)}^3$ of $ \includegraphics[width=47pt]{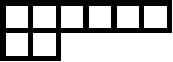}$ which are exchanged by triality.

\paragraph{{Twisted}}\label{twisted} {~}

{
\renewcommand{\arraystretch}{1.5}
\begin{longtable}{|c|c|c|}
\hline 
Irregular puncture&$(n_h,n_v)$&Flavour Symmetry\\
\hline
\endhead
$\bigl(\begin{matrix} \includegraphics[width=47pt]{D4twisted111111}\end{matrix}\,,Sp(2)\times SU(2)\bigr)$&$\left(112,\tfrac{223}{2}\right)$&${Sp(2)}_4\times{SU(2)}_0$\\
\hline 
$\bigl(\begin{matrix} \includegraphics[width=47pt]{D4twisted111111}\end{matrix}\,,Sp(2)\bigr)$&$\left(112,\tfrac{229}{2}\right)$&${Sp(2)}_4$\\
\hline 
$\bigl(\begin{matrix} \includegraphics[width=47pt]{D4twisted111111}\end{matrix}\,,SU(2)\times SU(2)\bigr)$&$\left(112,\tfrac{237}{2}\right)$&${SU(2)}_0\times{SU(2)}_0$\\
\hline 
$\bigl(\begin{matrix} \includegraphics[width=47pt]{D4twisted111111}\end{matrix}\,,SU(2)\bigr)$&$\left(112,\tfrac{243}{2}\right)$&${SU(2)}_0$\\
\hline 
$\Bigl(\begin{matrix} \includegraphics[width=39pt]{D4twisted21111}\end{matrix},Sp(2)\Bigr)$&$\left(122,\tfrac{243}{2}\right)$&${Sp(2)}_5$\\
\hline 
$\Bigl(\begin{matrix} \includegraphics[width=39pt]{D4twisted21111}\end{matrix},SU(2)\times SU(2)\Bigr)$&$\left(122,\tfrac{251}{2}\right)$&${SU(2)}_1\times{SU(2)}_1$\\
\hline 
$\Bigl(\begin{matrix} \includegraphics[width=39pt]{D4twisted21111}\end{matrix},SU(2)\Bigr)$&$\left(122,\tfrac{257}{2}\right)$&${SU(2)}_1$\\
\hline 
$\Bigl(\begin{matrix} \includegraphics[width=32pt]{D4twisted2211}\end{matrix},SU(2)\Bigr)$&$\left(130,\tfrac{269}{2}\right)$&${SU(2)}_2$\\
\hline 
$\biggl(\begin{matrix} \includegraphics[width=17pt]{D4twisted33}\end{matrix}\,,\emptyset\biggr)$&$\left(152,\tfrac{315}{2}\right)$&none\\
\hline 
$\Biggl(\begin{matrix} \includegraphics[width=24pt]{D4twisted411}\end{matrix},SU(2)\Biggr)$&$\left(155,\tfrac{315}{2}\right)$&${SU(2)}_3$\\
\hline 
$\Biggl(\begin{matrix} \includegraphics[width=17pt]{D4twisted42}\end{matrix},\emptyset\Biggr)$&$\left(155,\tfrac{315}{2}\right)$&none\\
\hline 
\end{longtable}
}

\subsubsection{{Cylinders}}\label{cylinders}

In addition to the untwisted cylinders of \cite{Chacaltana:2011ze}, we have

\begin{displaymath}
\begin{aligned}
\bigl(\begin{matrix} \includegraphics[width=62pt]{D4untwisted11111111G}\end{matrix}\,,Sp(2)\bigr)
&\xleftrightarrow{\qquad Sp(2)\qquad}
\begin{matrix} \includegraphics[width=62pt]{D4untwisted11111111}\end{matrix}\\
\bigl(\begin{matrix} \includegraphics[width=62pt]{D4untwisted11111111G}\end{matrix}\,,Sp(2)\bigr)
&\xleftrightarrow{\qquad SU(2)\qquad}\bigl(\begin{matrix} \includegraphics[width=62pt]{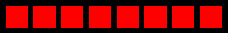}\end{matrix},Spin(7)\bigr)\\
\bigl(\begin{matrix} \includegraphics[width=62pt]{D4untwisted11111111G}\end{matrix}\,,Sp(2)\bigr)
&\xleftrightarrow{\qquad SU(2)\qquad}\bigl(\begin{matrix} \includegraphics[width=62pt]{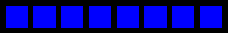}\end{matrix},Spin(7)\bigr)\\
\Bigl(\begin{matrix} \includegraphics[width=47pt]{D4untwisted221111G}\end{matrix},SU(2)\times SU(2)\Bigr)
&\xleftrightarrow{\quad SU(2)\times SU(2)\quad}
\begin{matrix} \includegraphics[width=47pt]{D4untwisted221111}\end{matrix}\\
\biggl(\begin{matrix} \includegraphics[width=47pt]{D4untwisted311111G}\end{matrix},SU(2)\times SU(2)\biggr)
&\xleftrightarrow{\quad SU(2)\times SU(2)\quad}
\begin{matrix} \includegraphics[width=47pt]{D4untwisted311111G}\end{matrix}\\
\Biggl(\begin{matrix} \includegraphics[width=32pt]{D4untwisted5111G}\end{matrix},SU(2)\Biggr)
&\xleftrightarrow{\qquad SU(2)\qquad}
\begin{matrix} \includegraphics[width=32pt]{D4untwisted5111G}\end{matrix}
\end{aligned}
\end{displaymath}
and the twisted sector adds the cylinders

\begin{displaymath}
\begin{aligned}
\begin{matrix} \includegraphics[width=47pt]{D4twisted111111}\end{matrix}
&\xleftrightarrow{\qquad Sp(3)\qquad}
\begin{matrix} \includegraphics[width=47pt]{D4twisted111111}\end{matrix}\\
\bigl(\begin{matrix} \includegraphics[width=47pt]{D4twisted111111}\end{matrix}\,,Sp(2)\times SU(2)\bigr)
&\xleftrightarrow{\quad Sp(2)\times SU(2)\quad}
\begin{matrix} \includegraphics[width=47pt]{D4twisted111111}\end{matrix}\\
\bigl(\begin{matrix} \includegraphics[width=47pt]{D4twisted111111}\end{matrix}\,,Sp(2)\bigr)
&\xleftrightarrow{\qquad Sp(2)\qquad}
\begin{matrix} \includegraphics[width=47pt]{D4twisted111111}\end{matrix}\\
\bigl(\begin{matrix} \includegraphics[width=47pt]{D4twisted111111}\end{matrix}\,,SU(2)\times SU(2)\bigr)
&\xleftrightarrow{\quad SU(2)\times SU(2)\quad}
\begin{matrix} \includegraphics[width=47pt]{D4twisted111111}\end{matrix}\\
\bigl(\begin{matrix} \includegraphics[width=47pt]{D4twisted111111}\end{matrix}\,,SU(2)\bigr)
&\xleftrightarrow{\qquad SU(2)\qquad}
\begin{matrix} \includegraphics[width=47pt]{D4twisted111111}\end{matrix}\\
\Bigl(\begin{matrix} \includegraphics[width=39pt]{D4twisted21111}\end{matrix},Sp(2)\Bigr)
&\xleftrightarrow{\qquad Sp(2)\qquad}
\begin{matrix} \includegraphics[width=39pt]{D4twisted21111}\end{matrix}\\
\Bigl(\begin{matrix} \includegraphics[width=39pt]{D4twisted21111}\end{matrix},SU(2)\times SU(2)\Bigr)
&\xleftrightarrow{\quad SU(2)\times SU(2)\quad}
\begin{matrix} \includegraphics[width=39pt]{D4twisted21111}\end{matrix}\\
\Bigl(\begin{matrix} \includegraphics[width=39pt]{D4twisted21111}\end{matrix},SU(2)\Bigr)
&\xleftrightarrow{\qquad SU(2)\qquad}
\begin{matrix} \includegraphics[width=39pt]{D4twisted21111}\end{matrix}\\
\Bigl(\begin{matrix} \includegraphics[width=32pt]{D4twisted2211}\end{matrix},SU(2)\Bigr)
&\xleftrightarrow{\qquad SU(2)\qquad}
\begin{matrix} \includegraphics[width=32pt]{D4twisted2211}\end{matrix}\\
\biggl(\begin{matrix} \includegraphics[width=17pt]{D4twisted33}\end{matrix}\,,\emptyset\biggr)
&\xleftrightarrow{\qquad \emptyset\qquad}
\begin{matrix} \includegraphics[width=17pt]{D4twisted33}\end{matrix}\\
\Biggl(\begin{matrix} \includegraphics[width=24pt]{D4twisted411}\end{matrix},SU(2)\Biggr)
&\xleftrightarrow{\qquad SU(2)\qquad}
\begin{matrix} \includegraphics[width=24pt]{D4twisted411}\end{matrix}\\
\Biggl(\begin{matrix} \includegraphics[width=17pt]{D4twisted42}\end{matrix},\emptyset\Biggr)
&\xleftrightarrow{\qquad \emptyset\qquad}
\begin{matrix} \includegraphics[width=17pt]{D4twisted42}\end{matrix}
\end{aligned}
\end{displaymath}

\subsection{{Fixtures}}\label{fixtures}

\subsubsection{{Free-field Fixtures}}\label{freefield_fixtures}

\begin{longtable}{|c|c|c|c|}
\hline
\#&Fixture&Number of hypers&Representation\\
\hline
\endhead
1&$\begin{matrix}\includegraphics[width=76pt]{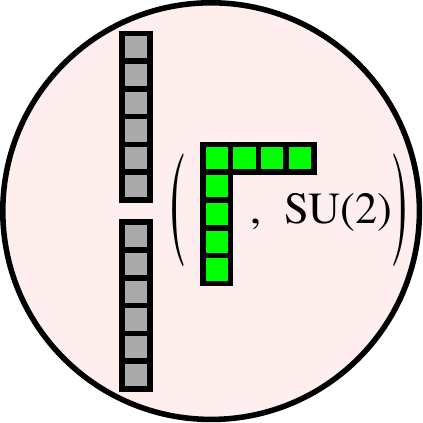}\end{matrix}$&0&empty\\
\hline
2&$\begin{matrix}\includegraphics[width=76pt]{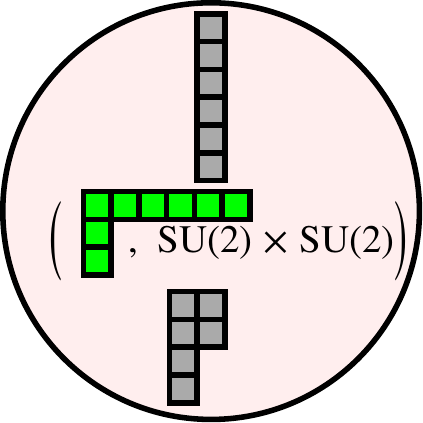}\end{matrix}$&0&empty\\
\hline
3&$\begin{matrix}\includegraphics[width=76pt]{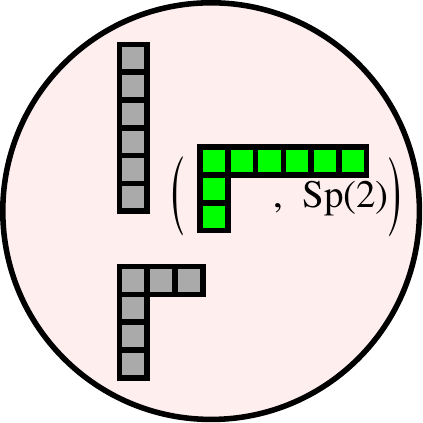}\end{matrix}$&5&$\tfrac{1}{2}(2,5)$\\
\hline
4&$\begin{matrix}\includegraphics[width=76pt]{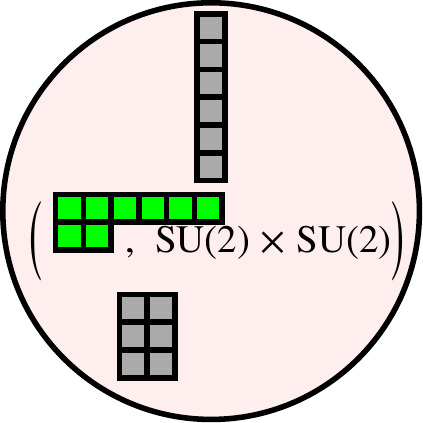}\end{matrix}$&0&empty\\
\hline
5&$\begin{matrix}\includegraphics[width=76pt]{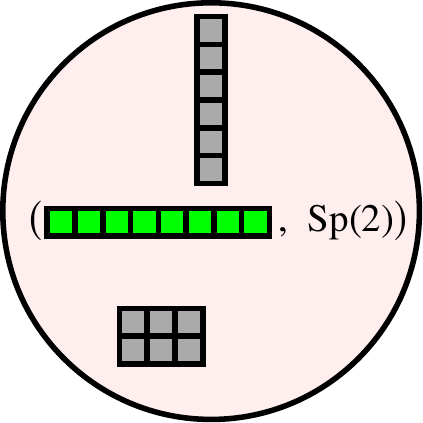}\end{matrix}$&0&empty\\
\hline
6&$\begin{matrix}\includegraphics[width=76pt]{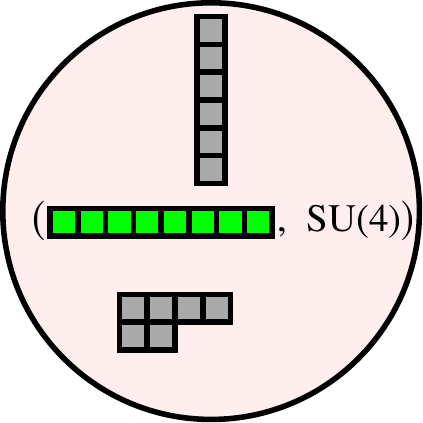}\end{matrix}$&6&$\tfrac{1}{2}(2,6)$\\
\hline
7&$\begin{matrix}\includegraphics[width=76pt]{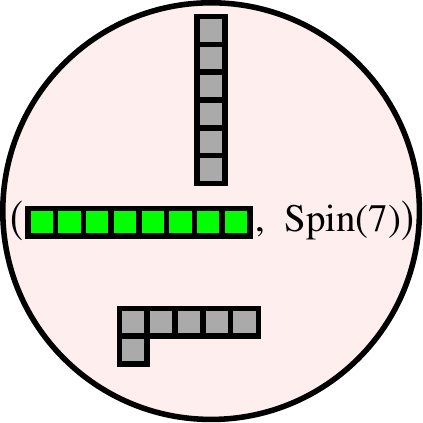}\end{matrix}$&14&$\tfrac{1}{2}(4,7)$\\
\hline
8&$\begin{matrix}\includegraphics[width=76pt]{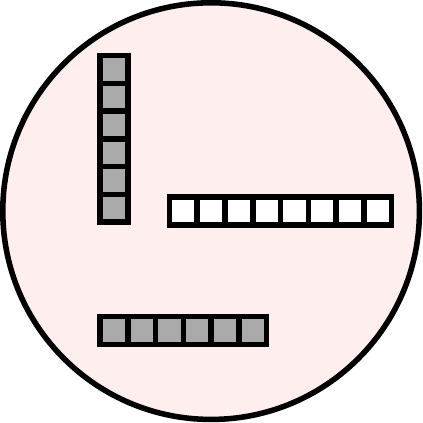}\end{matrix}$&24&$\tfrac{1}{2}(6,8_v)$\\
\hline
9&$\begin{matrix}\includegraphics[width=76pt]{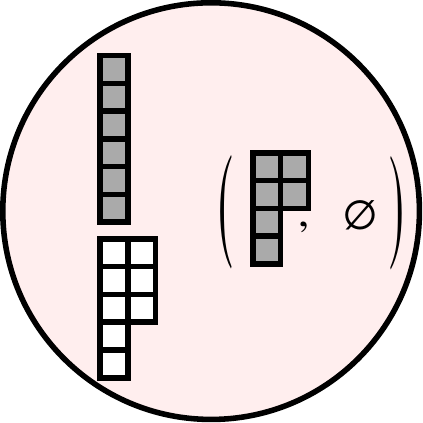}\end{matrix}$&0&empty\\
\hline
10&$\begin{matrix}\includegraphics[width=76pt]{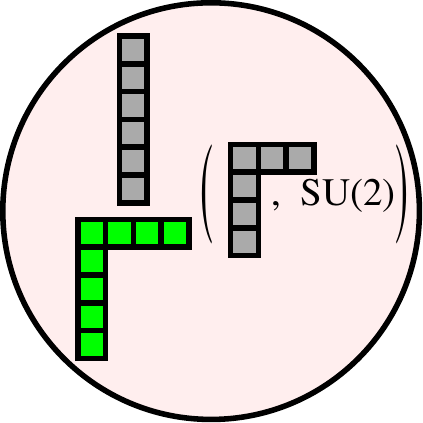}\end{matrix}$&3&$\tfrac{1}{2}(3,2)$\\
\hline
11&$\begin{matrix}\includegraphics[width=76pt]{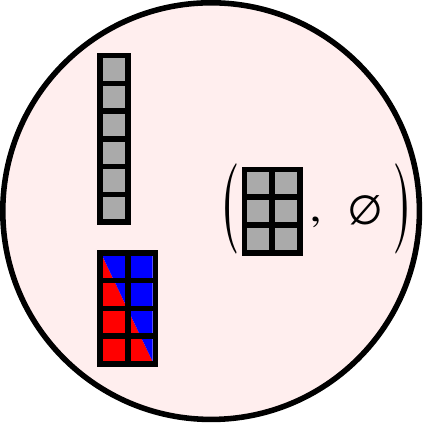}\end{matrix}$&0&empty\\
\hline
12&$\begin{matrix}\includegraphics[width=76pt]{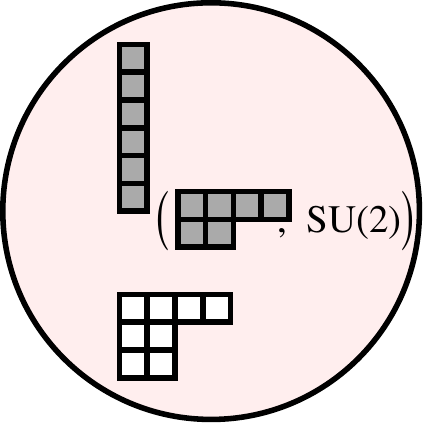}\end{matrix}$&2&$1(2)$\\
\hline
13&$\begin{matrix}\includegraphics[width=76pt]{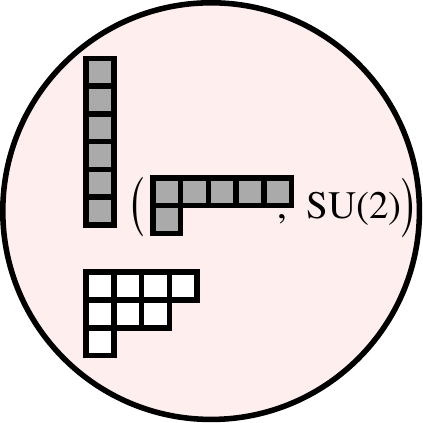}\end{matrix}$&1&$\tfrac{1}{2}(1,2)$\\
\hline
14&$\begin{matrix}\includegraphics[width=76pt]{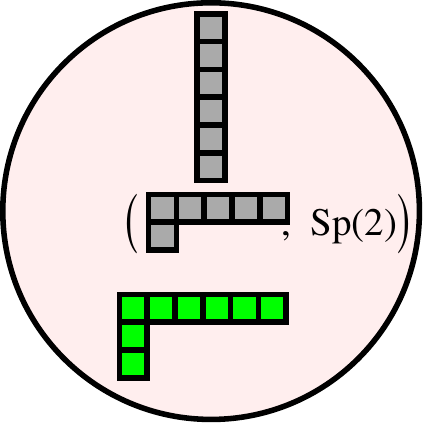}\end{matrix}$&10&$\tfrac{1}{2}(5,4)$\\
\hline
15&$\begin{matrix}\includegraphics[width=76pt]{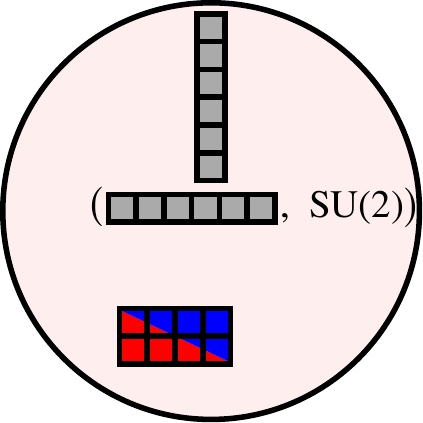}\end{matrix}$&0&empty\\
\hline
16&$\begin{matrix}\includegraphics[width=76pt]{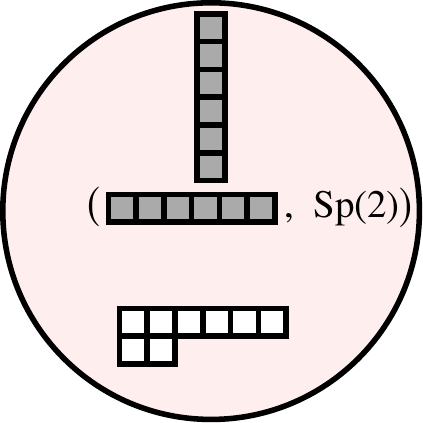}\end{matrix}$&8&$\tfrac{1}{2}(1,2,2;4)$\\
\hline
17&$\begin{matrix}\includegraphics[width=76pt]{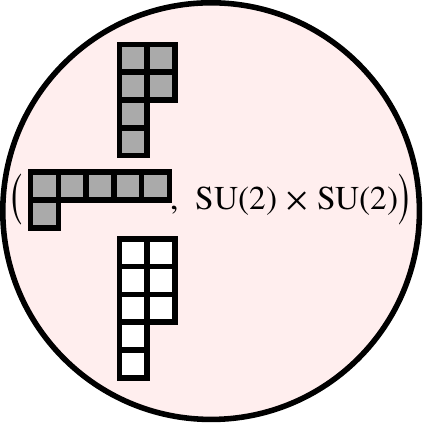}\end{matrix}$&2&$\tfrac{1}{2}(2,1)+\tfrac{1}{2}(1,2)$\\
\hline
18&$\begin{matrix}\includegraphics[width=76pt]{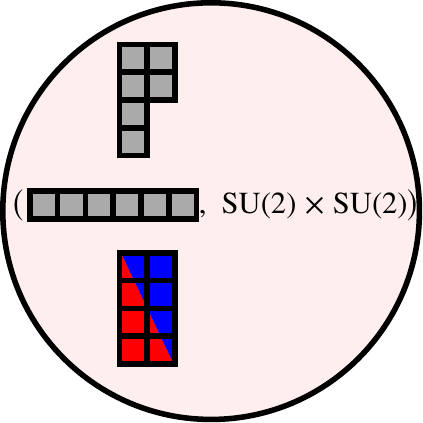}\end{matrix}$&0&empty\\
\hline
19&$\begin{matrix}\includegraphics[width=76pt]{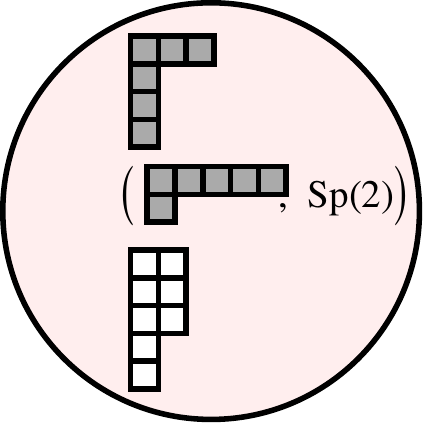}\end{matrix}$&7&$\tfrac{1}{2}(2,5)+\tfrac{1}{2}(1,4)$\\
\hline
20&$\begin{matrix}\includegraphics[width=76pt]{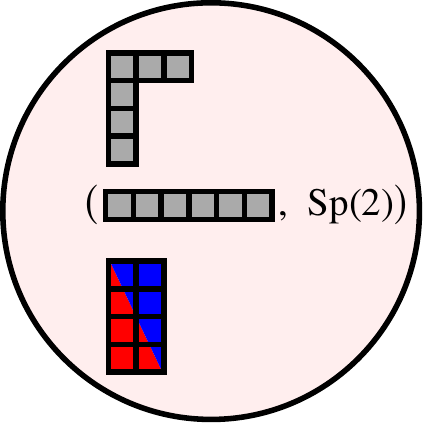}\end{matrix}$&5&$\tfrac{1}{2}(2,1,5)$\\
\hline
21&$\begin{matrix}\includegraphics[width=76pt]{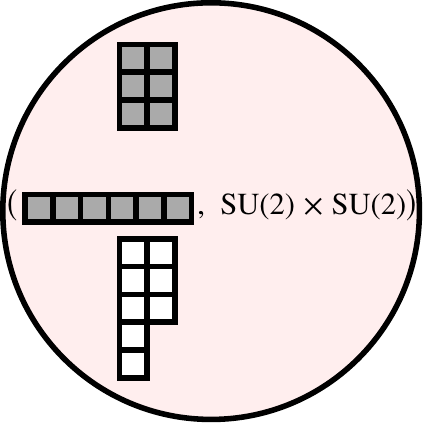}\end{matrix}$&0&empty\\
\hline
22&$\begin{matrix}\includegraphics[width=76pt]{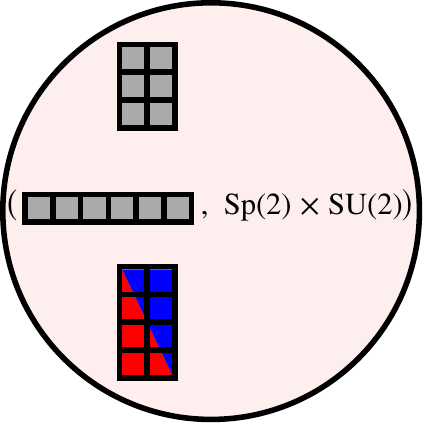}\end{matrix}$&8&$\tfrac{1}{2}(2,2;4,1)$\\
\hline
23&$\begin{matrix}\includegraphics[width=76pt]{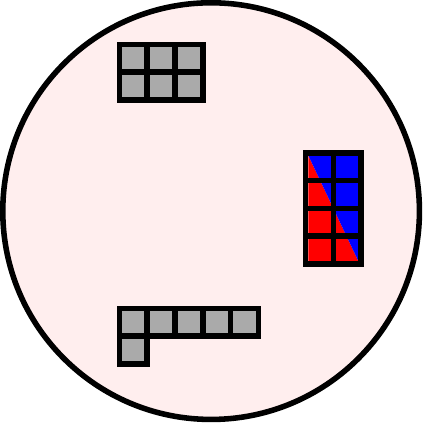}\end{matrix}$&14&$\tfrac{1}{2}(3,4,1)+\tfrac{1}{2}(1,5,2)+\tfrac{1}{2}(3,1,2)$\\
\hline
24&$\begin{matrix}\includegraphics[width=76pt]{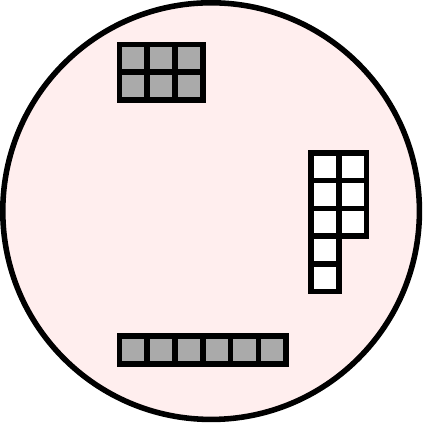}\end{matrix}$&16&$\tfrac{1}{2}(1,14')+\tfrac{1}{2}(3,6)$\\
\hline
\end{longtable}

\subsubsection{{Interacting Fixtures}}\label{interacting_fixtures}

\begin{longtable}{|c|c|c|c|c|}
\hline
\#&Fixture&$(d_2,d_3,d_4,d_5,d_6)$&$(n_h,n_v)$&$G_{\text{global}}$\\
\hline
\endhead
1&$\begin{matrix}\includegraphics[width=76pt]{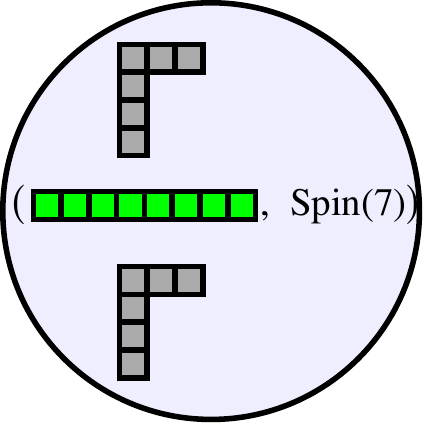}\end{matrix}$&$(0,0,2,0,0)$&$(26,14)$&${Spin(7)}_8\times {SU(2)}_{5}^2$\\
\hline
2&$\begin{matrix}\includegraphics[width=76pt]{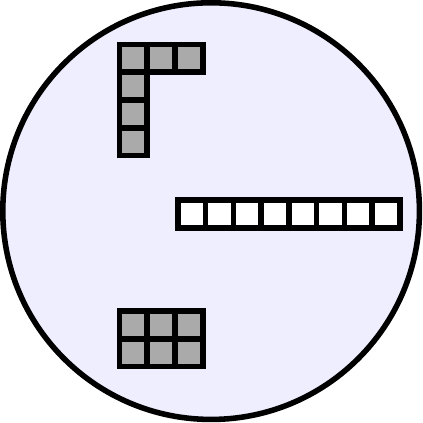}\end{matrix}$&$(0, 0, 2, 0, 1)$&$(45, 25)$&$Spin(11)_{12}\times SU(2)_5$\\
\hline
3&$\begin{matrix}\includegraphics[width=76pt]{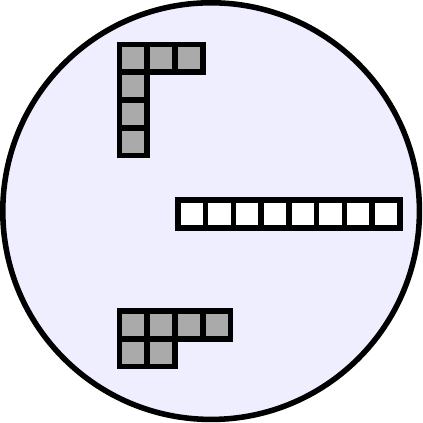}\end{matrix}$&$(0, 1, 2, 0, 1)$&$(51, 30)$&$Spin(10)_{12}\times SU(2)_6\times SU(2)_5$\\
\hline
4&$\begin{matrix}\includegraphics[width=76pt]{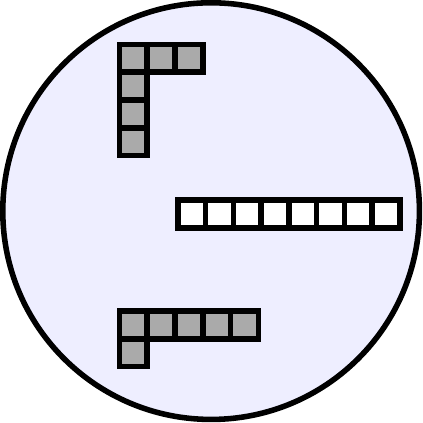}\end{matrix}$&$(0, 0, 2, 0, 2)$&$(59, 36)$&$Spin(9)_{12}\times Sp(2)_7\times SU(2)_5$\\
\hline
5&$\begin{matrix}\includegraphics[width=76pt]{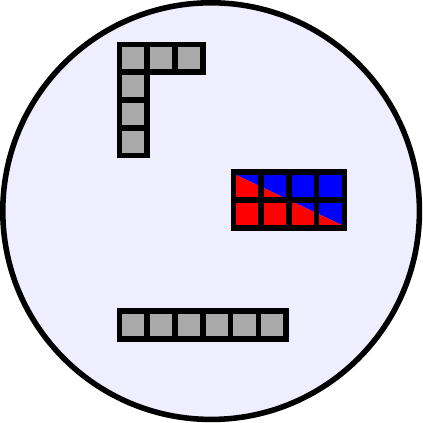}\end{matrix}$&$(0, 0, 2, 0, 1)$&$(45, 25)$&$Sp(5)_8\times SU(2)_5$\\
\hline
6&$\begin{matrix}\includegraphics[width=76pt]{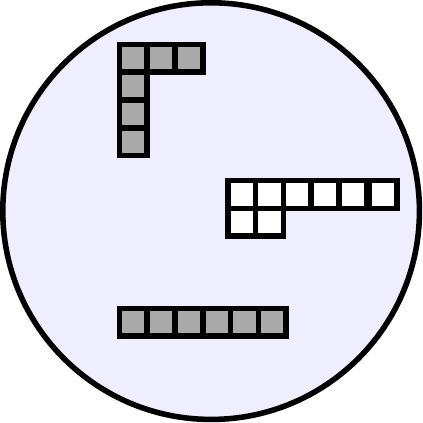}\end{matrix}$&$(0,0,3,0,1)$&$(53,32)$&$Sp(4)_8 \times SU(2)_8^2 \times SU(2)_5$\\
\hline
7&$\begin{matrix}\includegraphics[width=76pt]{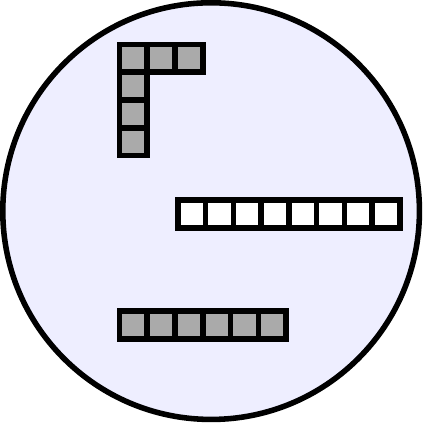}\end{matrix}$&$(0,0,3,0,2)$&$(69,43)$&$Spin(8)_{12} \times Sp(3)_8 \times SU(2)_5$\\
\hline
8&$\begin{matrix}\includegraphics[width=76pt]{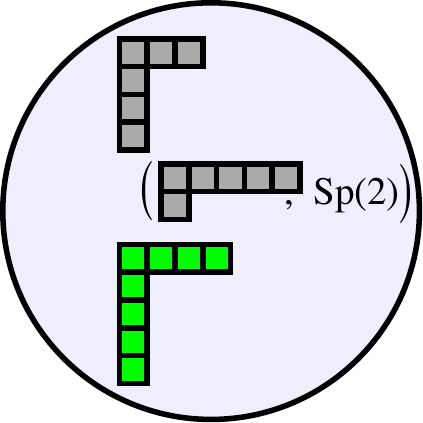}\end{matrix}$&$(0,0,1,0,0)$&(15,7)&$Sp(3)_5 \times SU(2)_8$\\
\hline
9&$\begin{matrix}\includegraphics[width=76pt]{D4interacting222_222_3311}\end{matrix}$&$(0, 1, 1, 0, 0)$&$(24, 12)$&${Sp(4)}_6\times {SU(2)}_8$\\
\hline
10&$\begin{matrix}\includegraphics[width=76pt]{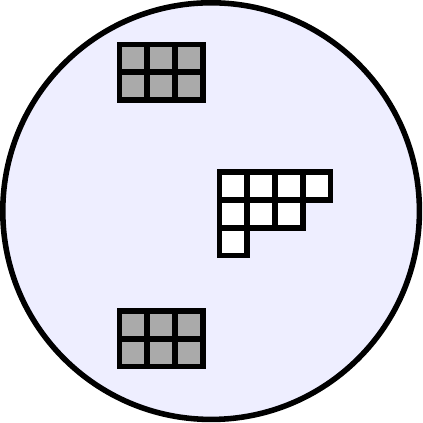}\end{matrix}$&$(0, 0, 1, 0, 1)$&$(31, 18)$&$SU(4)_{12}\times SU(2)_{7}\times U(1)$\\
\hline
11&$\begin{matrix}\includegraphics[width=76pt]{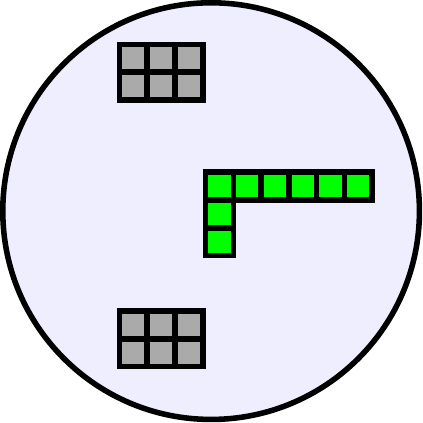}\end{matrix}$&$(0, 0, 2, 0, 1)$&$(40, 25)$&$SU(4)_{12}\times Sp(2)_{8}$\\
\hline
12&$\begin{matrix}\includegraphics[width=76pt]{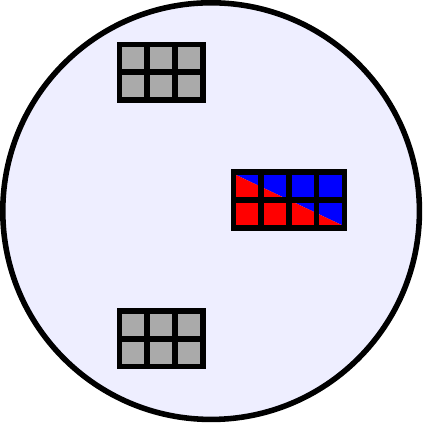}\end{matrix}$&$(0, 0, 2, 0, 1)$&$(40, 25)$&$SU(2)_{24}^2\times Sp(2)_{8}$\\
\hline
13&$\begin{matrix}\includegraphics[width=76pt]{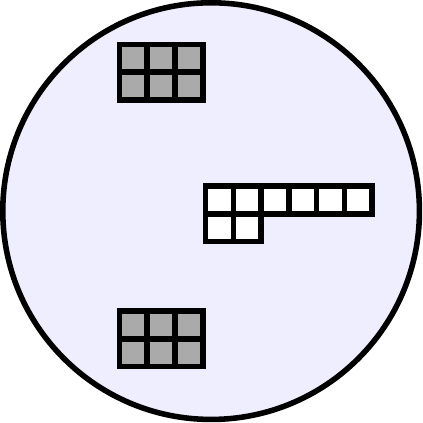}\end{matrix}$&$(0, 0, 3, 0, 1)$&$(48, 32)$&$SU(2)_24^2\times SU(2)_8^3$\\
\hline
14&$\begin{matrix}\includegraphics[width=76pt]{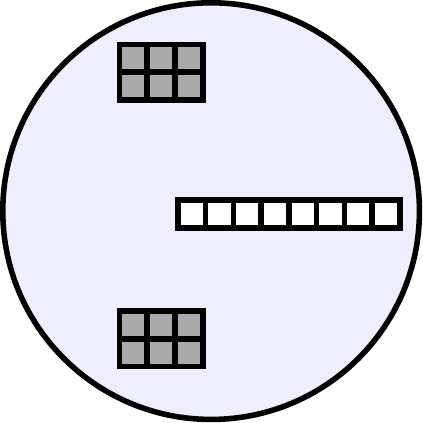}\end{matrix}$&$(0, 0, 3, 0, 2)$&$(64, 43)$&$Spin(8)_{12}\times (SU(2)_{24})^2$\\
\hline
15&$\begin{matrix}\includegraphics[width=76pt]{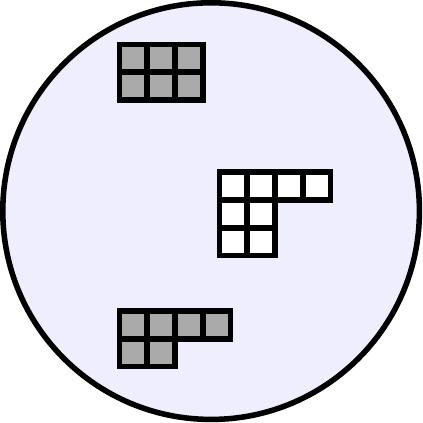}\end{matrix}$&$(0, 2, 1, 0, 0)$&$(30, 17)$&${Sp(2)}_6^2\times SU(2)_{6}\times U(1)$\\
\hline
16&$\begin{matrix}\includegraphics[width=76pt]{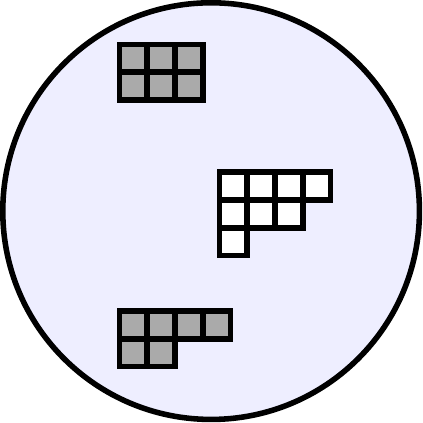}\end{matrix}$&$(0, 1, 1, 0, 1)$&$(37, 23)$&$Sp(2)_{12}\times SU(2)_7\times SU(2)_6$\\
\hline
17&$\begin{matrix}\includegraphics[width=76pt]{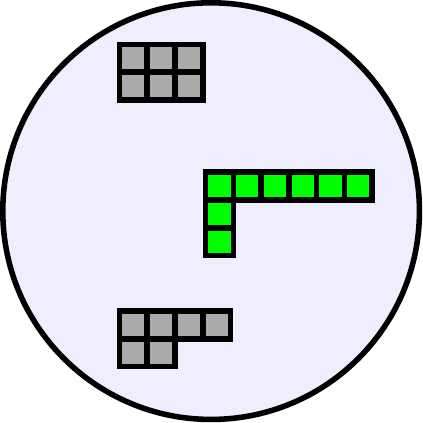}\end{matrix}$&$(0, 1, 2, 0, 1)$&$(46, 30)$&$Sp(2)_{12}\times Sp(2)_8\times SU(2)_6$\\
\hline
18&$\begin{matrix}\includegraphics[width=76pt]{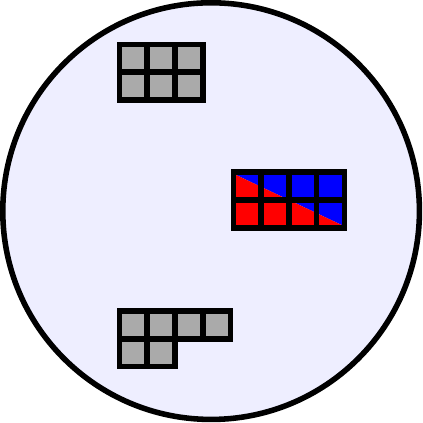}\end{matrix}$&$(0, 1, 2, 0, 1)$&$(46, 30)$&$Sp(2)_8\times SU(2)_{24}\times SU(2)_6\times U(1)$\\
\hline
19&$\begin{matrix}\includegraphics[width=76pt]{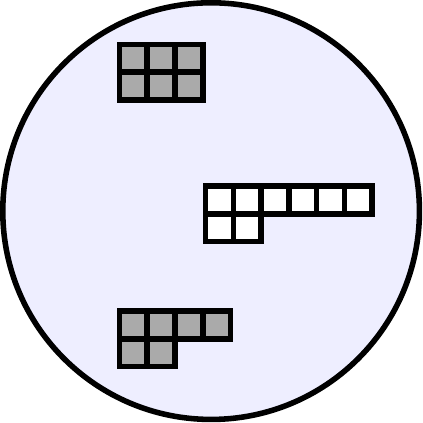}\end{matrix}$&$(0, 1, 3, 0, 1)$&$(54, 37)$&$SU(2)_{24}\times SU(2)^{3}_{8}\times SU(2)_{6}\times U(1)$\\
\hline
20&$\begin{matrix}\includegraphics[width=76pt]{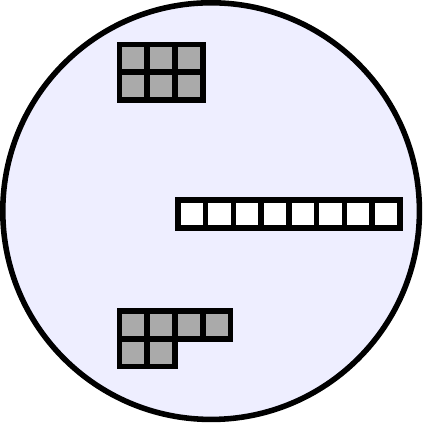}\end{matrix}$&$(0, 1, 3, 0, 2)$&$(70, 48)$&$Spin(8)_{12}\times SU(2)_{24}\times SU(2)_{6}\times U(1)$\\
\hline
21&$\begin{matrix}\includegraphics[width=76pt]{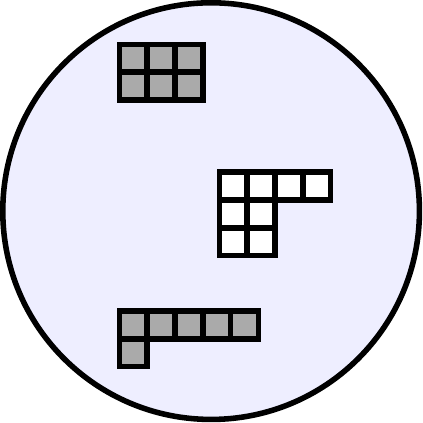}\end{matrix}$&$(0, 1, 1, 0, 1)$&$(38, 23)$&$Sp(2)_{12}\times Sp(2)_7\times U(1)$\\
\hline
22&$\begin{matrix}\includegraphics[width=76pt]{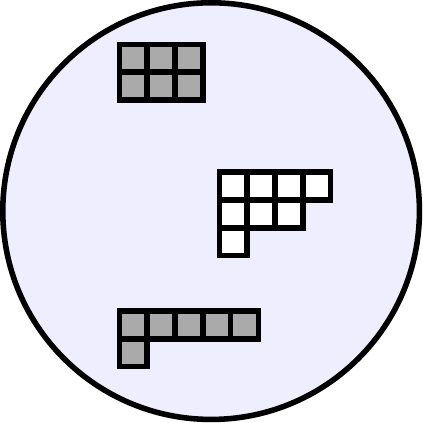}\end{matrix}$&$(0, 0, 1, 0, 2)$&$(45, 29)$&$Sp(2)_{7}\times SU(2)_7\times SU(2)_{12}^2$\\
\hline
23&$\begin{matrix}\includegraphics[width=76pt]{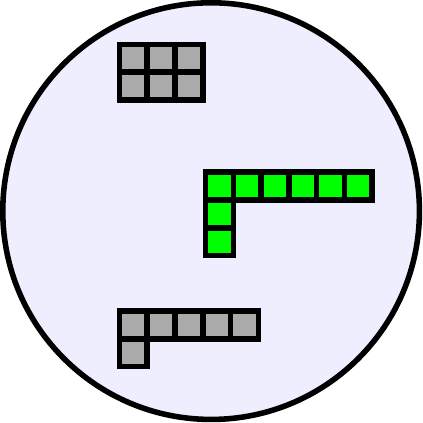}\end{matrix}$&$(0, 0, 2, 0, 2)$&$(54, 36)$&$Sp(2)_8\times Sp(2)_7\times (SU(2)_{12})^2$\\
\hline
24&$\begin{matrix}\includegraphics[width=76pt]{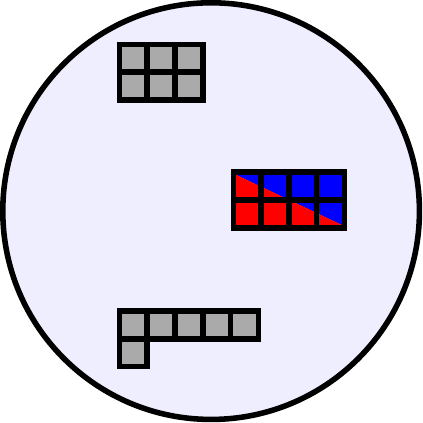}\end{matrix}$&$(0, 0, 2, 0, 2)$&$(54, 36)$&$Sp(2)_8\times Sp(2)_7\times SU(2)_{24}$\\
\hline
25&$\begin{matrix}\includegraphics[width=76pt]{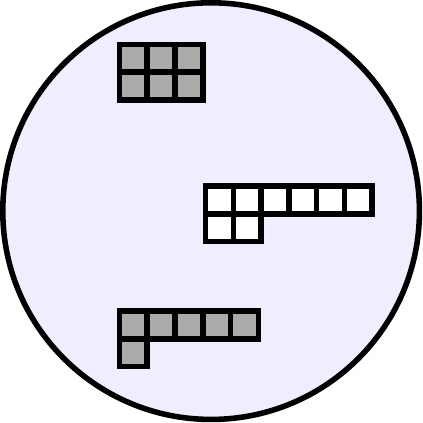}\end{matrix}$&$(0, 0, 3, 0, 2)$&$(62, 43)$&$Sp(2)_7\times SU(2)_{24}\times SU(2)^{3}_{8}$\\
\hline
26&$\begin{matrix}\includegraphics[width=76pt]{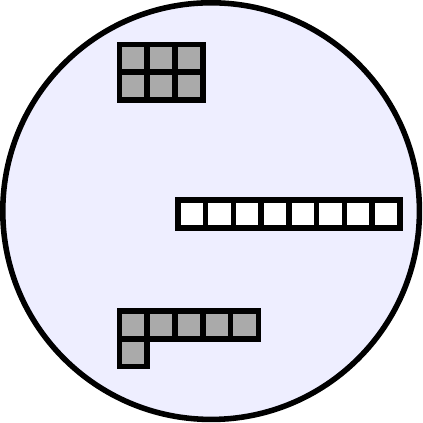}\end{matrix}$&$(0, 0, 3, 0, 3)$&$(78, 54)$&$Spin(8)_{12}\times Sp(2)_7\times SU(2)_{24}$\\
\hline
27&$\begin{matrix}\includegraphics[width=76pt]{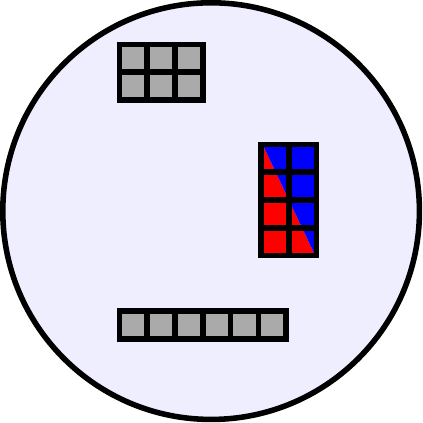}\end{matrix}$&$(0,0,1,0,0)$&$(24,7)$&$(E_7)_8$\\
\hline
28&$\begin{matrix}\includegraphics[width=76pt]{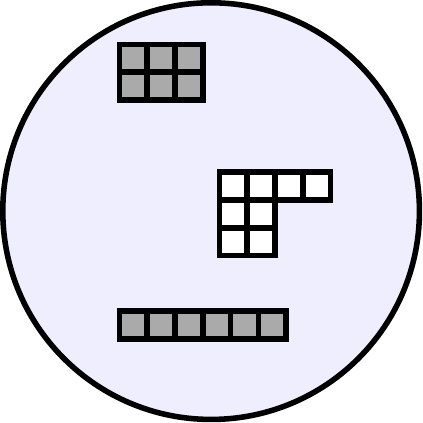}\end{matrix}$&$(0, 1, 2, 0, 1)$&$(48, 30)$&$Sp(3)_8 \times SU(2)_{24}\times U(1)^2$\\
\hline
29&$\begin{matrix}\includegraphics[width=76pt]{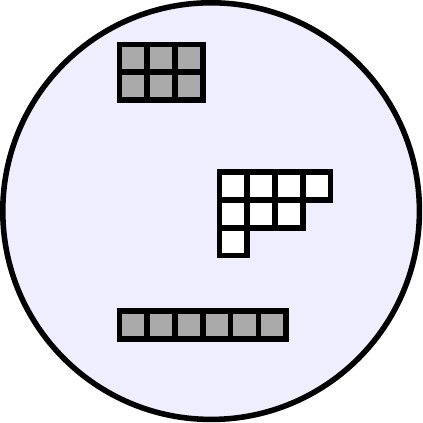}\end{matrix}$&$(0, 0, 2, 0, 2)$&$(55, 36)$&$Sp(3)_{8}\times SU(2)_{24}\times SU(2)_{7}$\\
\hline
30&$\begin{matrix}\includegraphics[width=76pt]{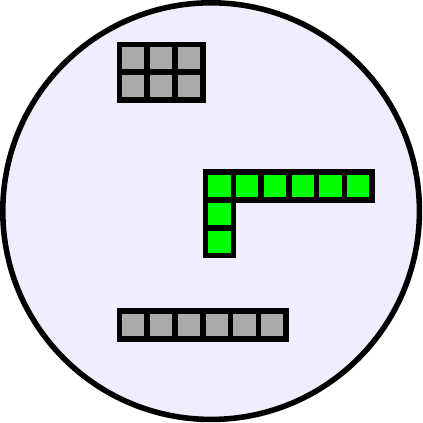}\end{matrix}$&$(0, 0, 3, 0, 2)$&$(64, 43)$&$Sp(3)_{8}\times Sp(2)_{8}\times SU(2)_{24}$\\
\hline
31&$\begin{matrix}\includegraphics[width=76pt]{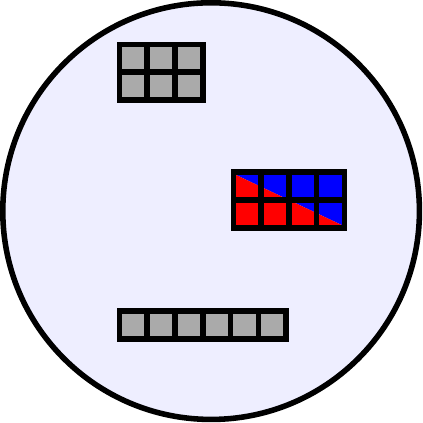}\end{matrix}$&$(0, 0, 3, 0, 2)$&$(64, 43)$&$Sp(3)_8\times Sp(2)_{8}\times SU(2)_{24}$\\
\hline
32&$\begin{matrix}\includegraphics[width=76pt]{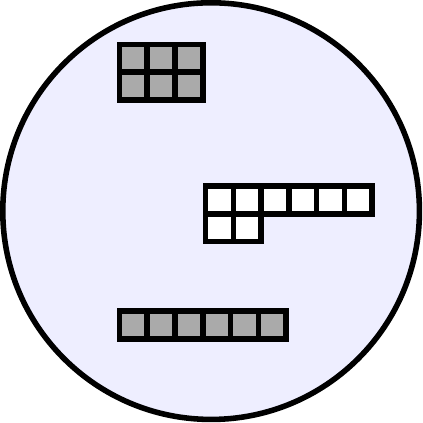}\end{matrix}$&$(0, 0, 4, 0, 2)$&$(72, 50)$&$Sp(3)_{8}\times SU(2)_{24}\times SU(2)^{3}_{8}$\\
\hline
33&$\begin{matrix}\includegraphics[width=76pt]{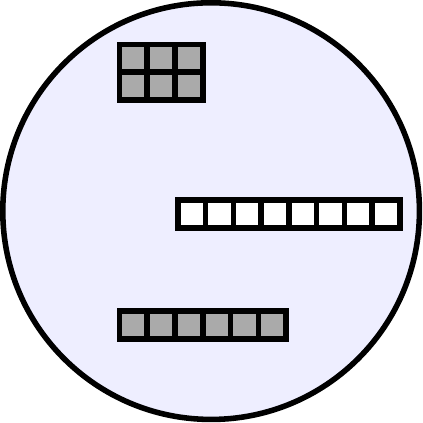}\end{matrix}$&$(0, 0, 4, 0, 3)$&$(88, 61)$&$Spin(8)_{12}\times Sp(3)_{8}\times SU(2)_{24}$\\
\hline
34&$\begin{matrix}\includegraphics[width=76pt]{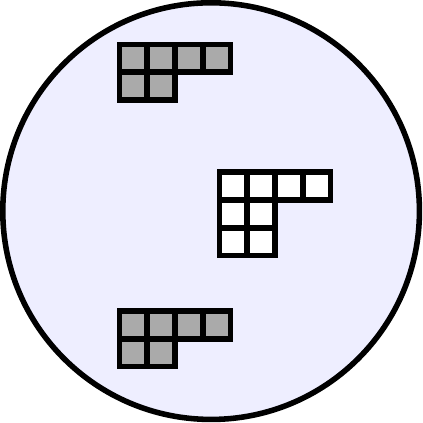}\end{matrix}$&$(0, 3, 1, 0, 0)$&$(36, 22)$&$SU(2)_{6}^6\times U(1)$\\
\hline
35&$\begin{matrix}\includegraphics[width=76pt]{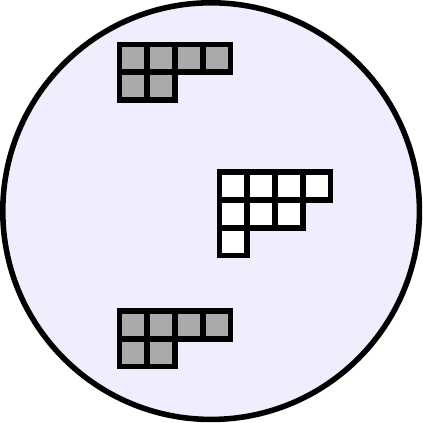}\end{matrix}$&$(0, 2, 1, 0, 1)$&$(43, 28)$&$SU(2)_{12}^2\times SU(2)_{6}^2\times SU(2)_{7}$\\
\hline
36&$\begin{matrix}\includegraphics[width=76pt]{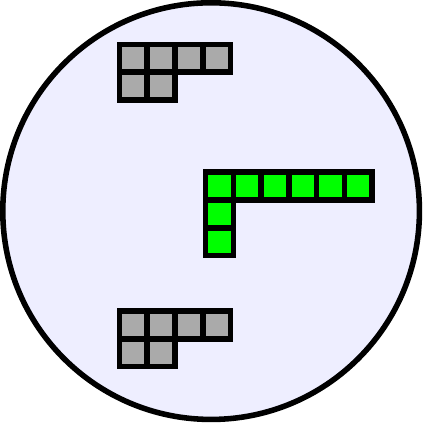}\end{matrix}$&$(0, 2, 2, 0, 1)$&$(52, 35)$&$Sp(2)_8\times SU(2)_{12}^2\times SU(2)_{6}^2$\\
\hline
37&$\begin{matrix}\includegraphics[width=76pt]{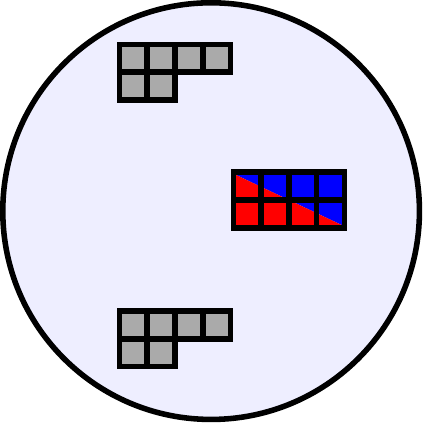}\end{matrix}$&$(0, 2, 2, 0, 1)$&$(52, 35)$&$Sp(2)_8 \times SU(2)_{6}^2\times U(1)^2$\\
\hline
38&$\begin{matrix}\includegraphics[width=76pt]{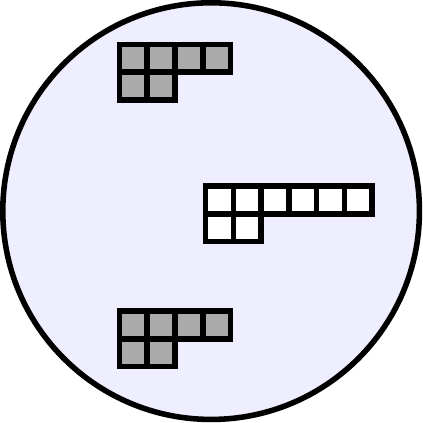}\end{matrix}$&$(0, 2, 3, 0, 1)$&$(60, 42)$&$SU(2)_{8}^3\times SU(2)_{6}^2\times U(1)^2$\\
\hline
39&$\begin{matrix}\includegraphics[width=76pt]{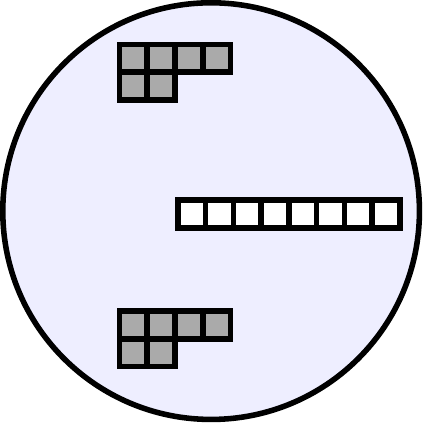}\end{matrix}$&$(0, 2, 3, 0, 2)$&$(76, 53)$&$Spin(8)_{12}\times SU(2)_{6}^2\times U(1)^2$\\
\hline
40&$\begin{matrix}\includegraphics[width=76pt]{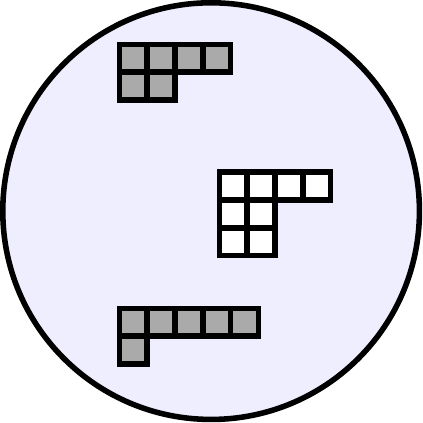}\end{matrix}$&$(0, 2, 1, 0, 1)$&$(44, 28)$&$Sp(2)_{7}\times SU(2)_{12}^{2}\times SU(2)_{6}\times U(1)$\\
\hline
41&$\begin{matrix}\includegraphics[width=76pt]{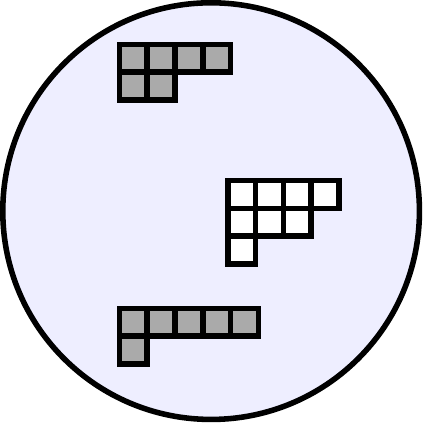}\end{matrix}$&$(0, 1, 1, 0, 2)$&$(51, 34)$&$Sp(2)_{7}\times SU(2)_{24}\times SU(2)_{7}\times SU(2)_{6}$\\
\hline
42&$\begin{matrix}\includegraphics[width=76pt]{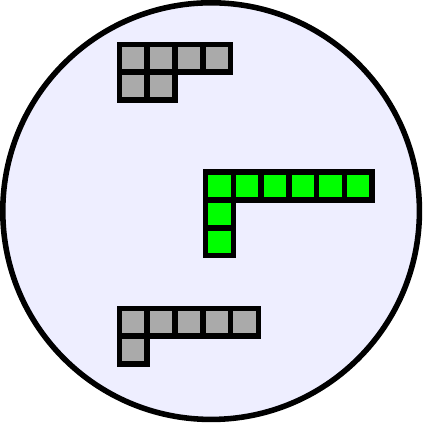}\end{matrix}$&$(0, 1, 2, 0, 2)$&$(60, 41)$&$Sp(2)_8\times Sp(2)_7\times SU(2)_{24}\times SU(2)_6$\\
\hline
43&$\begin{matrix}\includegraphics[width=76pt]{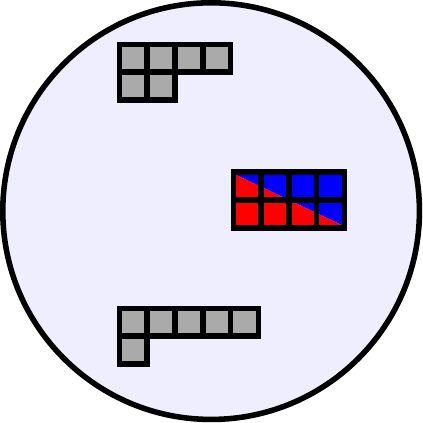}\end{matrix}$&$(0, 1, 2, 0, 2)$&$(60, 41)$&$Sp(2)_8\times Sp(2)_{7}\times SU(2)_{6}\times U(1)$\\
\hline
44&$\begin{matrix}\includegraphics[width=76pt]{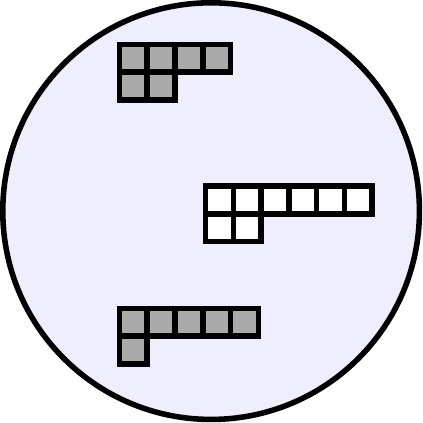}\end{matrix}$&$(0, 1, 3, 0, 2)$&$(68, 48)$&$Sp(2)_{7}\times SU(2)^{3}_{8}\times SU(2)_{6}\times U(1)$\\
\hline
45&$\begin{matrix}\includegraphics[width=76pt]{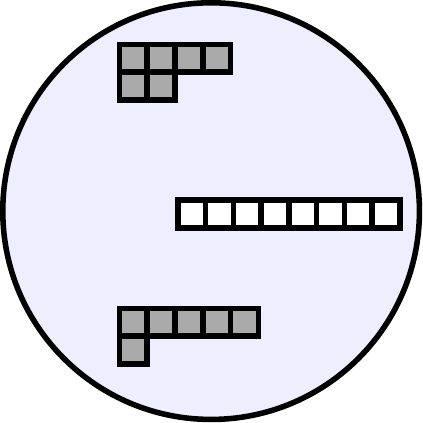}\end{matrix}$&$(0, 1, 3, 0, 3)$&$(84, 59)$&$Spin(8)_{12}\times Sp(2)_{7}\times SU(2)_{6}\times U(1)$\\
\hline
46&$\begin{matrix}\includegraphics[width=76pt]{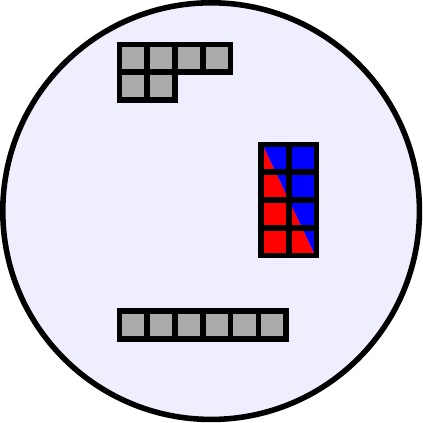}\end{matrix}$&$(0, 1, 1, 0, 0)$&$(30, 12)$&$SU(2)_{6}\times SU(8)_{8}$\\
\hline
47&$\begin{matrix}\includegraphics[width=76pt]{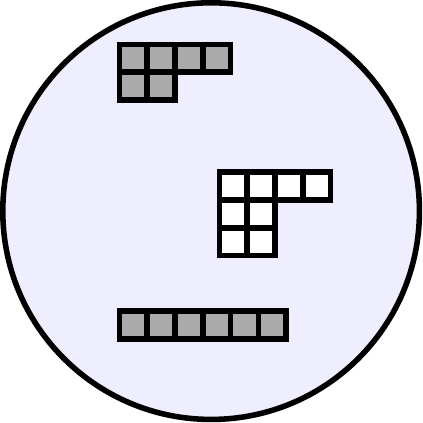}\end{matrix}$&$(0, 2, 2, 0, 1)$&$(54, 35)$&$Sp(3)_{8}\times SU(2)_{6}\times U(1)^3$\\
\hline
48&$\begin{matrix}\includegraphics[width=76pt]{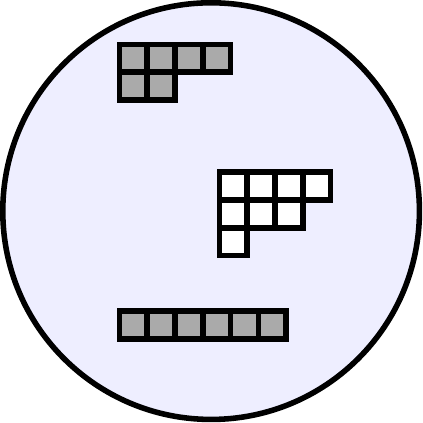}\end{matrix}$&$(0, 1, 2, 0, 2)$&$(61, 41)$&$Sp(3)_{8}\times SU(2)_{7}\times SU(2)_{6}\times U(1)$\\
\hline
49&$\begin{matrix}\includegraphics[width=76pt]{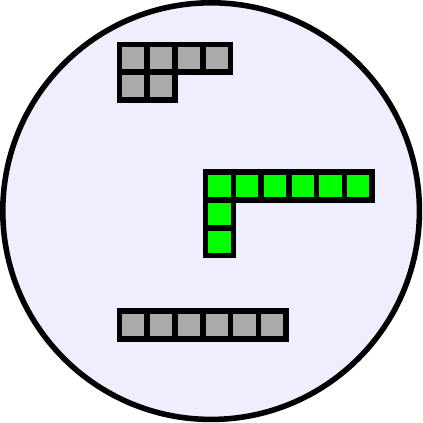}\end{matrix}$&$(0, 1, 3, 0, 2)$&$(70, 48)$&$Sp(3)_{8}\times Sp(2)_{8}\times SU(2)_{6}\times U(1)$\\
\hline
50&$\begin{matrix}\includegraphics[width=76pt]{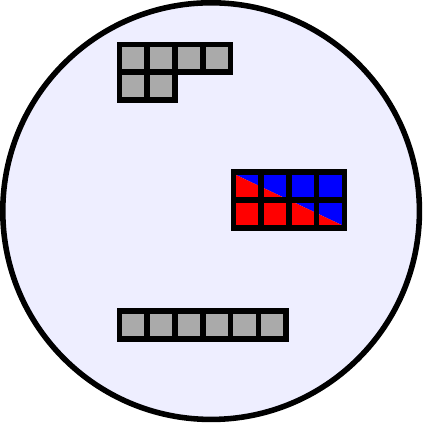}\end{matrix}$&$(0, 1, 3, 0, 2)$&$(70, 48)$&$Sp(3)_{8}\times Sp(2)_{8}\times SU(2)_{6}\times U(1)$\\
\hline
51&$\begin{matrix}\includegraphics[width=76pt]{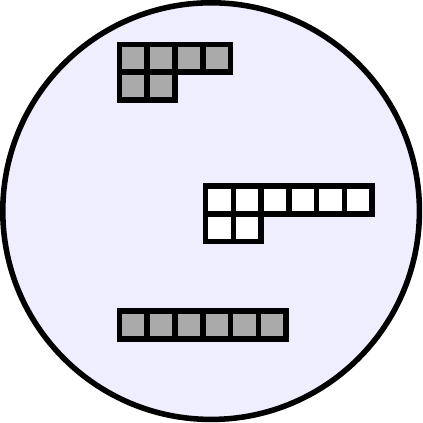}\end{matrix}$&$(0, 1, 4, 0, 2)$&$(78, 55)$&$Sp(3)_{8}\times SU(2)^{3}_{8}\times SU(2)_{6}\times U(1)$\\
\hline
52&$\begin{matrix}\includegraphics[width=76pt]{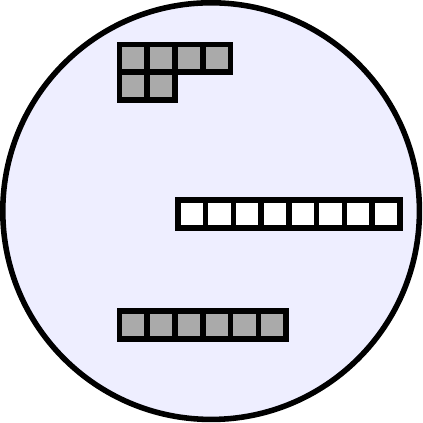}\end{matrix}$&$(0, 1, 4, 0, 3)$&$(94, 66)$&$Spin(8)_{12}\times Sp(3)_{8}\times SU(2)_{6}\times U(1)$\\
\hline
53&$\begin{matrix}\includegraphics[width=76pt]{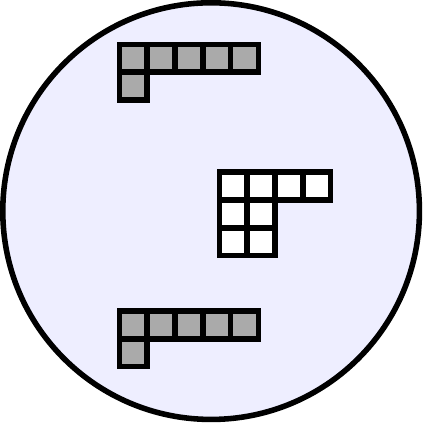}\end{matrix}$&$(0, 1, 1, 0, 2)$&$(52, 34)$&$Sp(2)_{7}^2\times SU(2)_{24}\times U(1)$\\
\hline
54&$\begin{matrix}\includegraphics[width=76pt]{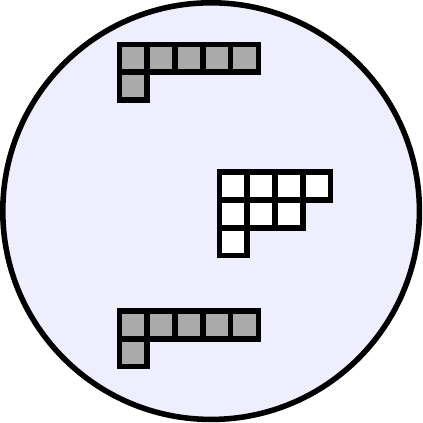}\end{matrix}$&$(0, 0, 1, 0, 3)$&$(59, 40)$&$Sp(2)_{7}^2\times SU(2)_{7}\times U(1)$\\
\hline
55&$\begin{matrix}\includegraphics[width=76pt]{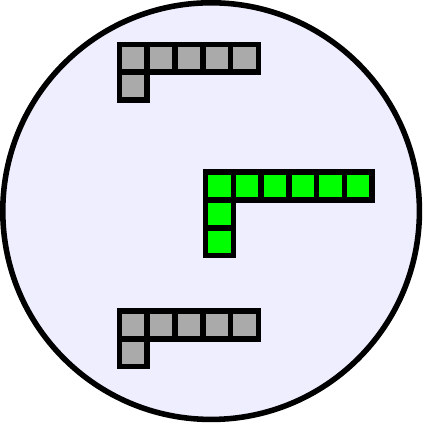}\end{matrix}$&$(0, 0, 2, 0, 3)$&$(68, 47)$&$Sp(2)_8\times Sp(2)_{7}^2\times U(1)$\\
\hline
56&$\begin{matrix}\includegraphics[width=76pt]{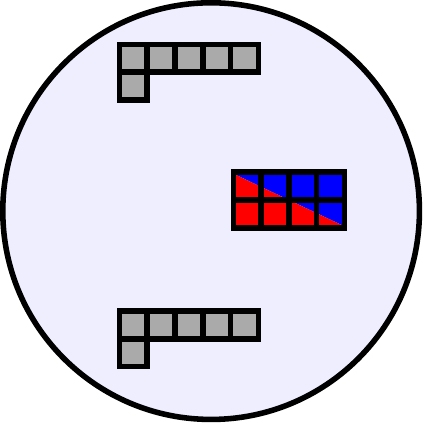}\end{matrix}$&$(0, 0, 2, 0, 3)$&$(68, 47)$&$Sp(2)_8\times Sp(2)_7^2$\\
\hline
57&$\begin{matrix}\includegraphics[width=76pt]{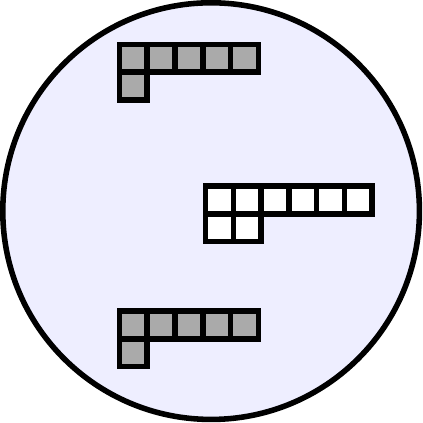}\end{matrix}$&$(0, 0, 3, 0, 3)$&$(76, 54)$&$Sp(2)_{7}^2\times SU(2)^{3}_{8}$\\
\hline
58&$\begin{matrix}\includegraphics[width=76pt]{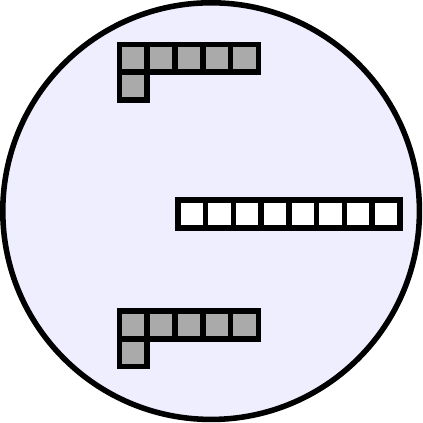}\end{matrix}$&$(0, 0, 3, 0, 4)$&$(92, 65)$&$Spin(8)_{12}\times Sp(2)_{7}^2$\\
\hline
59&$\begin{matrix}\includegraphics[width=76pt]{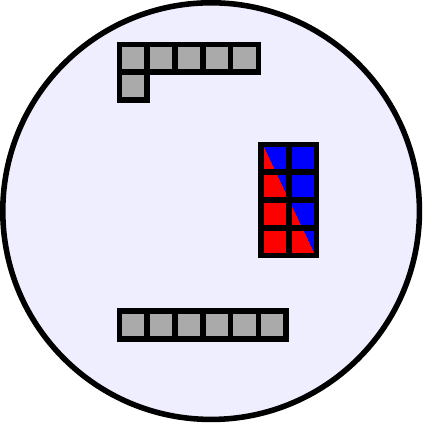}\end{matrix}$&$(0, 0, 1, 0, 1)$&$(38, 18)$&$Sp(4)_{8}\times Sp(2)_{7}$\\
\hline
60&$\begin{matrix}\includegraphics[width=76pt]{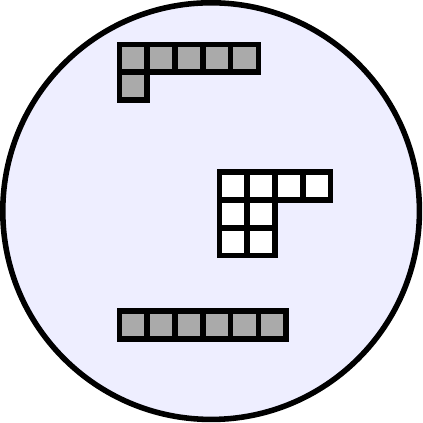}\end{matrix}$&$(0, 1, 2, 0, 2)$&$(62, 41)$&$Sp(3)_{8}\times Sp(2)_{7}\times U(1)^{2}$\\
\hline
61&$\begin{matrix}\includegraphics[width=76pt]{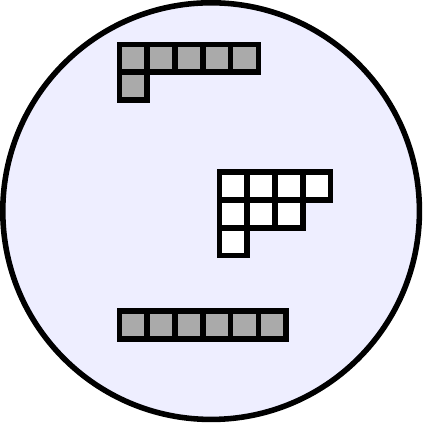}\end{matrix}$&$(0, 0, 2, 0, 3)$&$(69, 47)$&$Sp(3)_{8}\times Sp(2)_{7}\times SU(2)_{7}$\\
\hline
62&$\begin{matrix}\includegraphics[width=76pt]{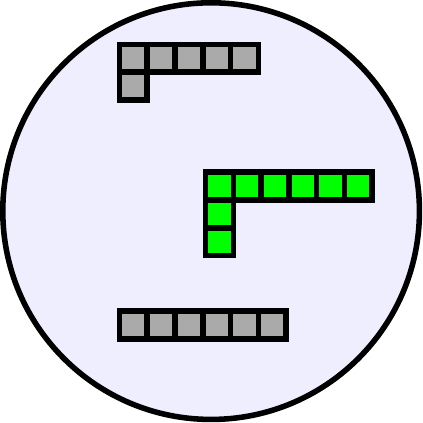}\end{matrix}$&$(0, 0, 3, 0, 3)$&$(78, 54)$&$Sp(3)_{8}\times Sp(2)_{8}\times Sp(2)_{7}$\\
\hline
63&$\begin{matrix}\includegraphics[width=76pt]{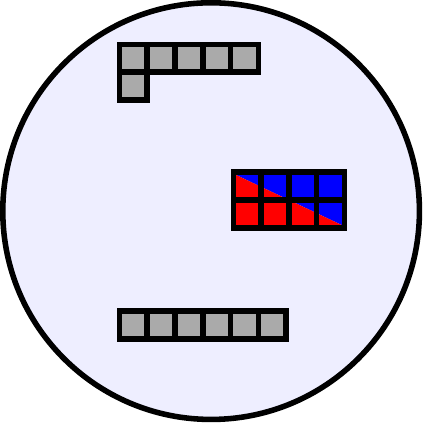}\end{matrix}$&$(0, 0, 3, 0, 3)$&$(78, 54)$&$Sp(3)_{8}\times Sp(2)_{8}\times Sp(2)_{7}$\\
\hline
64&$\begin{matrix}\includegraphics[width=76pt]{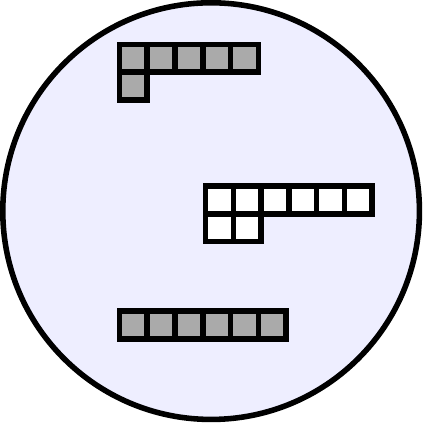}\end{matrix}$&$(0, 0, 4, 0, 3)$&$(86, 61)$&$Sp(3)_{8}\times Sp(2)_{7}\times SU(2)^{3}_{8}$\\
\hline
65&$\begin{matrix}\includegraphics[width=76pt]{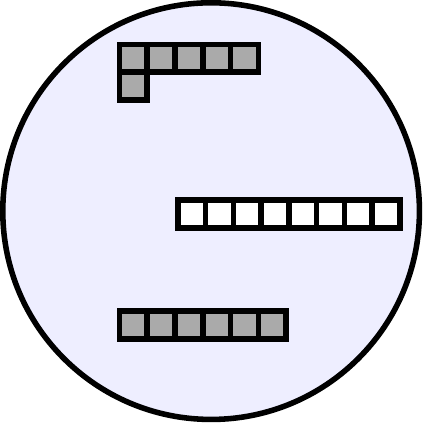}\end{matrix}$&$(0, 0, 4, 0, 4)$&$(102, 72)$&$Spin(8)_{12}\times Sp(3)_8\times Sp(2)_7$\\
\hline
66&$\begin{matrix}\includegraphics[width=76pt]{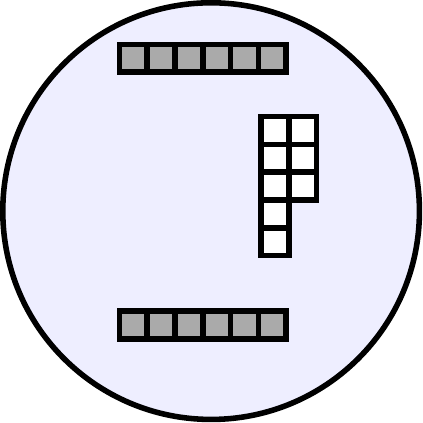}\end{matrix}$&$(0, 0, 1, 0, 1)$&$(40, 18)$&$Sp(6)_{8}$\\
\hline
67&$\begin{matrix}\includegraphics[width=76pt]{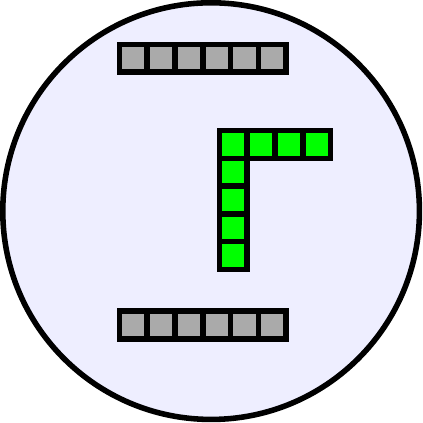}\end{matrix}$&$(0, 0, 2, 0, 1)$&$(48, 25)$&$Sp(6)_{8}\times SU(2)_{8}$\\
\hline
68&$\begin{matrix}\includegraphics[width=76pt]{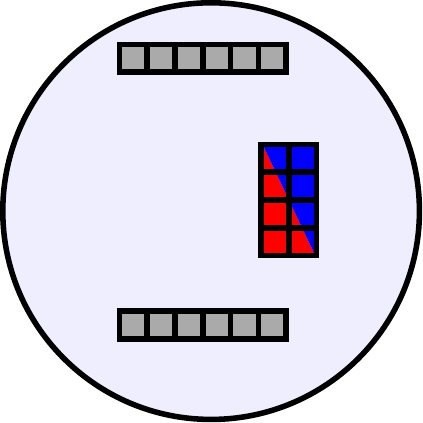}\end{matrix}$&$(0, 0, 2, 0, 1)$&$(48, 25)$&$Sp(3)_{8}^{2}\times SU(2)_{8}$\\
\hline
69&$\begin{matrix}\includegraphics[width=76pt]{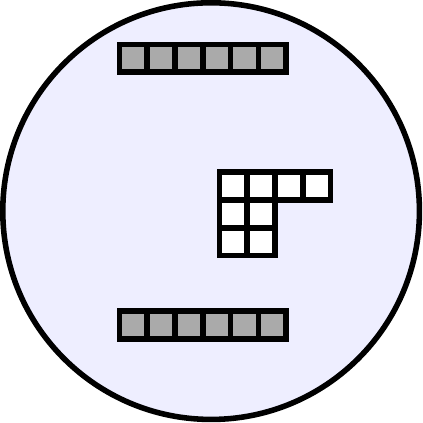}\end{matrix}$&$(0, 1, 3, 0, 2)$&$(72, 48)$&$Sp(3)_{8}^2\times U(1)^{2}$\\
\hline
70&$\begin{matrix}\includegraphics[width=76pt]{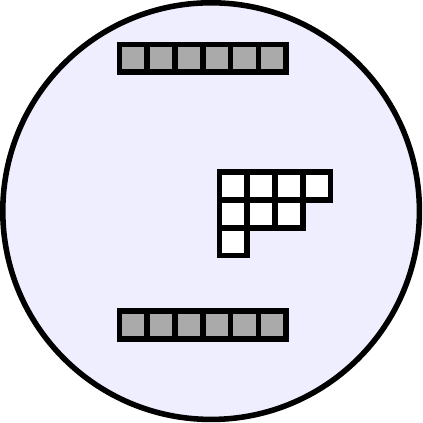}\end{matrix}$&$(0, 0, 3, 0, 3)$&$(79, 54)$&$Sp(3)_{8}^2\times SU(2)_{7}$\\
\hline
71&$\begin{matrix}\includegraphics[width=76pt]{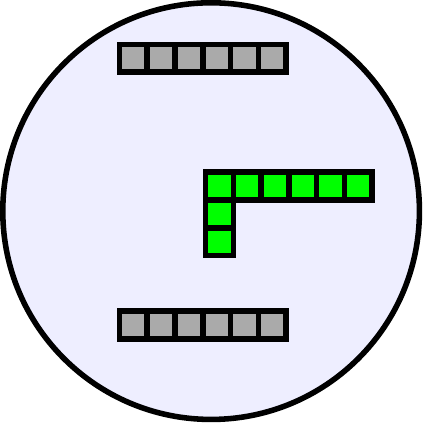}\end{matrix}$&$(0, 0, 4, 0, 3)$&$(88, 61)$&$Sp(3)_{8}^2\times Sp(2)_{8}$\\
\hline
72&$\begin{matrix}\includegraphics[width=76pt]{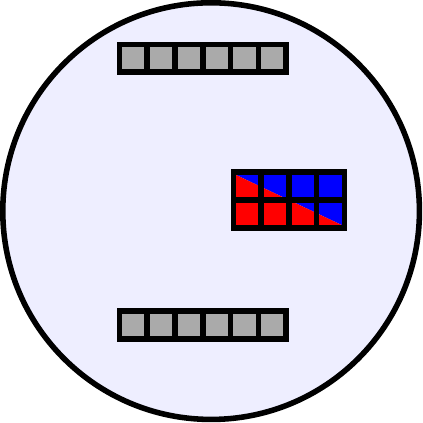}\end{matrix}$&$(0, 0, 4, 0, 3)$&$(88, 61)$&$Sp(3)_{8}^2\times Sp(2)_{8}$\\
\hline
73&$\begin{matrix}\includegraphics[width=76pt]{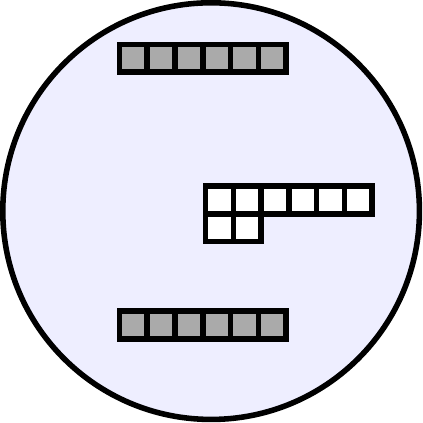}\end{matrix}$&$(0, 0, 5, 0, 3)$&$(96, 68)$&$Sp(3)_{8}^2\times SU(2)^{3}_{8}$\\
\hline
74&$\begin{matrix}\includegraphics[width=76pt]{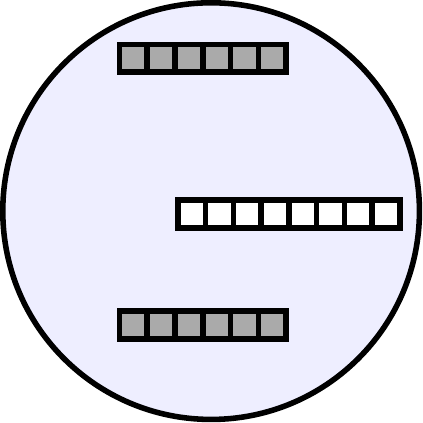}\end{matrix}$&$(0, 0, 5, 0, 4)$&$(112, 79)$&$Spin(8)_{12}\times Sp(3)_{8}^2$\\
\hline
\end{longtable}

\subsubsection{{Mixed Fixtures}}\label{mixed_fixtures}

Three new SCFTs make their appearance in the list of ``mixed'' fixtures (accompanied by some number of free hypermultiplets).

\begin{itemize}%
\item The $Sp(4)_7 \times SU(2)_5$ SCFT has Coulomb branch dimensions $(d_2,...,d_6)=(0,0,1,0,1)$ and $(n_h,n_v)=(33,18)$.
\item The $Sp(5)_7 \times SU(2)_8$ SCFT has Coulomb branch dimensions $(d_2,...,d_6)=(0,0,1,0,1)$ and $(n_h,n_v)=(35,18)$.
\item The $Sp(3)_7 \times Sp(2)_8 \times SU(2)_5$ SCFT has Coulomb branch dimensions $(d_2,...,d_6)=(0,0,2,0,1)$ and $(n_h,n_v)=(42,25)$.

\end{itemize}
The remaining SCFTs in our list of mixed fixtures include the venerable ${(E_6)}_6$ theory, the ${Sp(5)}_7$ theory (which appeared in the untwisted $D_4$ theory \cite{Chacaltana:2011ze}), two theories (${Sp(3)}_5\times{SU(2)}_8$ and ${Spin(7)}_8\times {SU(2)}_5^2$) which appear above (see also \cite{Chacaltana:2012ch}) and three more which appeared in the twisted $A_3$ theory \cite{Chacaltana:2012ch}.

\begin{longtable}{|c|c|c|}
\hline
\#&Fixture&Theory\\
\hline
\endhead
1&$\begin{matrix}\includegraphics[width=76pt]{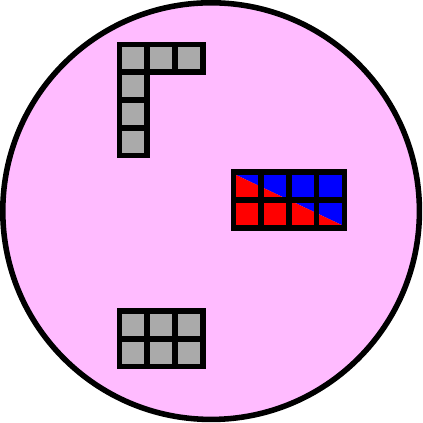}\end{matrix}$&$\tfrac{1}{2}(1,3,4) + Sp(3)_5\times SU(2)_8$\\
\hline 
2&$\begin{matrix}\includegraphics[width=76pt]{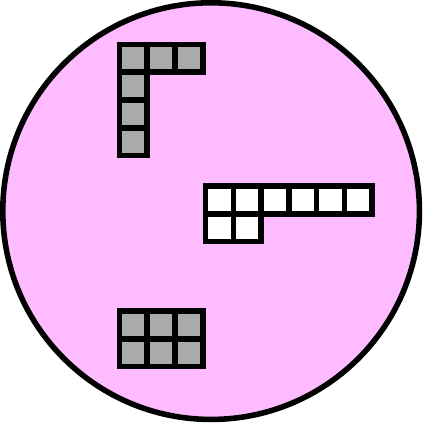}\end{matrix}$&$\tfrac{1}{2}(1,3;2,1,1) + SU(2)_5^2\times Spin(7)_8$\\
\hline 
3&$\begin{matrix}\includegraphics[width=76pt]{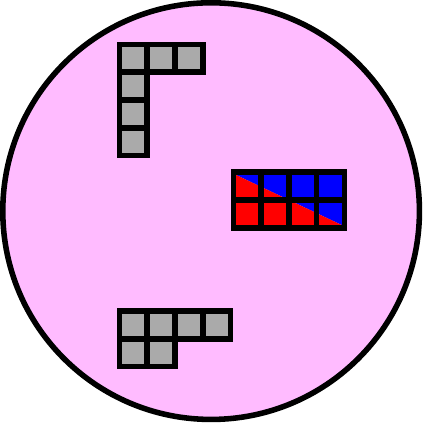}\end{matrix}$&$(1,1,4) + SU(2)_5\times Sp(3)_6\times U(1)$\\
\hline 
4&$\begin{matrix}\includegraphics[width=76pt]{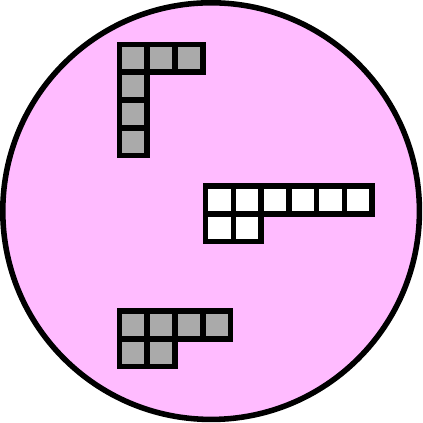}\end{matrix}$&$(1,1;2,1,1) + SU(2)_5\times SU(4)_8\times Sp(2)_6$\\
\hline 
5&$\begin{matrix}\includegraphics[width=76pt]{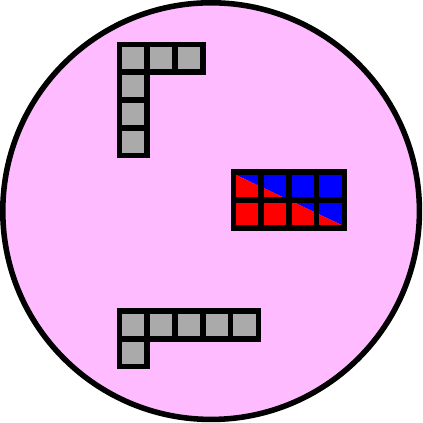}\end{matrix}$&$\tfrac{1}{2}(1,1,4) + Sp(4)_7 \times SU(2)_5$\\
\hline 
6&$\begin{matrix}\includegraphics[width=76pt]{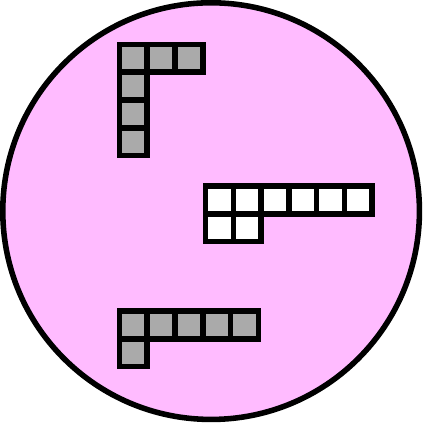}\end{matrix}$&$\tfrac{1}{2}(1,1;2,1,1) + {Sp(3)}_7\times {Sp(2)}_8\times {SU(2)}_5$\\
\hline 
7&$\begin{matrix}\includegraphics[width=76pt]{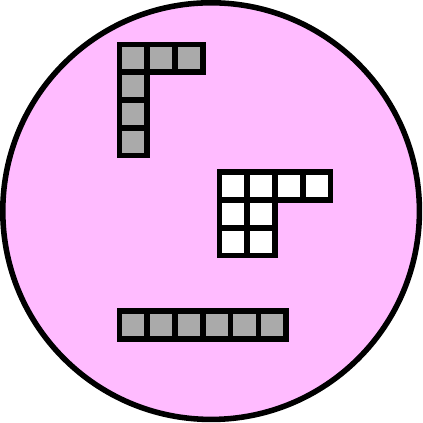}\end{matrix}$&$(1,6) + SU(2)_{5}\times Sp(3)_{6}\times U(1)$\\
\hline 
8&$\begin{matrix}\includegraphics[width=76pt]{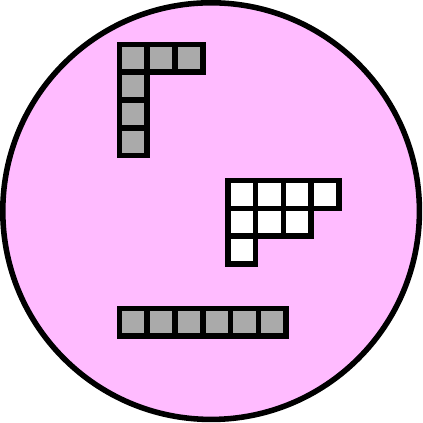}\end{matrix}$&$\tfrac{1}{2}(1,6,1) + Sp(4)_7 \times SU(2)_5$\\
\hline 
9&$\begin{matrix}\includegraphics[width=76pt]{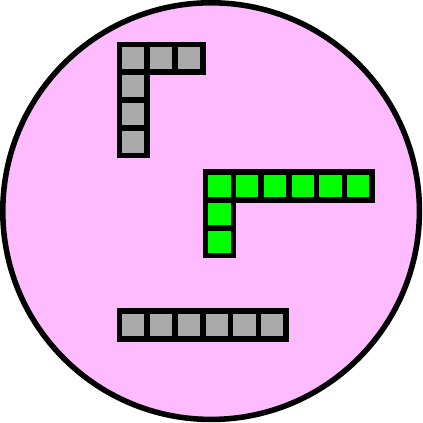}\end{matrix}$&$\tfrac{1}{2}(1,6,1) + Sp(3)_7 \times Sp(2)_8 \times SU(2)_5$\\
\hline 
10&$\begin{matrix}\includegraphics[width=76pt]{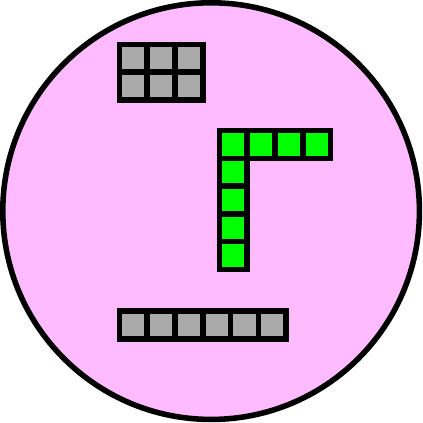}\end{matrix}$&$\tfrac{1}{2}(3,6,1) + Sp(3)_5\times SU(2)_8$\\
\hline 
11&$\begin{matrix}\includegraphics[width=76pt]{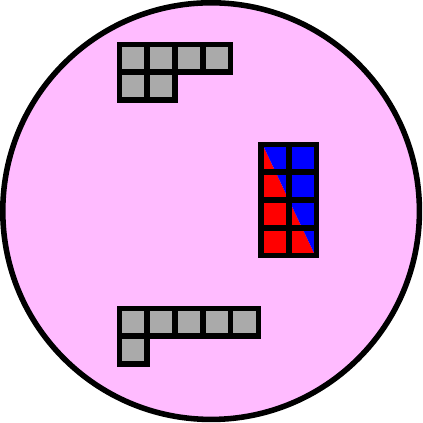}\end{matrix}$&$(1,1,2) + \tfrac{1}{2}(1,4,1) + (E_6)_6$\\
\hline 
12&$\begin{matrix}\includegraphics[width=76pt]{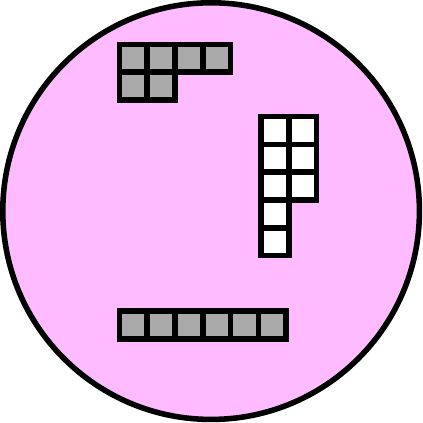}\end{matrix}$&$(1,6) + (E_6)_6$\\
\hline 
13&$\begin{matrix}\includegraphics[width=76pt]{D4mixed2211_111111_5111}\end{matrix}$&$(1,6,1) + Sp(4)_{6}\times SU(2)_{8}$\\
\hline 
14&$\begin{matrix}\includegraphics[width=76pt]{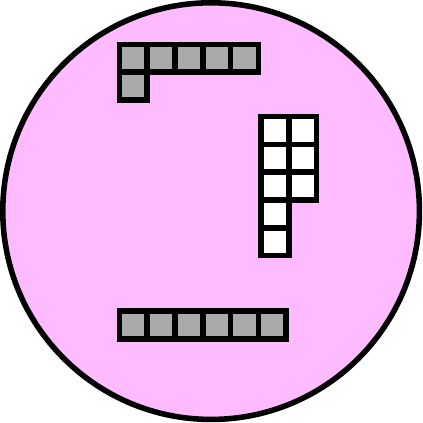}\end{matrix}$&$\tfrac{1}{2}(1,6) + Sp(5)_{7}$\\
\hline 
15&$\begin{matrix}\includegraphics[width=76pt]{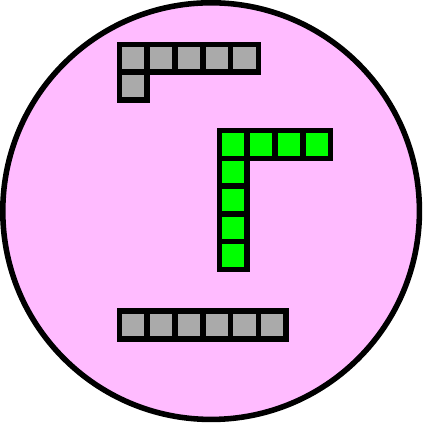}\end{matrix}$&$\tfrac{1}{2}(1,6,1) + Sp(5)_7\times SU(2)_8$\\
\hline 
16&$\begin{matrix}\includegraphics[width=76pt]{D4mixed21111_21111_44}\end{matrix}$&$\tfrac{1}{2}(1,1,2) + Sp(5)_{7}$\\
\hline 
\end{longtable}

\subsubsection{{Gauge Theory Fixtures}}\label{gaugetheory_fixtures}

For each gauge theory fixture, we list the gauge group, $G$, and the representation content of the hypermultiplets, $(R_{F_1},R_{F_2},R_{F_3};R_G)$. Here, $R_G$ is the representation of the gauge group and $R_{F_i}$ is the representation of the semisimple part of the flavour symmetry of the $i^{\mathrm{th}}$ puncture (where we work counterclockwise from the upper-left, and omit $F_i$ if it is abelian or empty).

{
\renewcommand{\arraystretch}{1.5}
\begin{longtable}{|c|c|c|c|c|c|}
\hline 
\#&Fixture&$(d_2,\dots,d_6)$&$G$&\# Hypers&Representation\\
\hline
\endhead
1&$\begin{matrix}\includegraphics[width=76pt]{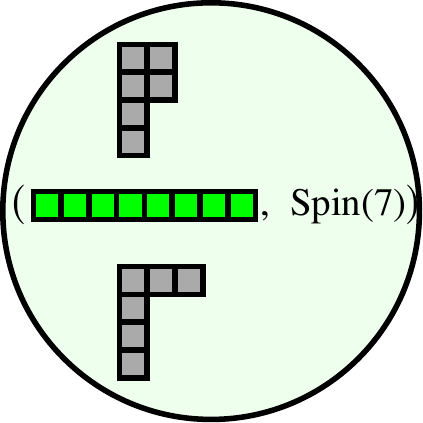}\end{matrix}$&$(1, 0, 1, 0, 0)$&$Sp(2)$&21&$\tfrac{1}{2}(1,8;4)+\tfrac{1}{2}(2,1;5)$\\
\hline
2&$\begin{matrix}\includegraphics[width=76pt]{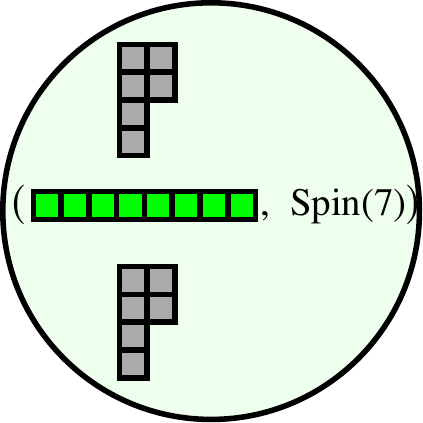}\end{matrix}$&$(2, 0, 0, 0, 0)$&$SU(2)\times SU(2)$&16&$\begin{gathered}\tfrac{1}{2}(R_1;2,1)\\ +\tfrac{1}{2}(R_2;1,2)\\ \text{where}\, R_i= 8\, \text{or}\, 1+7\end{gathered}$\\
\hline
3&$\begin{matrix}\includegraphics[width=76pt]{D4gauge42_33_11111111}\end{matrix}$&$(2, 0, 0, 0, 0)$&$SU(2)\times SU(2)$&24&${\begin{aligned}&\tfrac{1}{2}(2,8_v;1,1)\\&+\tfrac{1}{2}(1,8;2,1)\\&+\tfrac{1}{2}(1,8;1,2)\end{aligned}}$\\
\hline
4&$\begin{matrix}\includegraphics[width=76pt]{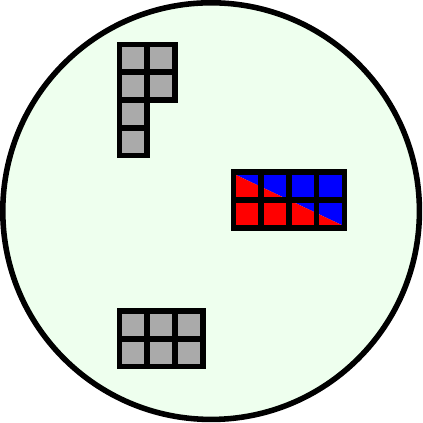}\end{matrix}$&$(1, 0, 0, 0, 0)$&$SU(2)$&16&${\begin{aligned}&\tfrac{1}{2}(1,5;2)\\&+\tfrac{1}{2}(1,4;1)\\&+\tfrac{1}{2}(3,4;1)\\&+\tfrac{1}{2}(3,1;2)\end{aligned}}$\\
\hline
5&$\begin{matrix}\includegraphics[width=76pt]{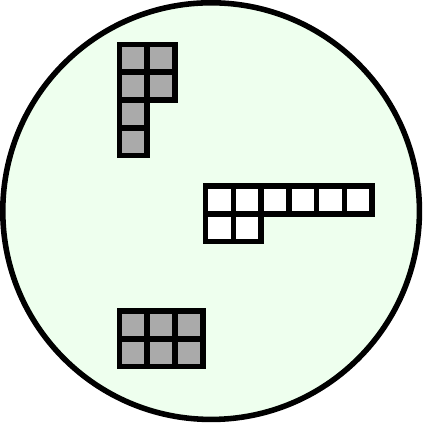}\end{matrix}$&$(1, 0, 1, 0, 0)$&$Sp(2)$&24&${\begin{aligned}&
\tfrac{1}{2}(2;1,2,1;4)\\&+\tfrac{1}{2}(2;1,1,2;4)\\&+\tfrac{1}{2}(1;2,1,1;5)\\&+\tfrac{1}{2}(3;2,1,1;1)\end{aligned}}$\\
\hline
6&$\begin{matrix}\includegraphics[width=76pt]{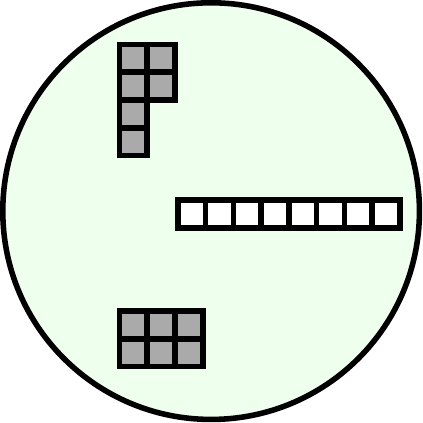}\end{matrix}$&$(1, 0, 1, 0, 1)$&$Sp(3)$&40&
${\begin{aligned}
 &\tfrac{1}{2}(1,8;6)\\
&+\tfrac{1}{2}(3,1;6)\\
&+\tfrac{1}{2}(1,1;14')
\end{aligned}}$\\
\hline
7&$\begin{matrix}\includegraphics[width=76pt]{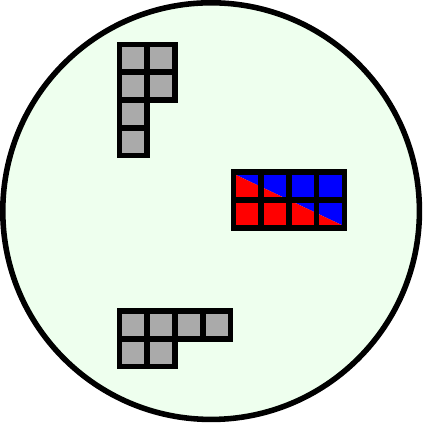}\end{matrix}$&$(1, 1, 0, 0, 0)$&$SU(3)$&22&${\begin{aligned}&(2,1;3)\\&+(1,4;3)\\&+(1,4;1)\end{aligned}}$\\
\hline
8&$\begin{matrix}\includegraphics[width=76pt]{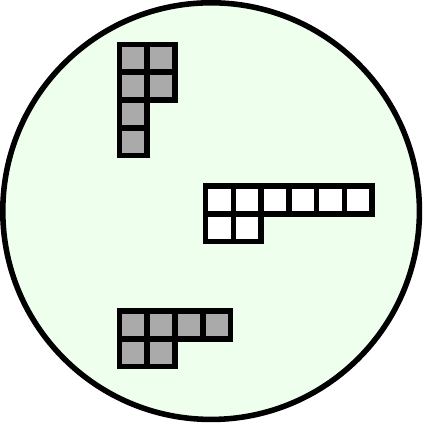}\end{matrix}$&$(1, 1, 1, 0, 0)$&$SU(4)$&30&${\begin{aligned}&\tfrac{1}{2}(2;1,1,1;6)\\&+\tfrac{1}{2}(1;2,1,1;6)\\&+(1;2,1,1;1)\\&+(1;1,2,1;4)\\&+(1;1,1,2;4)\end{aligned}}$\\
\hline
9&$\begin{matrix}\includegraphics[width=76pt]{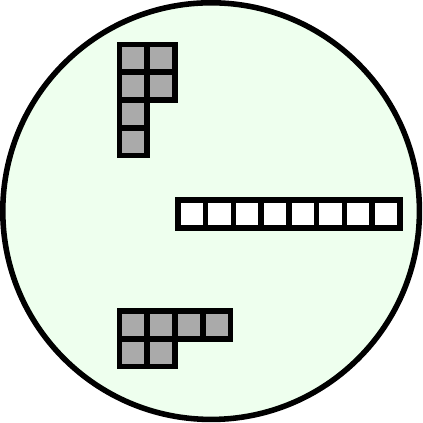}\end{matrix}$&$(1, 1, 1, 0, 1)$&$Sp(3)$&46&${\begin{aligned}&\tfrac{1}{2}(1,8;6)\\&+(1,1;6)\\&+(E_{6})_{6}\end{aligned}}$\\
\hline
10&$\begin{matrix}\includegraphics[width=76pt]{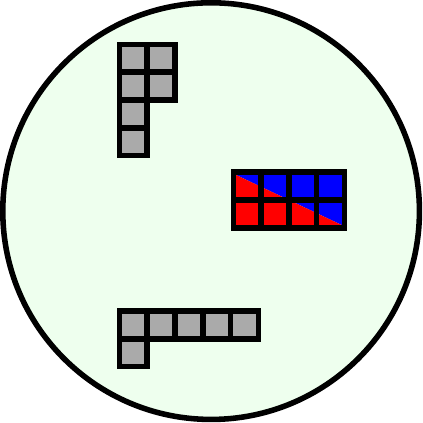}\end{matrix}$&$(1, 0, 0, 0, 1)$&$G_2$&30&${\begin{aligned}&\tfrac{1}{2}(4,1;7)\\&+\tfrac{1}{2}(1,4;7)\\&+\tfrac{1}{2}(1,4;1)\end{aligned}}$\\
\hline
11&$\begin{matrix}\includegraphics[width=76pt]{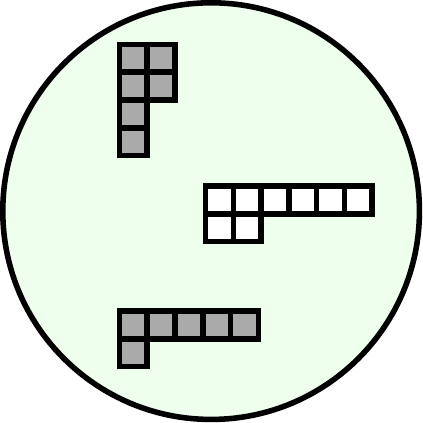}\end{matrix}$&$(1, 0, 1, 0, 1)$&$Spin(7)$&38&${\begin{aligned}&\tfrac{1}{2}(4;1,1,1;7)\\&+\tfrac{1}{2}(1;2,1,1;7)\\&+\tfrac{1}{2}(1;1,2,1;8)\\&+\tfrac{1}{2}(1;1,1,2;8)\\&+\tfrac{1}{2}(1;2,1,1;1)\end{aligned}}$\\
\hline
12&$\begin{matrix}\includegraphics[width=76pt]{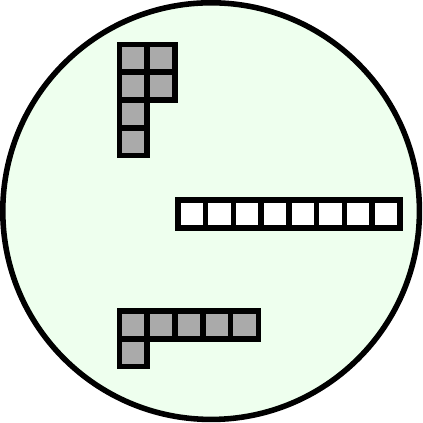}\end{matrix}$&$(1, 0, 1, 0, 2)$&$Spin(7)$&54&${\begin{aligned}&\tfrac{1}{2}(4,1;7)\\&+(E_8)_{12}\end{aligned}}$\\
\hline
13&$\begin{matrix}\includegraphics[width=76pt]{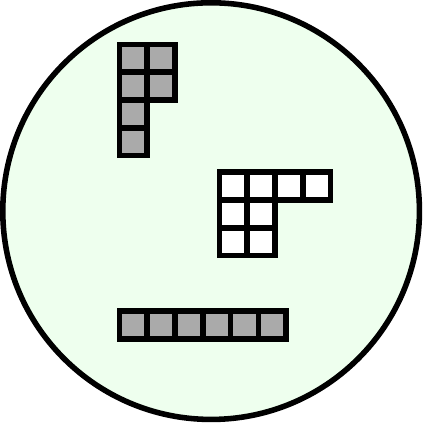}\end{matrix}$&$(1, 1, 0, 0, 0)$&$SU(3)$&24&${\begin{aligned}&(6;3)\\&+(6;1)\end{aligned}}$\\
\hline
14&$\begin{matrix}\includegraphics[width=76pt]{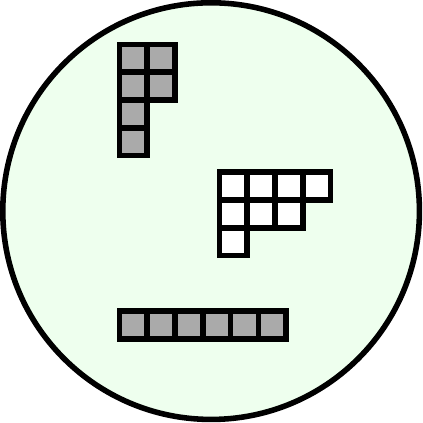}\end{matrix}$&$(1, 0, 0, 0, 1)$&$G_2$&31&${\begin{aligned}&\tfrac{1}{2}(1,2;7)\\&+\tfrac{1}{2}(6,1;7)\\&+\tfrac{1}{2}(6,1;1)\end{aligned}}$\\
\hline
15&$\begin{matrix}\includegraphics[width=76pt]{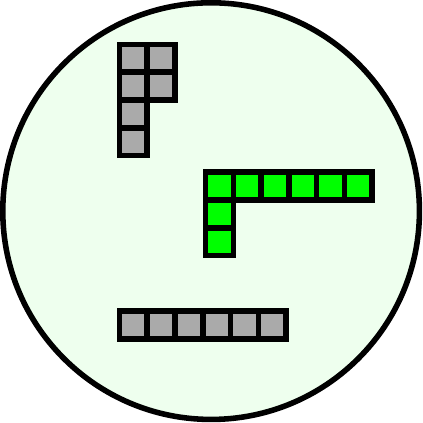}\end{matrix}$&$(1, 0, 1, 0, 1)$&$Spin(7)$&40&${\begin{aligned}&\tfrac{1}{2}(6,1;7)\\&+\tfrac{1}{2}(1,4;8)\\&+\tfrac{1}{2}(6,1;1)\end{aligned}}$\\
\hline
16&$\begin{matrix}\includegraphics[width=76pt]{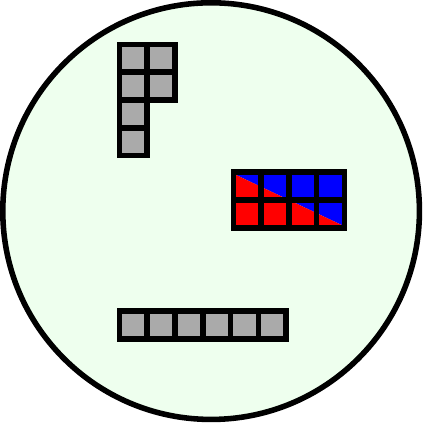}\end{matrix}$&$(1, 0, 1, 0, 1)$&$Spin(7)$&40&${\begin{aligned}&\tfrac{1}{2}(6,1;8)\\&+\tfrac{1}{2}(1,4;8)\end{aligned}}$\\
\hline
17&$\begin{matrix}\includegraphics[width=76pt]{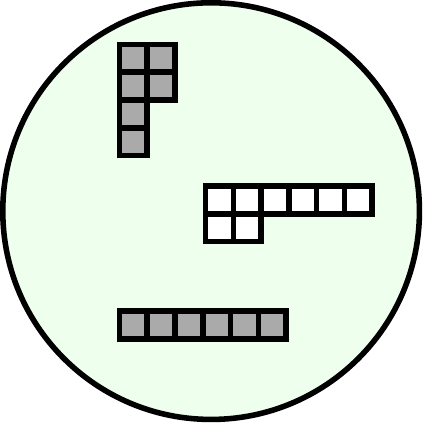}\end{matrix}$&$(1, 0, 2, 0, 1)$&$Spin(8)$&48&${\begin{aligned}&\tfrac{1}{2}(6;1,1,1;8_v)\\&+\tfrac{1}{2}(1;2,1,1;8_v)\\&+\tfrac{1}{2}(1;1,2,1;8_s)\\&+\tfrac{1}{2}(1;1,1,2;8_c)\end{aligned}}$\\
\hline
18&$\begin{matrix}\includegraphics[width=76pt]{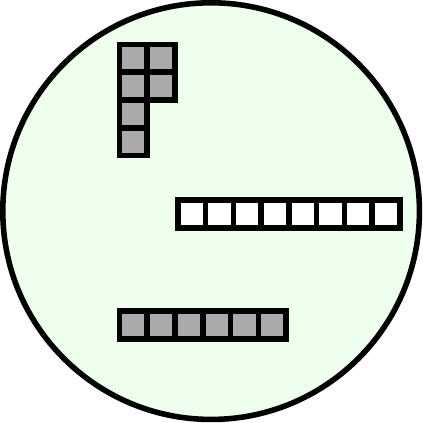}\end{matrix}$&$(1, 0, 2, 0, 2)$&$Spin(8)$&64&${\begin{aligned}&\tfrac{1}{2}(6,1;8)\\&+(E_8)_{12}\end{aligned}}$\\
\hline
19&$\begin{matrix}\includegraphics[width=76pt]{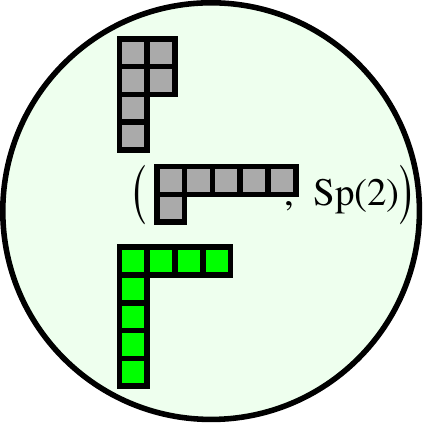}\end{matrix}$&$(1,0,0,0,0)$&$SU(2)$&$10$&${\begin{aligned}&\tfrac{1}{2}(2,4;2)\\&+\tfrac{1}{2}(1,4;1)\end{aligned}}$\\
\hline
20&$\begin{matrix}\includegraphics[width=76pt]{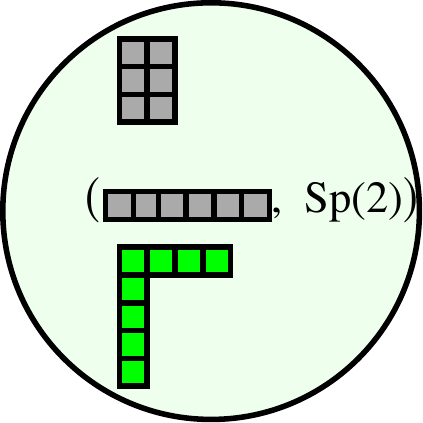}\end{matrix}$&$(1,0,0,0,0)$&$SU(2)$&$8$&$\begin{gathered}\tfrac{1}{2}(1,2,4;2)\\ \text{or}\\\tfrac{1}{2}(1,1,5;2)+\tfrac{1}{2}(1,3,1;2)\end{gathered}$\\
\hline
21&$\begin{matrix}\includegraphics[width=76pt]{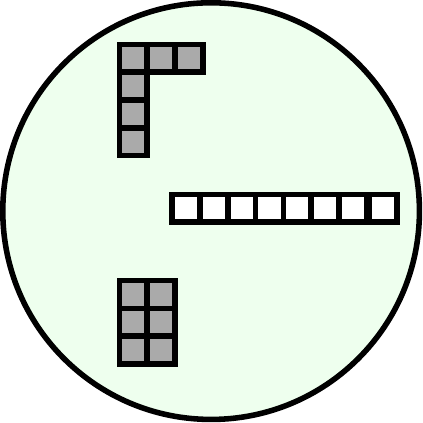}\end{matrix}$&$(1, 0, 1, 0, 0)$&$Sp(2)$&29&${\begin{aligned}&\tfrac{1}{2}(2,1,1;5)\\&+\tfrac{1}{2}(1,1,8;4)\\&+\tfrac{1}{2}(1,2,8_v;1)\end{aligned}}$\\
\hline
22&$\begin{matrix}\includegraphics[width=76pt]{D4gauge33_33_11111111}\end{matrix}$&$(2, 0, 1, 0, 0)$&$Sp(2)\times SU(2)$&32&${\begin{aligned}&\tfrac{1}{2}(2,2,1;4,1)\\&+\tfrac{1}{2}(1,1,8_v;4,1)\\&+\tfrac{1}{2}(1,1,8_v;1,2)\end{aligned}}$\\
\hline
23&$\begin{matrix}\includegraphics[width=76pt]{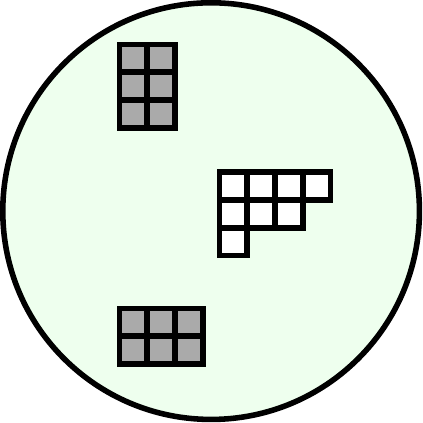}\end{matrix}$&$(1, 0, 0, 0, 0)$&$SU(2)$&15&${\begin{aligned}&\tfrac{1}{2}(2,1,2;2)\\&+\tfrac{1}{2}(1,3,1;2)\\&+\tfrac{1}{2}(1,1,1;2)\\&+\tfrac{1}{2}(2,3,1;1)\\&+\tfrac{1}{2}(1,3,2;1)\\&+\tfrac{1}{2}(2,1,1;1)\end{aligned}}$\\
\hline
24&$\begin{matrix}\includegraphics[width=76pt]{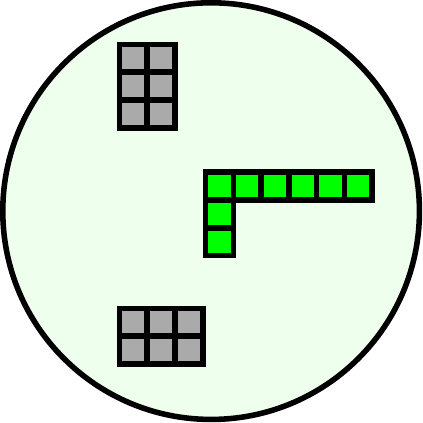}\end{matrix}$&$(1, 0, 1, 0, 0)$&$Sp(2)$&24&${\begin{aligned}&\tfrac{1}{2}(1,2,4;4)\\&+\tfrac{1}{2}(2,1,1;5)\\&+\tfrac{1}{2}(2,3,1;1)\end{aligned}}$\\
\hline
25&$\begin{matrix}\includegraphics[width=76pt]{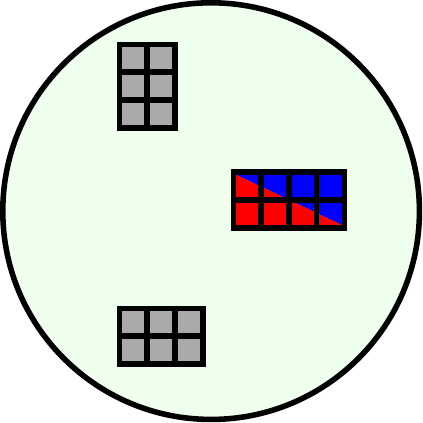}\end{matrix}$&$(1, 0, 1, 0, 0)$&$Sp(2)$&24&${\begin{aligned}&\tfrac{1}{2}(2,2,1;4)\\&+\tfrac{1}{2}(1,2,4;4)\end{aligned}}$\\
\hline
26&$\begin{matrix}\includegraphics[width=76pt]{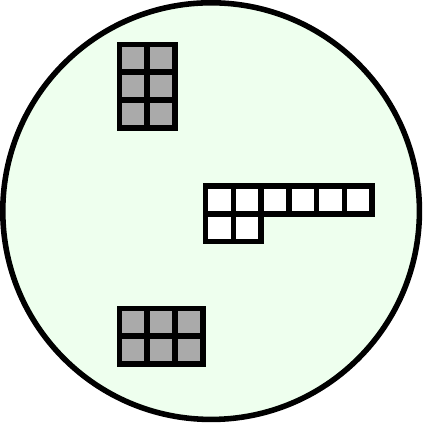}\end{matrix}$&$(1, 0, 2, 0, 0)$&$Sp(2)$&32&${\begin{aligned}&\tfrac{1}{2}(1,1;1,2,2;4)\\&+(E_7)_8\end{aligned}}$\\
\hline
27&$\begin{matrix}\includegraphics[width=76pt]{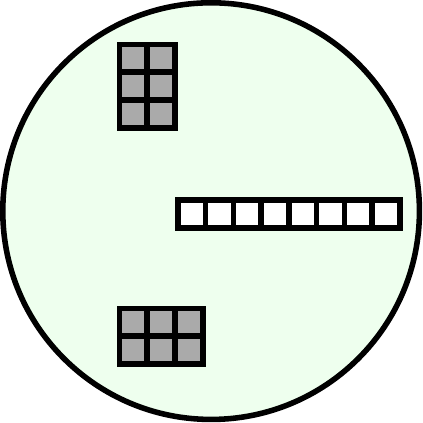}\end{matrix}$&$(1, 0, 2, 0, 1)$&$Sp(3)$&48&${\begin{aligned}&\tfrac{1}{2}(1,1,8;6)\\&+(E_7)_8\end{aligned}}$\\
\hline
28&$\begin{matrix}\includegraphics[width=76pt]{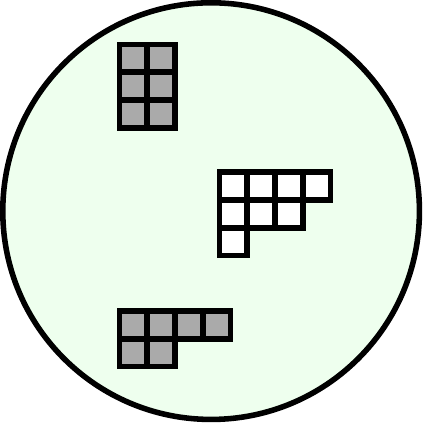}\end{matrix}$&$(1, 1, 0, 0, 0)$&$SU(3)$&21&${\begin{aligned}&(2,1,1;3)\\&+(1,2,1;3)\\&+(1,1,2;3)\\&+(2,1,1;1)\\&+\tfrac{1}{2}(1,1,2;1)\end{aligned}}$\\
\hline
29&$\begin{matrix}\includegraphics[width=76pt]{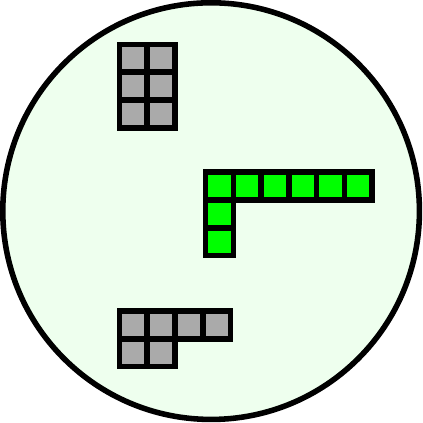}\end{matrix}$&$(1, 1, 1, 0, 0)$&$SU(4)$&30&${\begin{aligned}&\tfrac{1}{2}(1,2,1;6)\\&+\tfrac{1}{2}(2,1,1;6)\\&+(2,1,1;1)\\&+(1,1,4;4)\end{aligned}}$\\
\hline
30&$\begin{matrix}\includegraphics[width=76pt]{D4gauge33_2211_2222}\end{matrix}$&$(1, 1, 1, 0, 0)$&$SU(4)$&30&${\begin{aligned}&\tfrac{1}{2}(1,2,1;6)\\&+(2,1,1;4)\\&+(1,1,4;4)\end{aligned}}$\\
\hline
31&$\begin{matrix}\includegraphics[width=76pt]{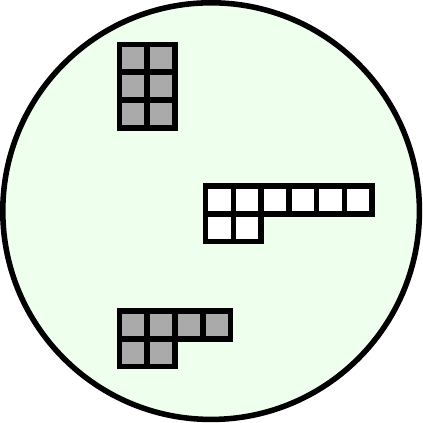}\end{matrix}$&$(1, 1, 2, 0, 0)$&$SU(4)$&38&${\begin{aligned}&(2,1;1,1,1;4)\\&+\tfrac{1}{2}(1,2;1,1,1;6)\\&+(E_{7})_{8}\end{aligned}}$\\
\hline
32&$\begin{matrix}\includegraphics[width=76pt]{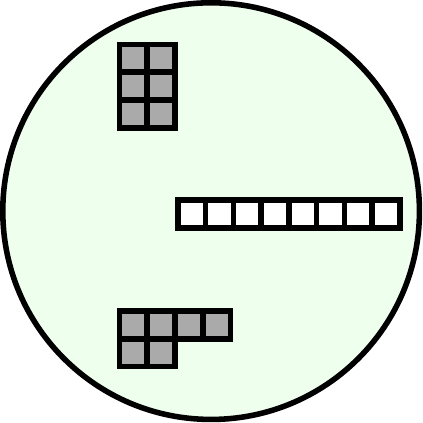}\end{matrix}$&$(1, 1, 2, 0, 1)$&$Sp(3)$&54&${\begin{aligned}&\tfrac{1}{2}(1,1,8;6)\\&+SU(2)_{6}\times SU(8)_{8}\end{aligned}}$\\
\hline
33&$\begin{matrix}\includegraphics[width=76pt]{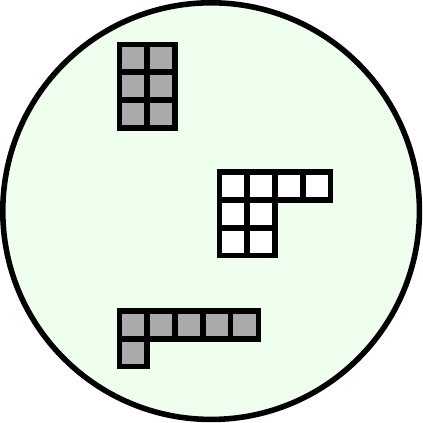}\end{matrix}$&$(1, 1, 0, 0, 0)$&$SU(3)$&22&${\begin{aligned}&(2,1;3)\\&+(1,4;3)\\&+(2,1;1)\\&+\tfrac{1}{2}(1,4;1)\end{aligned}}$\\
\hline
34&$\begin{matrix}\includegraphics[width=76pt]{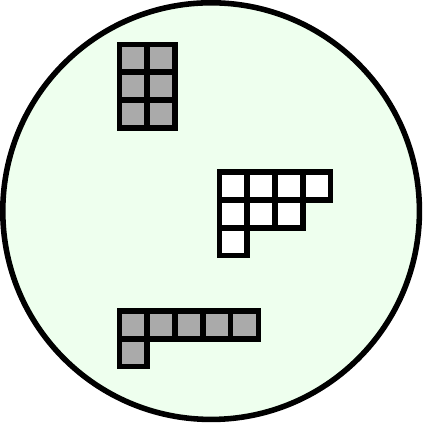}\end{matrix}$&$(1, 0, 0, 0, 1)$&$G_{2}$&29&${\begin{aligned}&\tfrac{1}{2}(1,1,2;7)\\&+\tfrac{1}{2}(1,4,1;7)\\&+\tfrac{1}{2}(2,1,1;7)\\&+\tfrac{1}{2}(2,1,1;1)\end{aligned}}$\\
\hline
35&$\begin{matrix}\includegraphics[width=76pt]{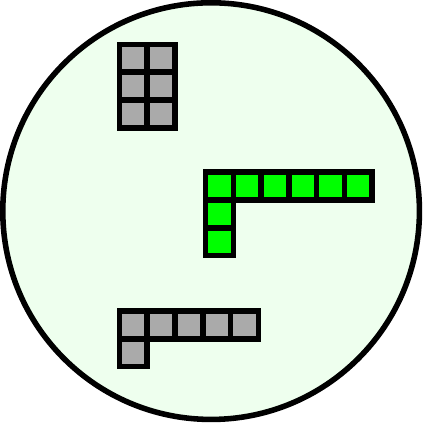}\end{matrix}$&$(1, 0, 1, 0, 1)$&$Spin(7)$&38&${\begin{aligned}&\tfrac{1}{2}(2,1,1;7)\\&+\tfrac{1}{2}(1,4,1;7)\\&+\tfrac{1}{2}(1,1,4;8)\\&+\tfrac{1}{2}(2,1,1;1)\end{aligned}}$\\
\hline
36&$\begin{matrix}\includegraphics[width=76pt]{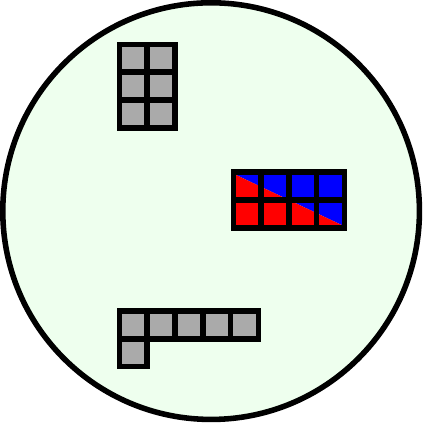}\end{matrix}$&$(1, 0, 1, 0, 1)$&$Spin(7)$&38&${\begin{aligned}&\tfrac{1}{2}(2,1,1;8)\\&+\tfrac{1}{2}(1,1,4;8)\\&+\tfrac{1}{2}(1,4,1;7)\end{aligned}}$\\
\hline
37&$\begin{matrix}\includegraphics[width=76pt]{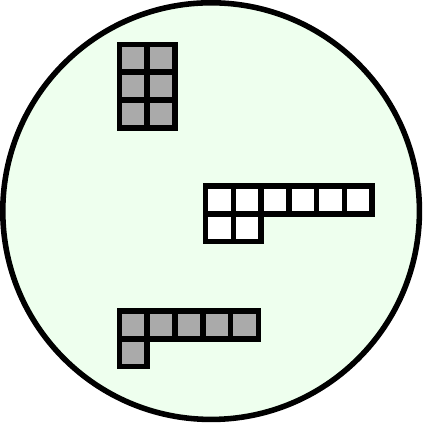}\end{matrix}$&$(1, 0, 2, 0, 1)$&$Spin(7)$&46&${\begin{aligned}&\tfrac{1}{2}(2,1;1,1,1;8)\\&+\tfrac{1}{2}(1,4;1,1,1;7)\\&+(E_{7})_{8}\end{aligned}}$\\
\hline
38&$\begin{matrix}\includegraphics[width=76pt]{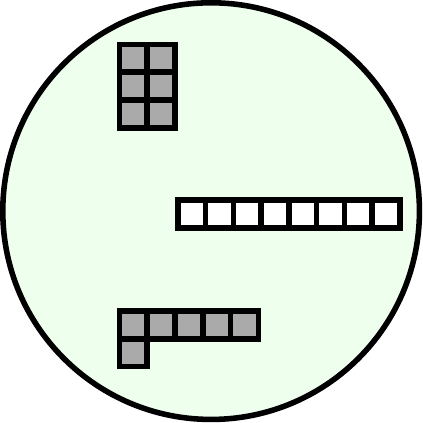}\end{matrix}$&$(1, 0, 2, 0, 2)$&$Spin(7)$&62&${\begin{aligned}&\tfrac{1}{2}(1,4,1;7)\\&+Spin(16)_{12}\times SU(2)_{8}\end{aligned}}$\\
\hline
39&$\begin{matrix}\includegraphics[width=76pt]{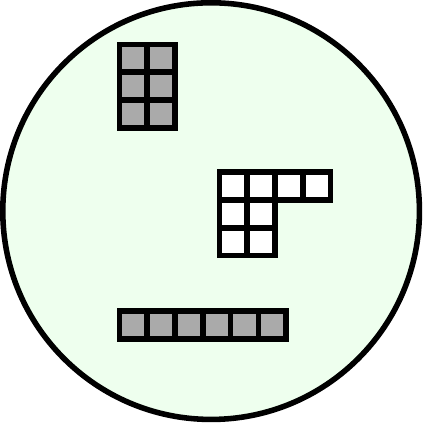}\end{matrix}$&$(1, 1, 1, 0, 0)$&$SU(4)$&32&${\begin{aligned}&(2,1;4)\\&+(1,6;4)\end{aligned}}$\\
\hline
40&$\begin{matrix}\includegraphics[width=76pt]{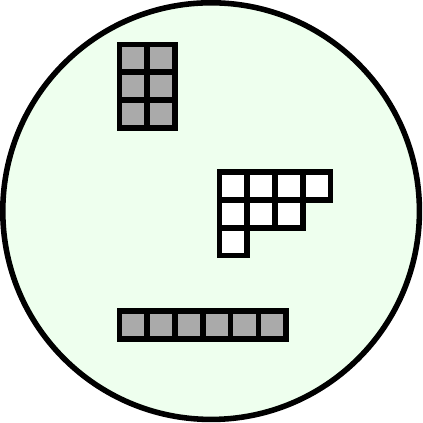}\end{matrix}$&$(1, 0, 1, 0, 1)$&$Spin(7)$&39&${\begin{aligned}&\tfrac{1}{2}(2,1,1;8)\\&+\tfrac{1}{2}(1,6,1;8)\\&+\tfrac{1}{2}(1,1,2;7)\end{aligned}}$\\
\hline
41&$\begin{matrix}\includegraphics[width=76pt]{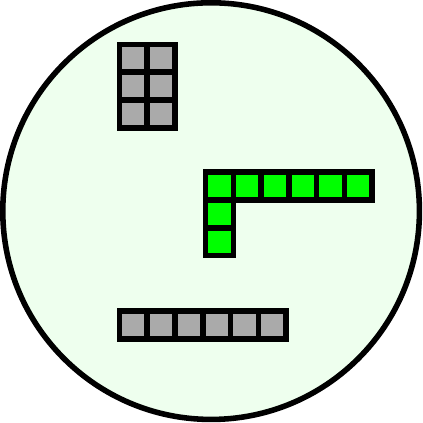}\end{matrix}$&$(1, 0, 2, 0, 1)$&$Spin(8)$&48&${\begin{aligned}&\tfrac{1}{2}(2,1,1;8_v)\\&+\tfrac{1}{2}(1,6,1;8_v)\\&+\tfrac{1}{2}(1,1,4;8_{s/c})\end{aligned}}$\\
\hline
42&$\begin{matrix}\includegraphics[width=76pt]{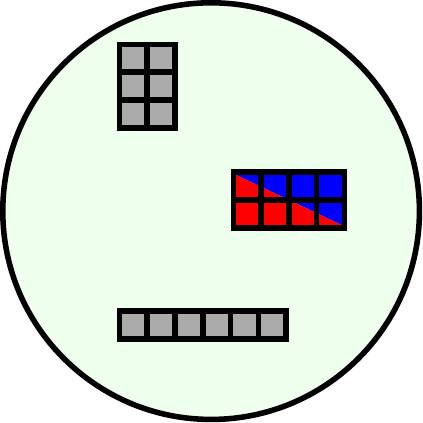}\end{matrix}$&$(1, 0, 2, 0, 1)$&$Spin(8)$&48&${\begin{aligned}&\tfrac{1}{2}(2,1,1;8_{c/s})\\&+\tfrac{1}{2}(1,6,1;8_v)\\&+\tfrac{1}{2}(1,1,4;8_{s/c})\end{aligned}}$\\
\hline
43&$\begin{matrix}\includegraphics[width=76pt]{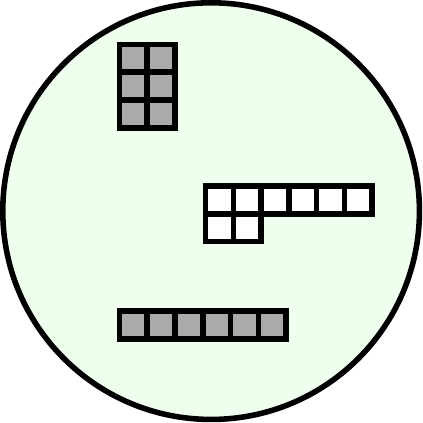}\end{matrix}$&$(1, 0, 3, 0, 1)$&$Spin(8)$&56&${\begin{aligned}&\tfrac{1}{2}(2,1;1,1,1;8_{s/c})\\&+\tfrac{1}{2}(1,6;1,1,1;8_v)\\&+(E_{7})_{8}\end{aligned}}$\\
\hline
44&$\begin{matrix}\includegraphics[width=76pt]{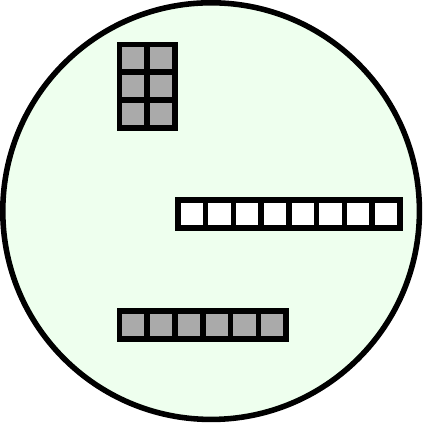}\end{matrix}$&$(1, 0, 3, 0, 2)$&$Spin(8)$&72&${\begin{aligned}&\tfrac{1}{2}(1,6,1;8_v)\\&+Spin(16)_{12}\times SU(2)_{8}\end{aligned}}$\\
\hline
\end{longtable}
}

\section{Applications}

\subsection{{$Spin(2N)$ and $Sp(N-1)$ Gauge Theory}}\label{_and__gauge_theory}

For general $N$, $SO(2N)$ gauge theory with $2(N-1)$ fundamental hypermultiplets, and $Sp(N-1)$ gauge theory with $2N$ fundamentals, are superconformal. Their construction is well-understood from the orientifold perspective \cite{Evans:1997hk,Landsteiner:1997vd,Brandhuber:1997cc,Hori:1998iv,Gimon:1998be}. In particular, the (2,0) theory of type $D_N$ is the theory on $2N$ coincident M5-branes at an orientifold singularity and, in that realization of these theories \cite{Tachikawa:2009rb}, the key building block is the fixture consisting of a twisted-sector minimal puncture, a twisted-sector full puncture and an untwisted-sector full puncture,

\begin{displaymath}
 \includegraphics[width=123pt]{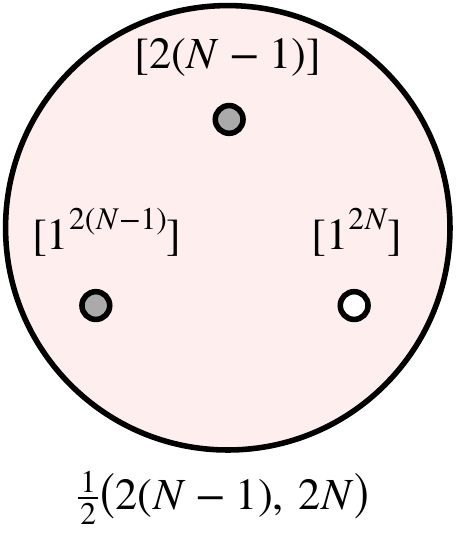}
\end{displaymath}
which is a free-field fixture transforming as a bifundamental half-hypermultiplet of ${Sp(N-1)\times SO(2N)}$. Taking two of these fixtures and connecting them with a $\quad\stackrel{\mathclap{[1^{2N}]}}{ \includegraphics[width=9pt]{untwistedPuncture}} \xleftrightarrow{\quad SO(2N)\quad}\stackrel{\mathclap{[1^{2N}]}}{ \includegraphics[width=9pt]{untwistedPuncture}}\quad$ cylinder yields the aformentioned $SO(2N)$ gauge theory. Connecting them, instead, with a $\quad\stackrel{\mathclap{[1^{2(N-1)}]}}{ \includegraphics[width=9pt]{twistedPuncture}} \xleftrightarrow{\,\quad Sp(N-1)\,\quad}\stackrel{\mathclap{[1^{2(N-1)}]}}{ \includegraphics[width=9pt]{twistedPuncture}}\quad$ cylinder yields the $Sp(N-1)$ gauge theory.

Here, we read off the S-dual strong-coupling descriptions. In the $SO(2N)$ case,

\begin{displaymath}
 \includegraphics[width=367pt]{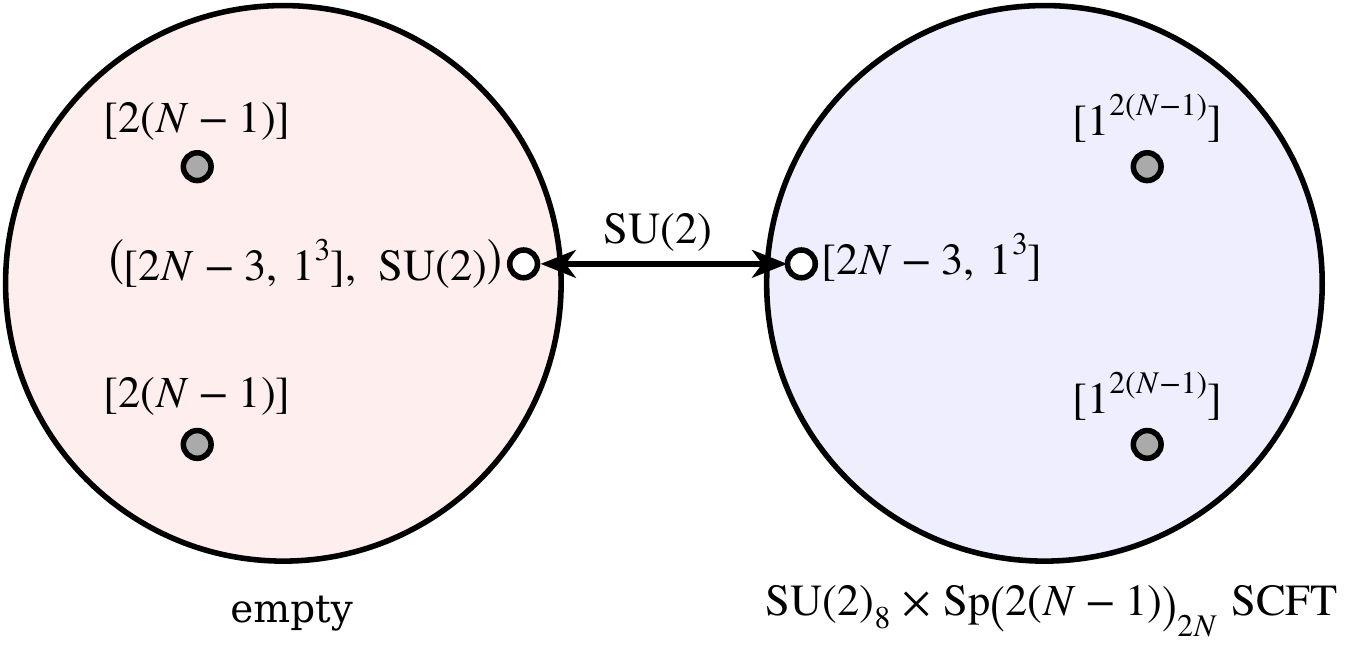}
\end{displaymath}
we have an $SU(2)$ gauging of the $SU(2)_8\times Sp\bigl(2(N-1)\bigr)_{2N}$ SCFT. In the $Sp\bigl(2(N-1)\bigr)$ case,

\begin{displaymath}
 \includegraphics[width=367pt]{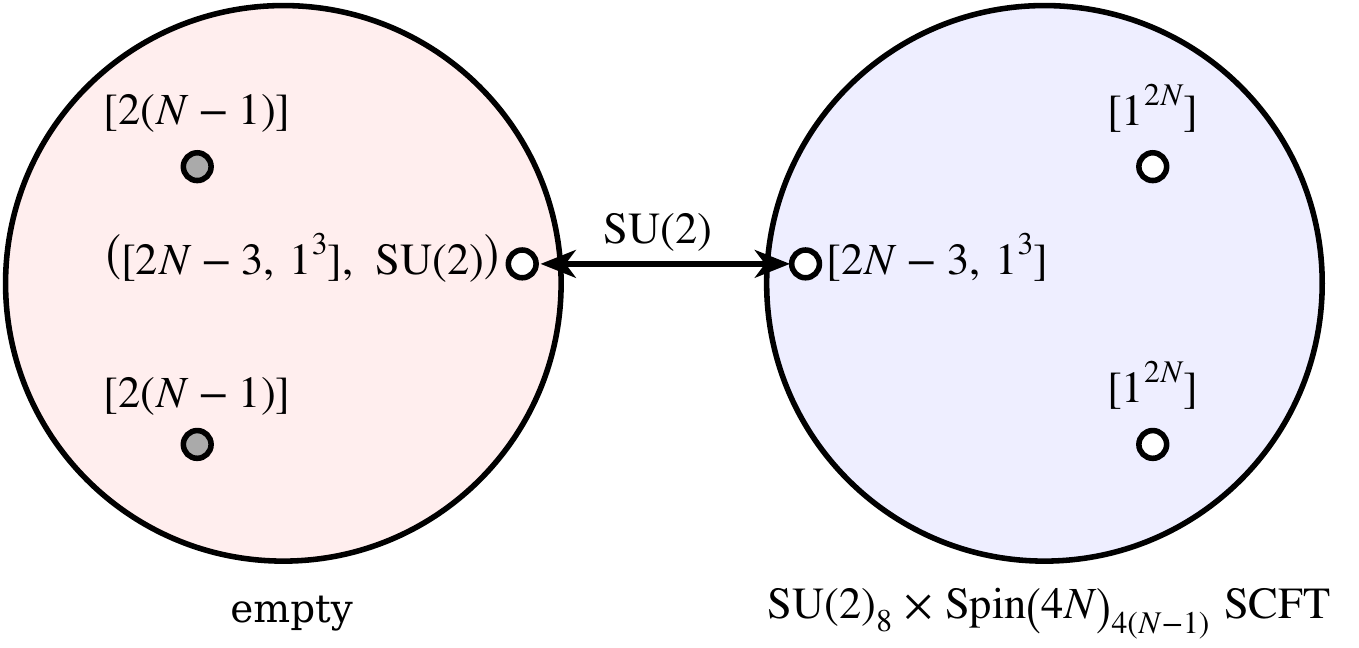}
\end{displaymath}
we have an $SU(2)$ gauging of the $SU(2)_8\times Spin(4N)_{4(N-1)}$ SCFT.

For completeness, let us note that the other $Sp(N)$ gauge theory which is superconformal for \emph{arbitrary} $N\gt 1$, namely the one with one hypermultiplet in the traceless antisymmetric tensor and four hypermultiplets in the fundamental representation, was already realized (with the addition of a single free hypermultiplet) in the untwisted sector of the $A_{2N-1}$ theory \cite{Chacaltana:2010ks}

\begin{displaymath}
 \includegraphics[width=267pt]{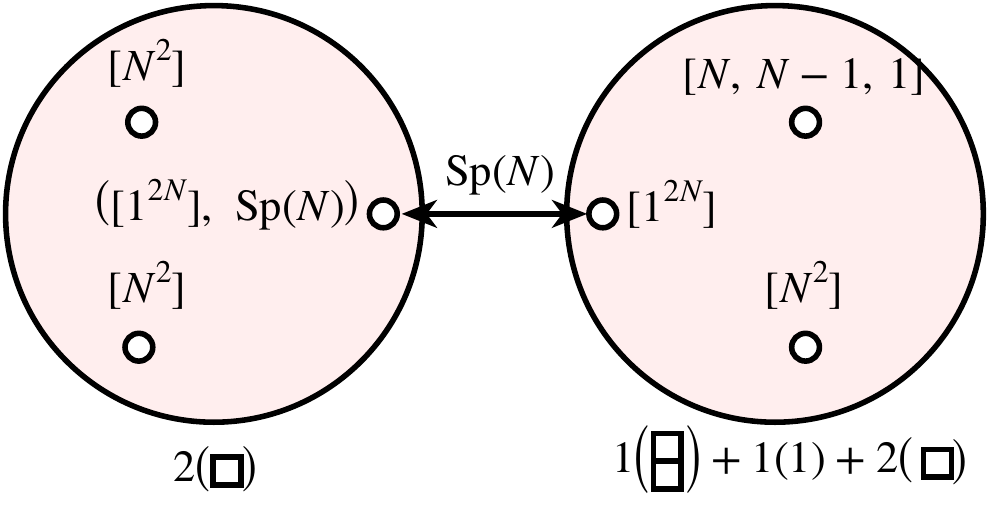}
\end{displaymath}
For this theory, by contrast, all the degeneration limits are (isomorphic) weakly-coupled Lagrangian field theories. The flavour symmetry group for this family of field theories is $F= {SU(2)}_{2N^2-N-1} \times {Spin(8)}_{2N}$. As is the case for $SU(2)$, $N_f=4$, the S-duality, which acts as an $S_3$ symmetry on $\mathcal{M}_{0,4}$, acts as outer automorphisms of the $Spin(8)$ flavour symmetry. Moreover, the Seiberg-Witten curve takes the absurdly simple form

\begin{displaymath}
0 = \lambda^{2N} +\sum_{k=1}^{N} u_{2k}\, \eta^k\lambda^{2(N-k)}
\end{displaymath}
where the quadratic differential

\begin{displaymath}
\eta(z) = \frac{z_{1 3}z_{2 4}{(d z)}^2}{(z-z_1)(z-z_2)(z-z_3)(z-z_4)}
\end{displaymath}

\subsection{{$Spin(8)$, $Spin(7)$ and $Sp(3)$ Gauge Theory}}\label{__and__gauge_theory}

\subsubsection{{$Spin(8)$ Gauge Theory}}\label{_gauge_theory}

$Spin(8)$ gauge theory, with matter in the $n_v(8_v)+n_s(8_s)+n_c(8_c)$, is superconformal for $n_v+n_s+n_c=6$. Up to permutations, related to triality, the list of possible values for $n_v,n_s,n_c$ is quite short and we discussed most of them in \cite{Chacaltana:2011ze}. There were, however, two cases which were not realizable with only untwisted sector punctures.

One is $n_v=6$, which is a special case of the construction in \S\ref{_and__gauge_theory}. The other case is $n_v=5,\, n_s=1$ (which, as we shall presently see, lies in the same moduli space as $n_v=5,\, n_c=1$).

Consider the 4-punctured sphere

\begin{displaymath}
 \includegraphics[width=318pt]{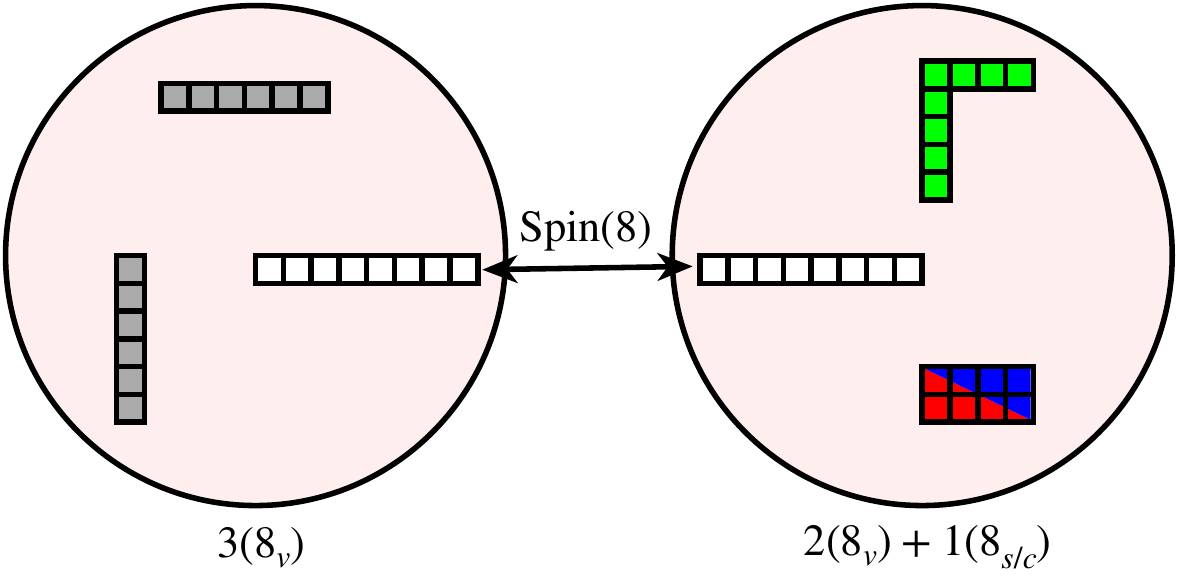}
\end{displaymath}
This is a weakly-coupled $Spin(8)$ gauge theory with matter in either the $5(8_v)+1(8_s)$ or the $5(8_v)+1(8_c)$. The two realizations are exchanged by dragging the $ \includegraphics[width=30pt]{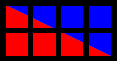}$ puncture around one of the twisted-sector punctures and returning it to its original location.

The strong coupling limits are $SU(2)$ gauge theories

\begin{displaymath}
 \includegraphics[width=318pt]{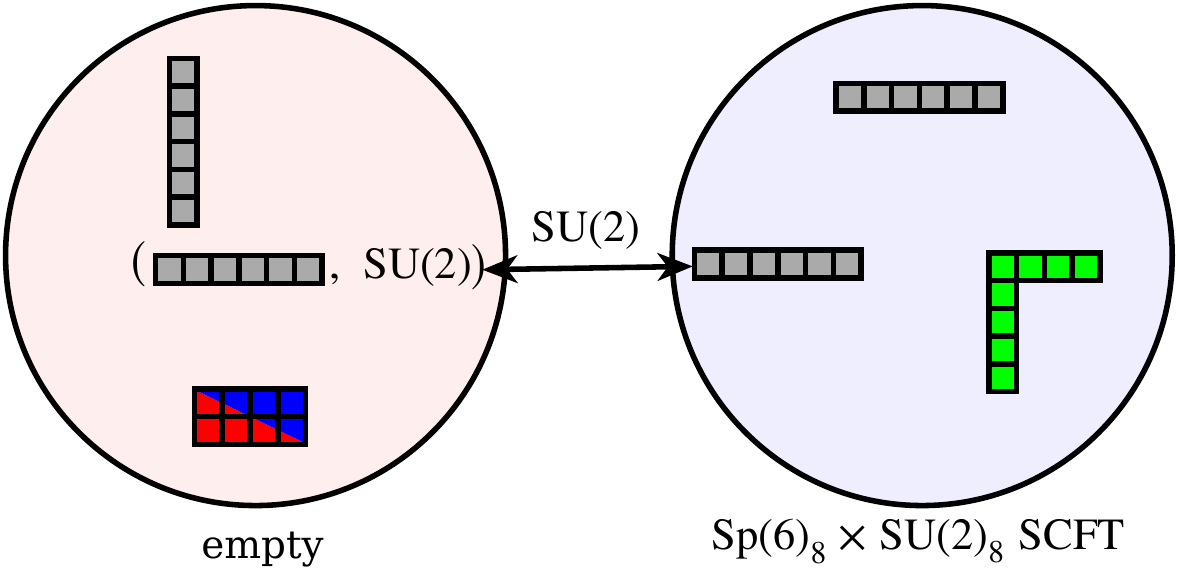}
\end{displaymath}
(where we gauge an $SU(2)$ subgroup of ${Sp(6)}_8$) and

\begin{displaymath}
 \includegraphics[width=318pt]{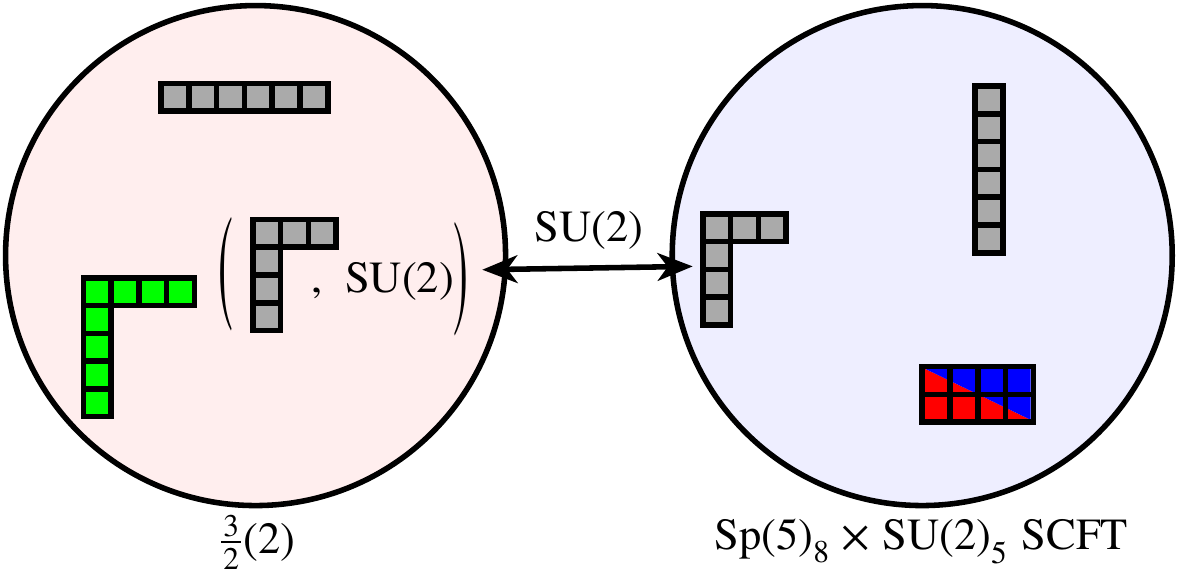}
\end{displaymath}
where the ${SU(2)}_5$ is gauged.

\subsubsection{{$Spin(7)$ Gauge Theory}}\label{spin7_gauge_theory}

Similar to the case of $Spin(8)$ gauge theory, realizations of most cases of conformally-invariant $Spin(7)$ gauge theory were already discussed in \cite{Chacaltana:2011ze}. Here we show realizations of the missing two cases.

\paragraph{{$5(7)$}}\label{_7}{~}

With the addition of three free hypermultiplets, we have a realization of the theory with 5 hypermultiplets in the vector representation as

\begin{displaymath}
 \includegraphics[width=318pt]{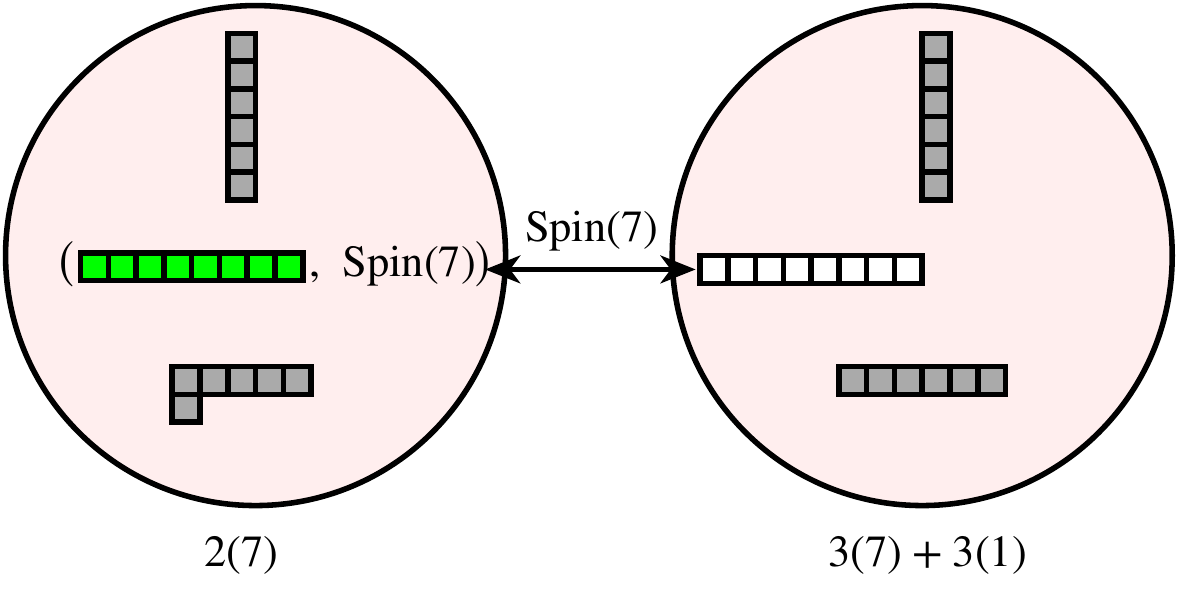}
\end{displaymath}
The S-dual theory is an $SU(2)$ gauging of the $Sp(5)_7\times SU(2)_8$ SCFT, plus 3 free hypermultiplets.

\begin{displaymath}
 \includegraphics[width=318pt]{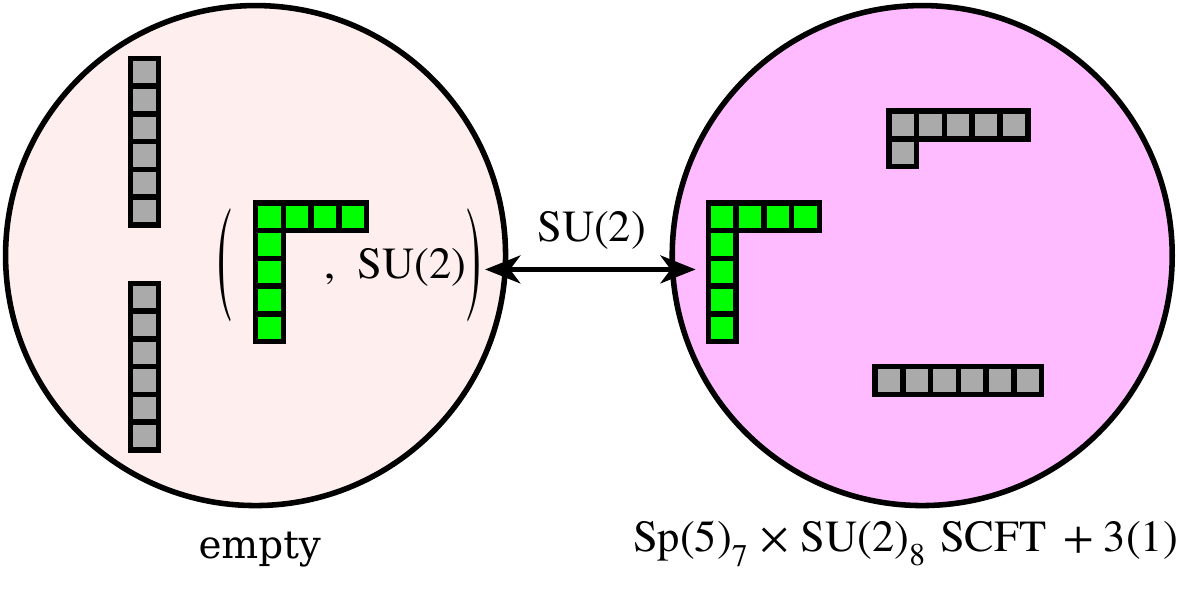}
\end{displaymath}

\paragraph{{$1(8)+4(7)$}}\label{_8} {~}

The $Spin(7)$ gauge theory, with one spinor and four vectors, can be realized in a couple of different ways. With the addition of three free hypermultiplets, we have

\begin{displaymath}
 \includegraphics[width=318pt]{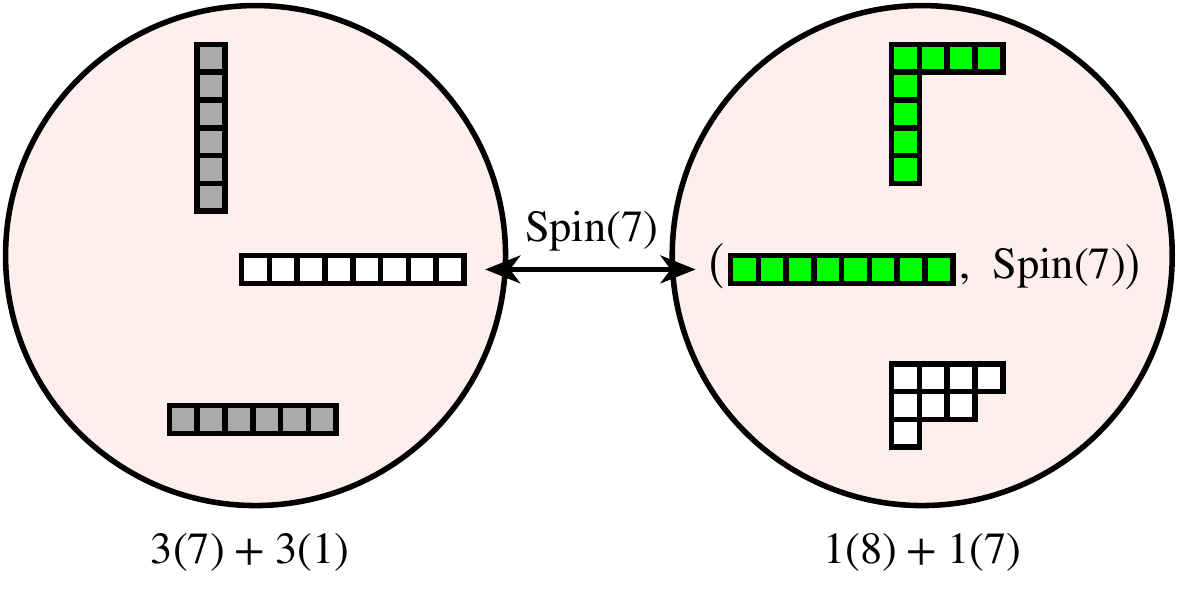}
\end{displaymath}
There are two S-dual descriptions. Both are $SU(2)$ gauge theories; one with a half-hypermultiplet in the fundamental, gauging an $SU(2)$ subgroup of the $Sp(5)$ symmetry of the $Sp(5)_7\times SU(2)_8$ SCFT,

\begin{displaymath}
 \includegraphics[width=318pt]{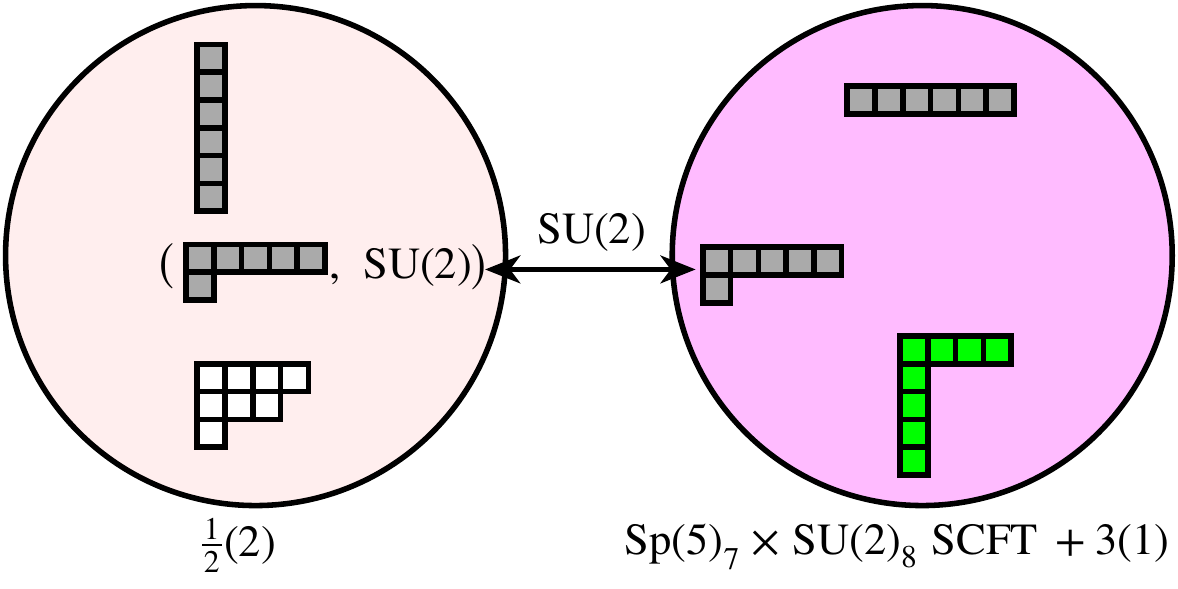}
\end{displaymath}
the other with three half-hypermultiplets in the fundamental, gauging the ${SU(2)}_5$ of the ${Sp(4)}_7\times {SU(2)}_5$ SCFT

\begin{displaymath}
 \includegraphics[width=318pt]{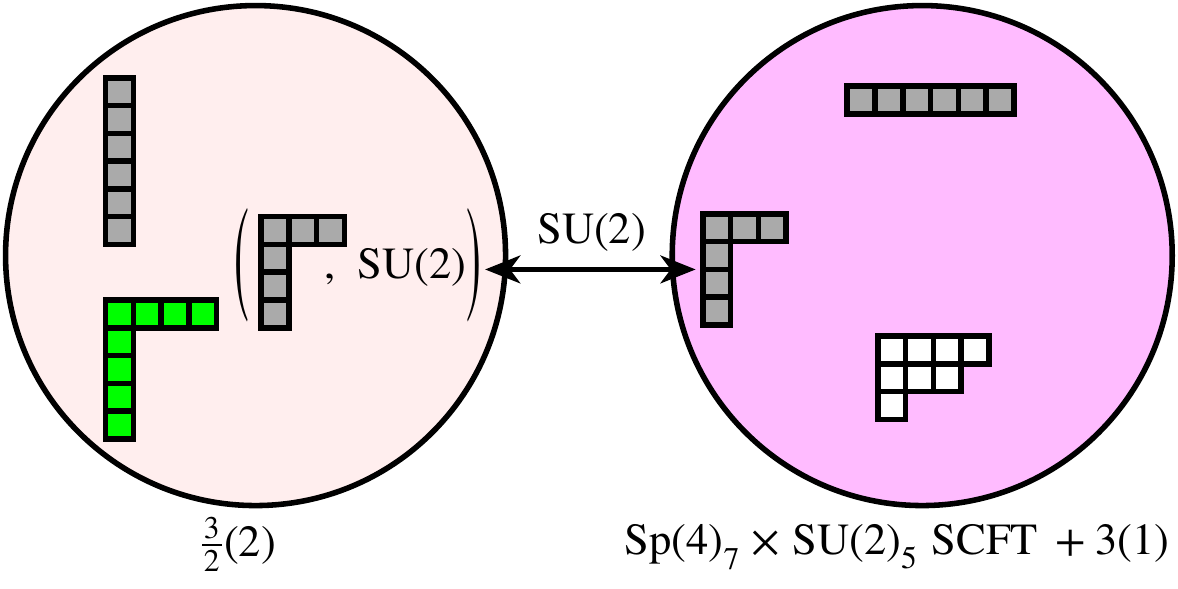}
\end{displaymath}
Another realization, with the addition of only two free hypermultiplets, is

\begin{displaymath}
 \includegraphics[width=318pt]{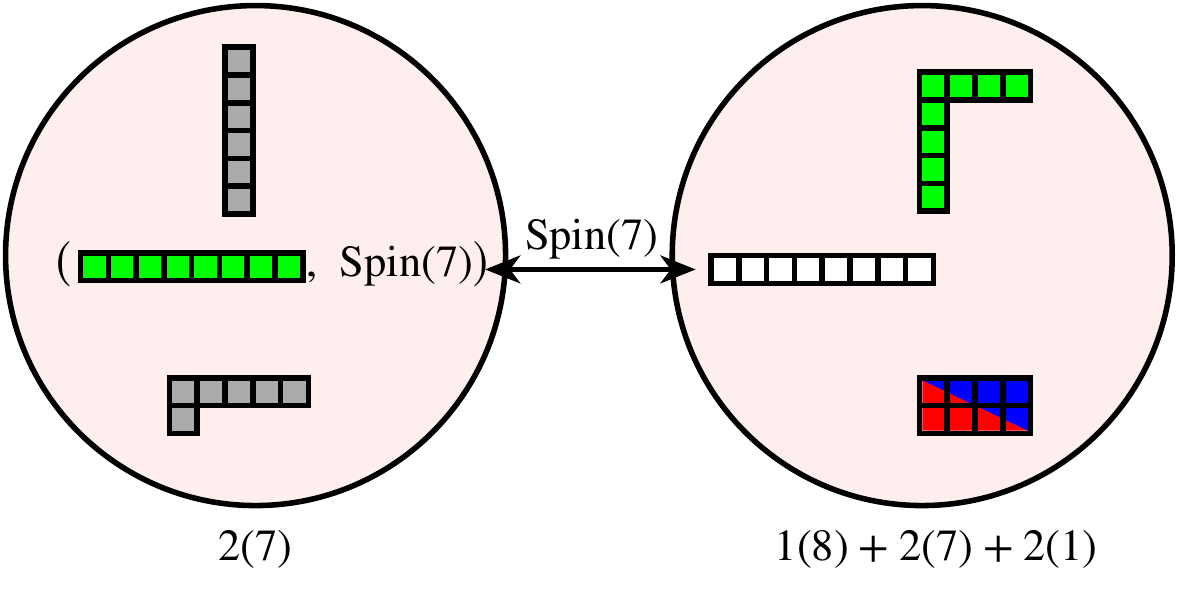}
\end{displaymath}
where the S-dual theories are

\begin{displaymath}
 \includegraphics[width=318pt]{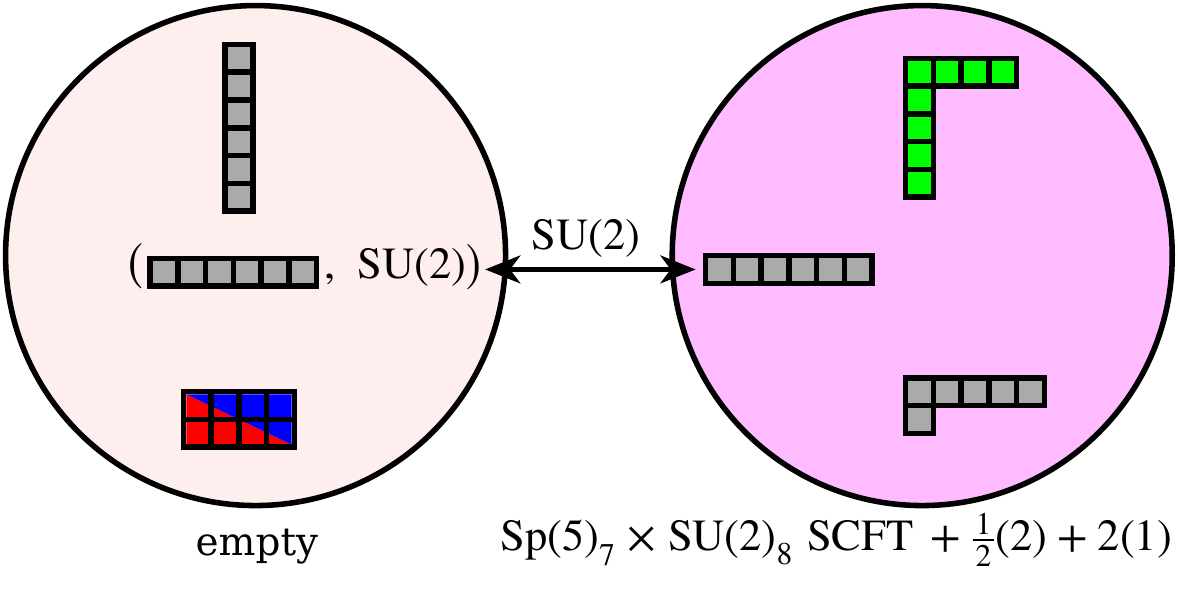}
\end{displaymath}
and
\begin{displaymath}
 \includegraphics[width=318pt]{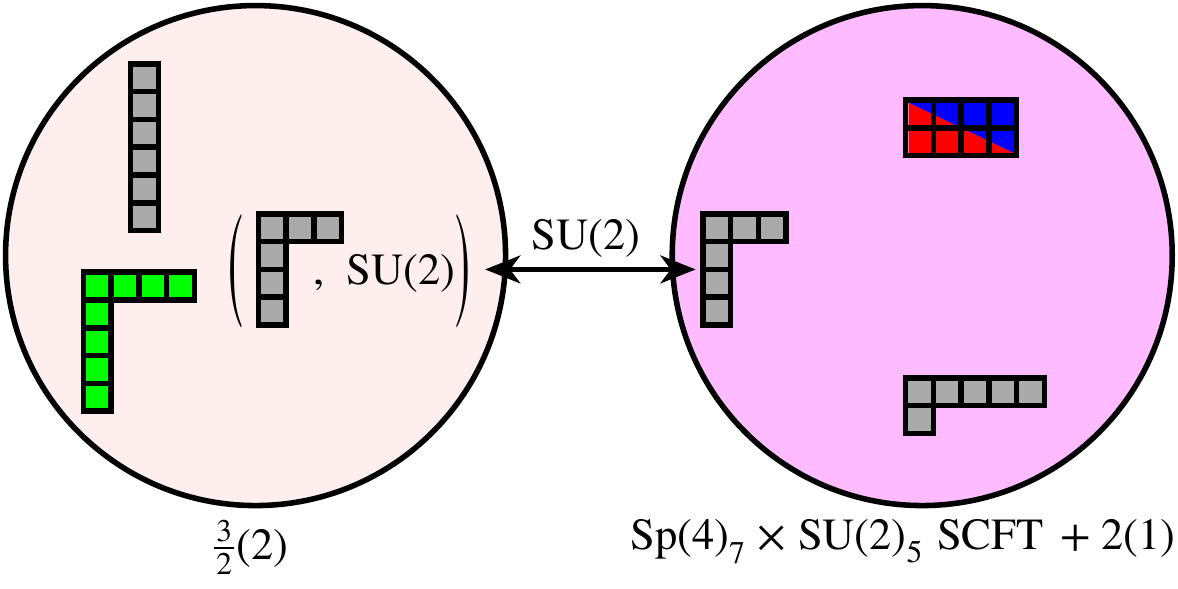}
\end{displaymath}

\subsubsection{{$Sp(3)$ Gauge Theory}}\label{sp3_gauge_theory}

In this section, we will consider various cases of $Sp(3)$ gauge theory, with vanished $\beta$-function. We have already discussed the theory with $8(6)$ and the theory with $1(14)+4(6)$ (special cases of the discussion of \S\ref{_and__gauge_theory}).

The $14'$, the traceless 3-index antisymmetric tensor representation, is pseudoreal and has index $\ell=5$. So we can replace five fundamental (half-)hypermultiplets with a $14'$ (half-)hypermultiplet.

\paragraph{\underline{$\tfrac{11}{2}(6) +\tfrac{1}{2}(14')$}}\label{_9}

With one half-hypermultiplet in the $14'$, we have

\begin{equation}
 \includegraphics[width=318pt]{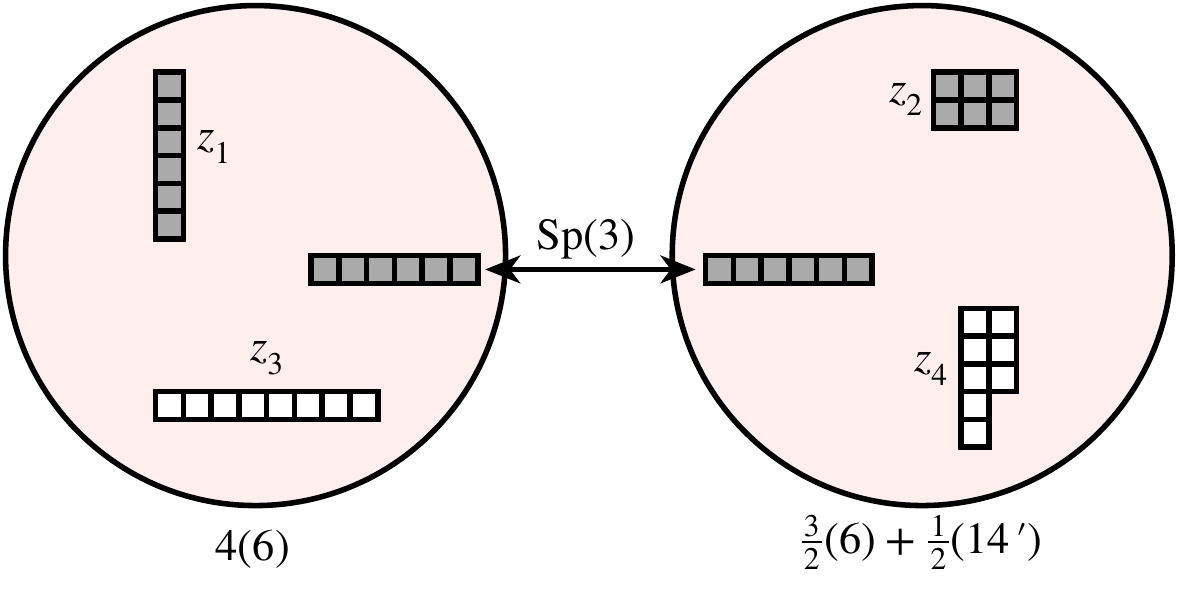}
\label{Sp3half14prime}\end{equation}
At strong coupling, we have an $Sp(2)$ gauging of the ${(E_8)}_{12}$ SCFT

\begin{displaymath}
 \includegraphics[width=318pt]{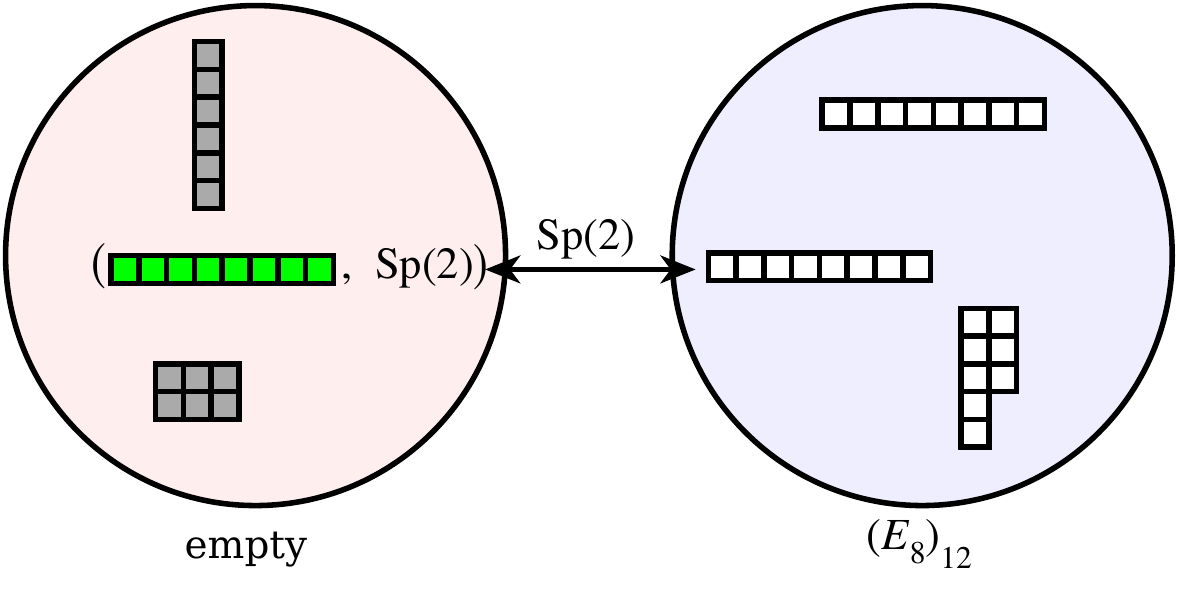}
\end{displaymath}
The third boundary point involves a gauge-theory fixture

\begin{displaymath}
 \includegraphics[width=318pt]{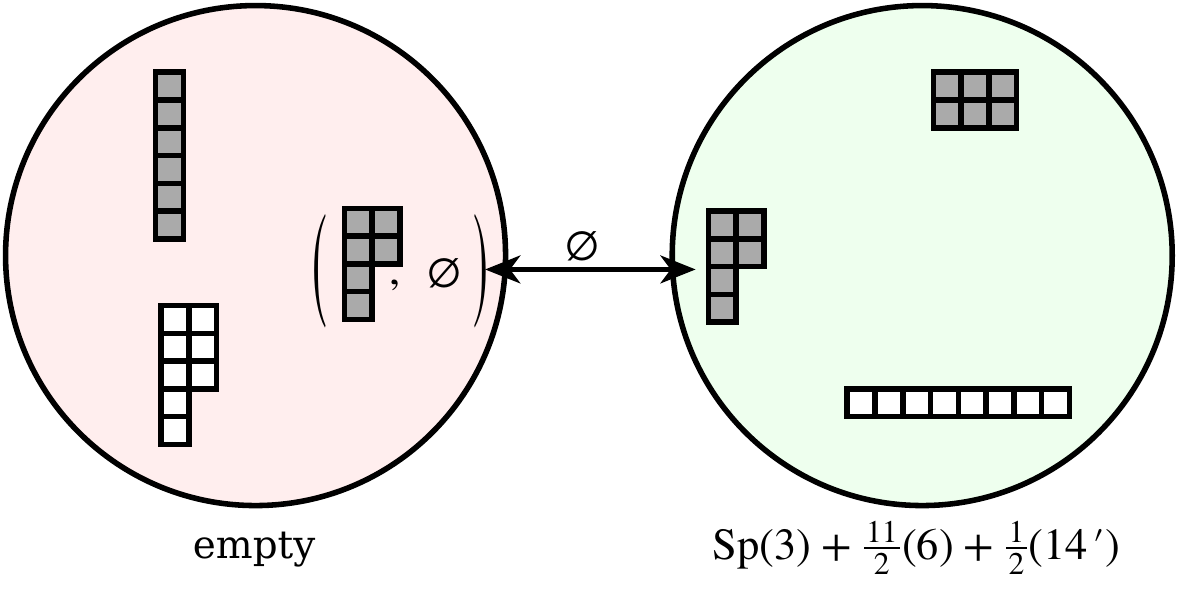}
\end{displaymath}

\paragraph{\underline{$3(6)+1(14')$}}\label{_10}

With two half- or one full-hypermultiplet in the $14'$, we have

\begin{equation}
 \includegraphics[width=318pt]{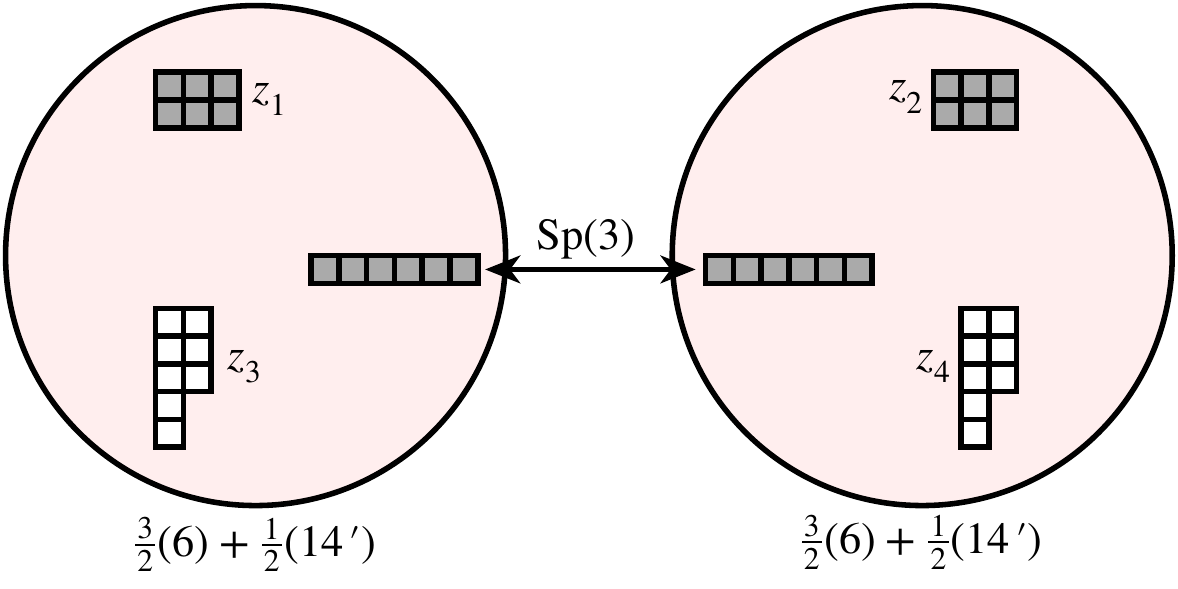}
\label{Sp3full14prime}\end{equation}
whose S-dual is an $SU(2)$ gauging of the ${SU(4)}_{12}\times {SU(2)}_7\times U(1)$ SCFT, with an additional half-hypermultiplet in the fundamental:

\begin{displaymath}
 \includegraphics[width=318pt]{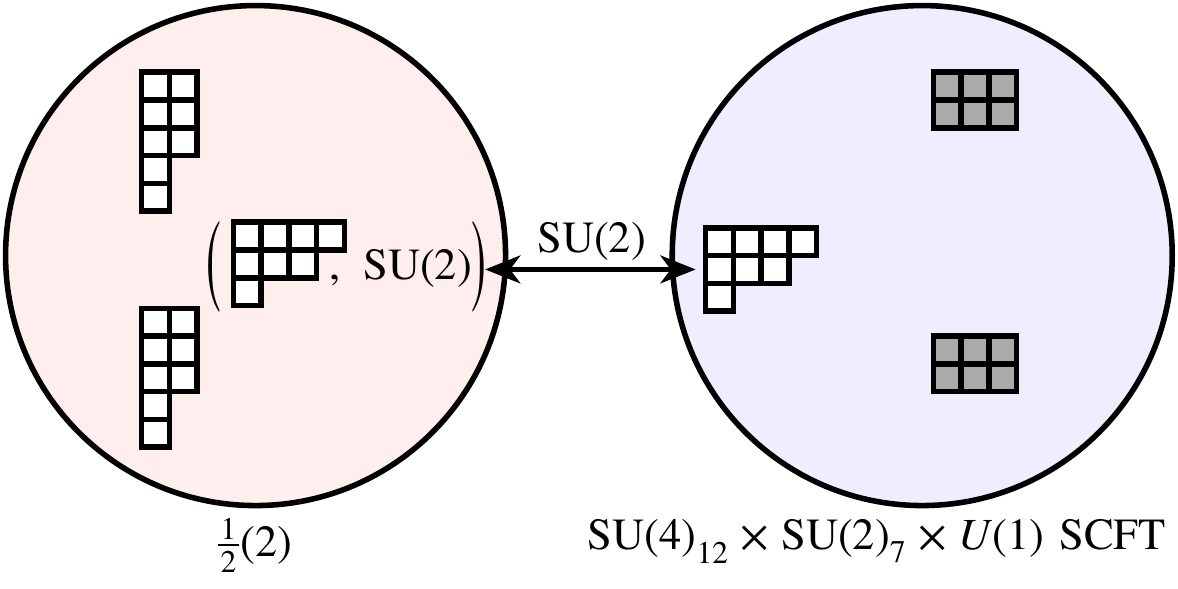}
\end{displaymath}
Because, to our knowledge, the Seiberg-Witten solution to this theory has not been studied in the literature, let us present some of the details, here. Setting the locations of the punctures on $C=\mathbb{C}\mathrm{P}^1$ as in \eqref{Sp3full14prime}, the Seiberg-Witten curve is the locus in $T^*C$ given by the equation

\begin{equation}
0 = \lambda^8 +\sum_{k=1}^3 \lambda^{8-2k}\phi_{2k}(z) +{(\tilde{\phi}(z))}^2
\label{SWSp3full14prime}\end{equation}
where $\lambda=y d z$ is the Seiberg-Witten differential. In the case at hand,

\begin{displaymath}
\begin{split}
\phi_2(z) &= \frac{u_2 z_{1 4} z_{2 3} {(d z)}^2}{(z-z_1)(z-z_2)(z-z_3)(z-z_4)}\\
\phi_4(z) &= \frac{z_{1 4} z_{2 3}\left[\tfrac{1}{4}u_2^2 (z-z_1)(z-z_2)z_{1 4}z_{2 3} + u_4 (z-z_3)(z-z_4)z_{1 2}^2
\right] {(d z)}^4}{{(z-z_1)}^3{(z-z_2)}^3{(z-z_3)}^2{(z-z_4)}^2}\\
\phi_6(z) &= \frac{u_6 z_{1 4}^2 z_{2 3}^2 z_{1 2}^2 {(d z)}^6}{{(z-z_1)}^4{(z-z_2)}^4{(z-z_3)}^2{(z-z_4)}^2}\\
\tilde{\phi}(z) &= 0
\end{split}
\end{displaymath}
Setting $(z_1,z_2,z_3,z_4)\to (0,\infty,x,1)$, \eqref{SWSp3full14prime} simplifies to

\begin{multline}
0 = y^2\left[
y^6 + y^4 \frac{u_2}{z(z-1)(z-x)}
\right.\\
\left.
+ y^2\frac{1}{z(z-1)(z-x)}
  \left(
  \frac{\tfrac{1}{4}u_2^2}{(z-1)(z-x)}+\frac{u_4}{z^2}
  \right)
+\frac{u_6}{z^4{(z-1)}^2{(z-x)}^2}
\right]
\end{multline}
The S-duality group of this theory is $\Gamma(2)$, and we have $f(\tau)=x$.

Repeating the analysis for \eqref{Sp3half14prime}, we find the Seiberg-Witten curve for $Sp(3)$ with $\tfrac{11}{2}(6)+\tfrac{1}{2}(14')$ to be

\begin{multline}
0=y^2\left[y^6+y^4 \frac{u_2}{z(z-1)(z-x)}
\right.\\
\left.
+ y^2\frac{1}{z(z-1){(z-x)}^3}\left(\tfrac{1}{4}u_2^2 \frac{(x-1)}{(z-1)}+u_4\right)+\frac{u_6(x-1)}{z{(z-1)}^2{(z-x)}^5}\right]
\end{multline}
In this case, the moduli space is the branched double-cover of $\mathcal{M}_{0,4}$, parametrized by $w^2=x$. The gauge coupling is

\begin{displaymath}
f(\tau) = \frac{2w}{1+w}
\end{displaymath}
In particular, the S-duality group is the $\Gamma_0(2)$, generated by
\begin{displaymath}
T:\,\tau\mapsto\tau+1,\quad S T^2 S:\,\tau\mapsto\frac{\tau}{1-2\tau}\quad .
\end{displaymath}
Here, $T$ acts as the deck transformation, $w\mapsto -w$, and $S T^2 S$ acts trivially on the $w$-plane. The theory at $f(\tau)=0$ is the Lagrangian field theory; at $f(\tau)=1,\infty$ (which project to $x=1$) we have the $Sp(2)$ gauging of the ${(E_8)}_{12}$ SCFT. The gauge theory fixture, at $x=\infty$, is the theory at the $\mathbb{Z}_2$-invariant interior point of the moduli space, $f(\tau)=2$.

\paragraph{\underline{Other cases}}
The remaining cases of $Sp(3)$ with vanishing $\beta$-function have matter in the
\begin{itemize}
\item $2(14)$
\item $\tfrac{3}{2}(6) + 1(14) + \tfrac{1}{2}(14')$
\item $\tfrac{1}{2}(6) + \tfrac{3}{2}(14')$
\end{itemize}
Unfortunately, we don't know how to realize these theories as compactifications from 6 dimensions. Presumably, the methods of \cite{Tachikawa:2011yr} can be applied, to recover these cases as well.

\subsection{{Higher Genus}}\label{higher_genus}

In almost all of the discussion in this paper, we have taken $C$ to be genus-zero. We should close with at least one example of higher-genus, so that we can see the effect of twists around handles of $C$.

Consider a genus-one curve, with one minimal puncture, in the $D_4$ theory.

\begin{displaymath}
 \includegraphics[width=227pt]{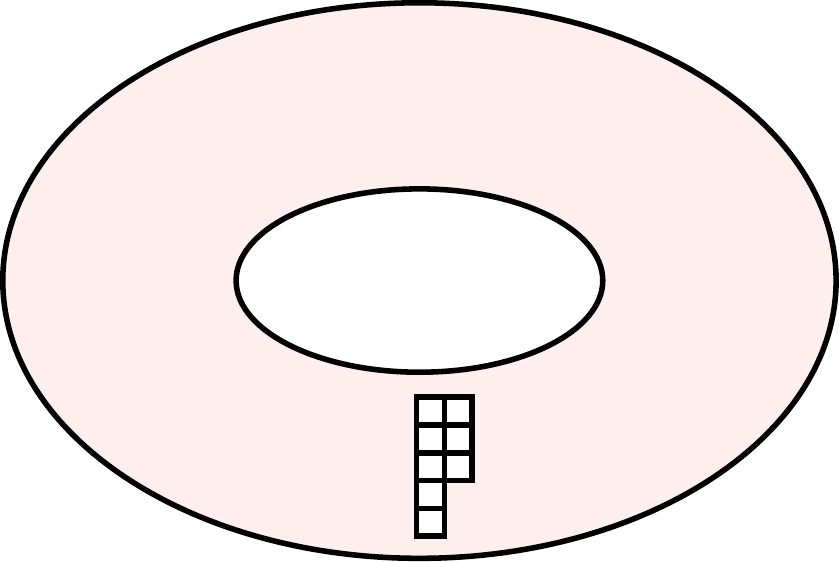}
\end{displaymath}
$H^1(T^2-{p},\mathbb{Z}_2) = {(\mathbb{Z}_2)}^2$. Under the action of the modular group, $H^1(T^2-{p},\mathbb{Z}_2)$ breaks up into two orbits: the zero orbit (the ``untwisted theory'') and the nonzero orbit (``the twisted theory'').

The untwisted theory is a $Spin(8)$ gauging of the ${(E_8)}_{12}$ SCFT. There are three inequivalent index-2 embeddings of $Spin(8)$ in $E_8$. They can be characterized by how the $248$ decomposes (up to outer automorphisms of $Spin(8)$). Either

\begin{subequations}
\begin{equation}
248 = 3(1) +5(28) +{35}_v + {35}_s +{35}_c
\label{E8decomp1}\end{equation}
or
\begin{equation}
248 = 1 +2(8_v)+3(28) +{35}_v + 2({56}_v)
\label{E8decomp2}\end{equation}
or
\begin{equation}
248 = 8_v+8_s+8_c+2(28) +{56}_v + {56}_s +{56}_c
\label{E8decomp3}\end{equation}
\end{subequations}

The untwisted theory corresponds to \eqref{E8decomp1}. The twisted theory, depending on the S-duality frame chosen, corresponds either to a $Spin(8)$ gauging of the ${(E_8)}_{12}$ SCFT using the embedding \eqref{E8decomp2}, or to an $Sp(3)$ gauging of the ${Sp(6)}_8$ SCFT.

For the untwisted theory, the gauge theory moduli space is the fundamental domain for $PSL(2,\mathbb{Z})$ in the UHP, and $\tau$ is the modular parameter of the torus. For the twisted theory, the moduli space of the gauge theory is the moduli space of pairs $(C,\gamma)$, where $\gamma$ is a nonzero element of $H^1(C,\mathbb{Z}_2)$. This is the fundamental domain of $\Gamma_0(2)$, as discussed in \S\ref{gauge_couplings}.

\section*{Acknowledgements}\label{Acknowledgements}
\addcontentsline{toc}{section}{Acknowledgements}

We would like to thank S.~Katz, D.~Morrison, A.~Neitzke, R.~Plesser and Y.~Tachikawa for helpful discussions. J.~D.~and O.~C.~would like to thank the Aspen Center for Physics (supported, in part, by the National Science Foundation under Grant PHY-1066293) for their hospitality during the writing of this manuscript. O.~C.~would further like to thank the Simons Foundation for partial support in Aspen. The work of J.~D.~and A.~T.~was supported in part by the National Science Foundation under Grants PHY-0969020 and PHY-1316033. The work of O.~C.~was supported in part by the INCT-Matem\'atica and the ICTP-SAIFR in Brazil through a Capes postdoctoral fellowship.

%\appendix
\begin{appendices}
\section{{Tables of Properties of Twisted Sectors}}\label{appendix_tables_of_properties_of_twisted_sectors}

\subsection{{$D_{5}$ Twisted Sector}}\label{_twisted_sector}

{
\renewcommand{\arraystretch}{1.5}
\begin{longtable}{|c|c|c|c|c|c|}
\hline
\mbox{\shortstack{\\Nahm\\C-partition}}&\mbox{\shortstack{\\Hitchin\\B-partition}}&Pole structure&Constraints&Flavour group&$(\delta n_{h},\delta n_{v})$\\
\hline 
\endhead
$\begin{matrix}\includegraphics[width=62pt]{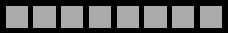}\end{matrix}$&$[9]$&$\{1,3,5,7;\frac{9}{2}\}$&$-$&${Sp(4)}_{10}$&$(240,\tfrac{449}{2})$\\
\hline
$\begin{matrix} \includegraphics[width=62pt]{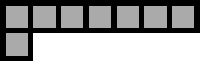}\end{matrix}(\text{ns})$&$([7,1^{2}],\mathbb{Z}_2)$&$\{1,3,5,7;\frac{7}{2}\}$&$-$&${Sp(3)}_9$&$(227,\tfrac{431}{2})$\\
\hline
$\begin{matrix}\includegraphics[width=47pt]{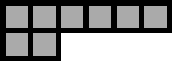}\end{matrix}$&$[7,1^{2}]$&$\{1,3,5,7;\frac{7}{2}\}$&$c^{(8)}_{7}=(a^{(4)}_{7/2})^{2}$&${Sp(2)}_8\times U(1)$&$(216,\tfrac{415}{2})$\\
\hline 
$\begin{matrix} \includegraphics[width=39pt]{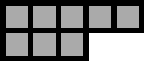}\end{matrix}(\text{ns})$&$([5,3,1],\mathbb{Z}_2)$&$\{1,3,5,6;\frac{7}{2}\}$&$-$&${SU(2)}_{32}\times {SU(2)}_7$&$(207,\tfrac{401}{2})$\\
\hline 
$\begin{matrix}\includegraphics[width=32pt]{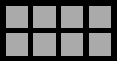}\end{matrix}$&$[5,3,1]$&$\{1,3,5,6;\frac{7}{2}\}$&$c^{(6)}_{5}=(a^{(3)}_{5/2})^{2}$&${SU(2)}_{16}^2$&$(200,\tfrac{389}{2})$\\
\hline 
$\begin{matrix}\includegraphics[width=32pt]{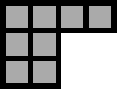}\end{matrix}$&$[5,2^{2}]$&$\{1,3,5,6;\frac{7}{2}\}$&$\begin{gathered}c^{(6)}_{5}=(a^{(3)}_{5/2})^{2}\\c^{(8)}_6=2a^{(3)}_{5/2}\tilde{c}^{(5)}_{7/2}\end{gathered}$&${SU(2)}_{10}\times {SU(2)}_6$&$(184,\tfrac{359}{2})$\\
\hline 
$\begin{matrix} \includegraphics[width=39pt]{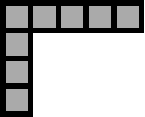}\end{matrix}(\text{ns})$&$([5,1^{4}],\mathbb{Z}_2)$&$\{1,3,5,5;\frac{5}{2}\}$&$-$&${Sp(2)}_7$&$(182,\tfrac{353}{2})$\\
\hline 
$\begin{matrix}\includegraphics[width=32pt]{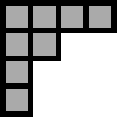}\end{matrix}$&$[5,1^{4}]$&$\{1,3,5,5;\frac{5}{2}\}$&$c^{(6)}_{5}=(a^{(3)}_{5/2})^{2}$&${SU(2)}_6$&$(174,\tfrac{341}{2})$\\
\hline 
$\begin{matrix}\includegraphics[width=24pt]{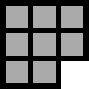}\end{matrix}$&$[3^{3}]$&$\{1,3,4,5;\frac{7}{2}\}$&$-$&${SU(2)}_{10}$&$(178,\tfrac{349}{2})$\\
\hline 
$\begin{matrix}\includegraphics[width=24pt]{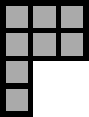}\end{matrix}$&$[3^{2},1^{3}]$&$\{1,3,4,5;\frac{5}{2}\}$&$-$&$U(1)$&$(168,\tfrac{331}{2})$\\
\hline 
$\begin{matrix}\includegraphics[width=17pt]{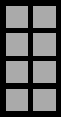}\end{matrix}$&$[3,2^{2},1^{2}]$&$\{1,3,4,5;\frac{5}{2}\}$&$\begin{gathered}c^{(4)}_{3}=(a^{(2)}_{3/2})^{2}\\c^{(6)}_{4}=2a^{(2)}_{3/2}a^{(4)}_{5/2}\\c^{(8)}_{5}=(a^{(4)}_{5/2})^{2}\end{gathered}$&$U(1)$&$(144,\tfrac{285}{2})$\\
\hline 
$\begin{matrix} \includegraphics[width=24pt]{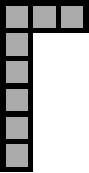}\end{matrix}(\text{ns})$&$([3,1^{6}],\mathbb{Z}_2)$&$\{1,3,3,3;\frac{3}{2}\}$&$-$&${SU(2)}_5$&$(117,\tfrac{231}{2})$\\
\hline 
$\begin{matrix}\includegraphics[width=17pt]{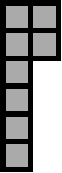}\end{matrix}$&$[3,1^{6}]$&$\{1,3,3,3;\frac{3}{2}\}$&$c^{(4)}_{3}=(a^{(2)}_{3/2})^{2}$&none&$(112,\tfrac{223}{2})$\\
\hline 
$\begin{matrix}\includegraphics[width=9pt]{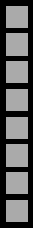}\end{matrix}$&$[1^{9}]$&$\{1,1,1,1;\frac{1}{2}\}$&$-$&none&$(40,\tfrac{81}{2})$\\
\hline 
\end{longtable}
}

\subsection{{$D_6$ Twisted Sector}}\label{_twisted_sector_2}

{
\renewcommand{\arraystretch}{1.5}
\begin{longtable}{|c|c|c|c|c|c|}
\hline 
\mbox{\shortstack{\\Nahm\\C-partition}}&\mbox{\shortstack{\\Hitchin\\B-partition}}&\mbox{\shortstack{\\Pole\\structure}}&Constraints&Flavour group&$(\delta n_{h},\delta n_{v})$\\
\hline
\endhead
$\begin{matrix}\includegraphics[width=51pt]{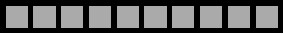}\end{matrix}$&$[11]$&$\{1,3,5,7,9;\tfrac{11}{2}\}$&$-$&${Sp(5)}_{12}$&$(440,\tfrac{831}{2})$\\
\hline 
$\begin{matrix} \includegraphics[width=46pt]{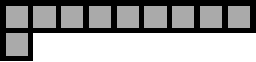}\end{matrix}(\text{ns})$&$([9,1^2],\mathbb{Z}_2)$&$\{1,3,5,7,9;\tfrac{9}{2}\}$&$-$&${Sp(4)}_{11}$&$(424,\tfrac{809}{2})$\\
\hline 
$\begin{matrix}\includegraphics[width=41pt]{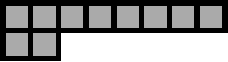}\end{matrix}$&$[9,1^2]$&$\{1,3,5,7,9;\tfrac{9}{2}\}$&$c^{(10)}_9=(a^{(5)}_{9/2})^2$&${Sp(3)}_{10}\times U(1)$&$(410,\tfrac{789}{2})$\\
\hline 
$\begin{matrix} \includegraphics[width=36pt]{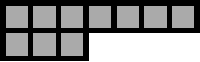}\end{matrix}(\text{ns})$&($[7,3,1],\mathbb{Z}_2)$&$\{1,3,5,7,8;\tfrac{9}{2}\}$&$-$&${Sp(2)}_9\times {SU(2)}_{40}$&$(398,\tfrac{771}{2})$\\
\hline 
$\begin{matrix}\includegraphics[width=31pt]{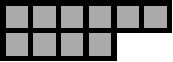}\end{matrix}$&$[7,3,1]$&$\{1,3,5,7,8;\tfrac{9}{2}\}$&$c^{(8)}_7=(a^{(4)}_{7/2})^2$&${SU(2)}_{20}^2\times{SU(2)}_8$&$(388,\tfrac{755}{2})$\\
\hline 
$\begin{matrix}\includegraphics[width=26pt]{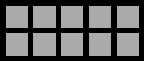}\end{matrix}$&$[5^2,1]$&$\{1,3,5,6,8;\tfrac{9}{2}\}$&$-$&${Sp(2)}_{12}$&$(380,\tfrac{741}{2})$\\
\hline 
$\begin{matrix}\includegraphics[width=31pt]{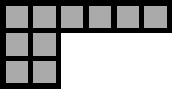}\end{matrix}$&$[7,2^2]$&$\{1,3,5,7,8;\tfrac{9}{2}\}$&$\begin{gathered}c^{(8)}_7 =(a^{(4)}_{7/2})^2\\c^{(10)}_8 =2a^{(4)}_{7/2}\tilde{c}^{(6)}_{9/2}\end{gathered}$&${Sp(2)}_8\times {SU(2)}_{12}$&$(368,\tfrac{717}{2})$\\
\hline 
$\begin{matrix} \includegraphics[width=26pt]{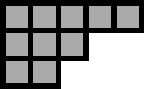}\end{matrix}(\text{ns})$&$([5,3^2],\mathbb{Z}_2)$&$\{1,3,5,6,7;\tfrac{9}{2}\}$&$-$&${SU(2)}_{12}\times{SU(2)}_7$&$(359,\tfrac{703}{2})$\\
\hline 
$\begin{matrix}\includegraphics[width=21pt]{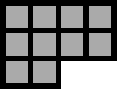}\end{matrix}$&$[5,3^2]$&$\{1,3,5,6,7;\tfrac{9}{2}\}$&$c^{(6)}_5=(a^{(3)}_{5/2})^2$&${SU(2)}_{12}\times U(1)$&$(352,\tfrac{691}{2})$\\
\hline 
$\begin{matrix} \includegraphics[width=36pt]{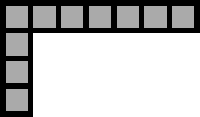}\end{matrix}(\text{ns})$&$([7,1^4],\mathbb{Z}_2)$&$\{1,3,5,7,7;\tfrac{7}{2}\}$&$-$&${Sp(3)}_9$&$(367,\tfrac{711}{2})$\\
\hline 
$\begin{matrix}\includegraphics[width=31pt]{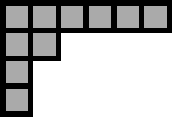}\end{matrix}$&$[7,1^4]$&$\{1,3,5,7,7;\tfrac{7}{2}\}$&$c^{(8)}_7=(a^{(4)}_{7/2})^2$&${Sp(2)}_8$&$(356,\tfrac{695}{2})$\\
\hline 
$\begin{matrix} \includegraphics[width=26pt]{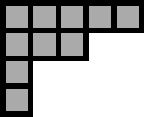}\end{matrix}(\text{ns})$&$([5,3,1^3],\mathbb{Z}_2)$&$\{1,3,5,6,7;\tfrac{7}{2}\}$&$-$&${SU(2)}_7\times U(1)$&$(347,\tfrac{681}{2})$\\
\hline 
$\begin{matrix}\includegraphics[width=21pt]{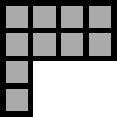}\end{matrix}$&$[5,3,1^3]$&$\{1,3,5,6,7;\tfrac{7}{2}\}$&$c^{(6)}_5=(a^{(3)}_{5/2})^2$&${SU(2)}_{32}$&$(340,\tfrac{669}{2})$\\
\hline 
$\begin{matrix}\includegraphics[width=21pt]{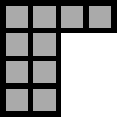}\end{matrix}$&$[5,2^2,1^2]$&$\{1,3,5,6,7;\tfrac{7}{2}\}$&$\begin{gathered}c^{(6)}_5 =(a^{(3)}_{5/2})^2\\c^{(8)}_6 =2a^{(3)}_{5/2}a^{(5)}_{7/2}\\c^{(10)}_7 =(a^{(5)}_{7/2})^2\end{gathered}$&${SU(2)}_6\times U(1)$&$(314,\tfrac{619}{2})$\\
\hline 
$\begin{matrix} \includegraphics[width=16pt]{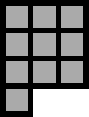}\end{matrix}(\text{ns})$&$([3^3,1^2],\mathbb{Z}_2)$&$\{1,3,4,5,7;\tfrac{7}{2}\}$&$-$&${SU(2)}_{11}$&$(319,\tfrac{629}{2})$\\
\hline 
$\begin{matrix}\includegraphics[width=16pt]{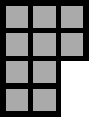}\end{matrix}$&$[3^3,1^2]$&$\{1,3,4,5,7;\tfrac{7}{2}\}$&$c^{(10)}_7=(a^{(5)}_{7/2})^2$&$U(1)$&$(308,\tfrac{609}{2})$\\
\hline 
$\begin{matrix}\includegraphics[width=11pt]{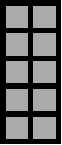}\end{matrix}$&$[3,2^4]$&$\{1,3,4,5,6;\tfrac{7}{2}\}$&$\begin{gathered}c^{(4)}_3 =(a^{(2)}_{3/2})^2\\c^{(6)}_4 =2a^{(2)}_{3/2}a^{(4)}_{5/2}\\c^{(8)}_5=(a^{(4)}_{5/2})^2\\+2a^{(2)}_{3/2}\tilde{c}^{(6)}_{7/2}\\c^{(10)}_6 =2a^{(4)}_{5/2}\tilde{c}^{(6)}_{7/2}\end{gathered}$&${SU(2)}_{12}$&$(256,\tfrac{507}{2})$\\
\hline 
$\begin{matrix} \includegraphics[width=26pt]{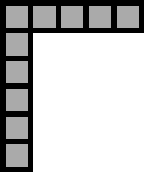}\end{matrix}(\text{ns})$&$([5,1^6],\mathbb{Z}_2)$&$\{1,3,5,5,5;\tfrac{5}{2}\}$&$-$&${Sp(2)}_7$&$(282,\tfrac{553}{2})$\\
\hline 
$\begin{matrix}\includegraphics[width=21pt]{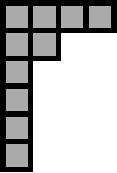}\end{matrix}$&$[5,1^6]$&$\{1,3,5,5,5;\tfrac{5}{2}\}$&$c^{(6)}_5=(a^{(3)}_{5/2})^2$&${SU(2)}_6$&$(274,\tfrac{541}{2})$\\
\hline 
$\begin{matrix}\includegraphics[width=16pt]{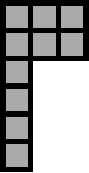}\end{matrix}$&$[3^2,1^5]$&$\{1,3,4,5,5;\tfrac{5}{2}\}$&$-$&$U(1)$&$(268,\tfrac{531}{2})$\\
\hline 
$\begin{matrix}\includegraphics[width=11pt]{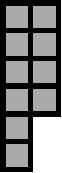}\end{matrix}$&$[3,2^2,1^4]$&$\{1,3,4,5,5;\tfrac{5}{2}\}$&$\begin{gathered}c^{(4)}_3 =(a^{(2)}_{3/2})^2\\c^{(6)}_4 =2a^{(2)}_{3/2}a^{(4)}_{5/2}\\c^{(8)}_5=(a^{(4)}_{5/2})^2\end{gathered}$&none&$(244,\tfrac{485}{2})$\\
\hline
$\begin{matrix} \includegraphics[width=16pt]{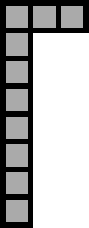}\end{matrix}(\text{ns})$&$([3,1^8],\mathbb{Z}_2)$&$\{1,3,3,3,3;\tfrac{3}{2}\}$&$-$&${SU(2)}_5$&$(177,\tfrac{351}{2})$\\
\hline 
$\begin{matrix}\includegraphics[width=11pt]{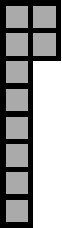}\end{matrix}$&$[3,1^8]$&$\{1,3,3,3,3;\tfrac{3}{2}\}$&$c^{(4)}_3=(a^{(2)}_{3/2})^2$&none&$(172,\tfrac{343}{2})$\\
\hline 
$\begin{matrix}\includegraphics[width=6pt]{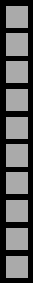}\end{matrix}$&$[1^{11}]$&$\{1,1,1,1,1;\tfrac{1}{2}\}$&$-$&none&$(60,\tfrac{121}{2})$\\
\hline 
\end{longtable}
}
\end{appendices}

\bibliographystyle{utphys}
%\small\baselineskip=.93\baselineskip
%\let\bbb\bibitem\def\bibitem{\itemsep1pt\bbb}
\bibliography{ref}

\end{document}